\documentclass[a4paper, amsfonts, amssymb, amsmath, reprint, showkeys, nofootinbib, twoside, prd]{revtex4-2}
\usepackage[english]{babel}
\usepackage[utf8]{inputenc}
\usepackage{graphicx}
\usepackage{bbm}
\usepackage{appendix}
\usepackage{amsthm}
\usepackage{mathtools}
\usepackage{physics}
\usepackage{xcolor}
\usepackage{graphicx}
\usepackage[left=23mm,right=13mm,top=35mm,columnsep=15pt]{geometry} 
\usepackage{adjustbox}
\usepackage{placeins}
\usepackage[T1]{fontenc}
\usepackage{lipsum}
\usepackage{csquotes}
\usepackage{color}
\usepackage{braket}
\usepackage{amsmath}
\usepackage{float}
\usepackage{hyperref}
\usepackage[caption=false]{subfig}

\begin{document}

\title{Holographic boundary states and dimensionally-reduced braneworld spacetimes}

\author{Stefano Antonini}
    \email[Correspondence email address: ]{santonin@umd.edu}
    \affiliation{Maryland Center for Fundamental Physics and Department of Physics, University of Maryland, College Park MD 20742, USA}
\author{Brian Swingle}
    \affiliation{Brandeis University, Waltham, MA 02453, USA}
    \affiliation{University of Maryland, College Park MD 20742, USA}

\date{\today} 

\begin{abstract}
Recently it was proposed that microscopic models of braneworld cosmology could be realized in the context of AdS/CFT using black hole microstates containing an end-of-the-world brane. Motivated by a desire to establish the microscopic existence of such microstates, which so far have been discussed primarily in bottom-up models, we have studied similar microstates in a simpler version of AdS/CFT. On one side, we define and study boundary states in the charged Sachdev-Ye-Kitaev model and show that these states typically look thermal with a certain pattern of symmetry breaking. On the other side, we study the dimensional reduction of microstates in Einstein-Maxwell theory featuring an end-of-the-world brane and show that they have an equivalent description in terms of 2D Jackiw-Teitelboim gravity coupled to an end-of-the-world particle. In particular, the same pattern of symmetry breaking is realized in both sides of the proposed duality. These results give significant evidence that such black hole microstates have a sensible microscopic realization.  
\end{abstract}


\maketitle

\section{Introduction}






The Anti-de Sitter/Conformal Field Theory (AdS/CFT) correspondence \cite{Maldacena:1997re,Witten:1998qj,Gubser:1998bc,aharony} is one of the most promising frameworks for the formulation of a quantum theory of gravity. As a realization of the holographic principle \cite{thooft,susskind}, it relates the degrees of freedom of a quantum gravity theory in $d+1$ dimensions to those of a $d$-dimensional conformal field theory living on the conformal boundary of the spacetime manifold. However, while AdS/CFT provides a powerful microscopic description of certain classes of spacetimes, an important open question in quantum gravity is how to describe cosmological universes such as the phase of cosmic expansion we observe in the sky.

Inspired by the AdS/BCFT proposal \cite{Takayanagi:2011zk,Fujita:2011fp,karch}, one possible answer was provided by the holographic braneworld cosmology models recently proposed in \cite{Cooper:2018cmb,Antonini:2019qkt}. These bottom-up models describe an asymptotically $AdS$ black hole spacetime with the second asymptotic region cut off by a spherically symmetric, constant tension, codimension-1 end-of-the-world (ETW) brane. The fields satisfy Neumann boundary conditions on the brane, which emerges from the past horizon and falls into the future horizon (in Lorentzian signature). From the point of view of a comoving observer, the radial motion of the brane is equivalent to the evolution of a closed FLRW universe. When the brane is far from the black hole horizon, gravity is also localized on the brane \cite{Antonini:2019qkt} via a Randall-Sundrum mechanism \cite{RS1,RS2,Karch:2000ct}, and the setup describes, to a good approximation, a cosmological model with expansion and contraction. In the Einstein-Maxwell model of \cite{Antonini:2019qkt}, this only happened for large near-extremal charged black holes and near-critical branes. One possible microscopic realization of such models has been recently proposed by Van Raamsdonk in the context of $\mathcal{N}=4$ super-Yang-Mills (SYM) theory \cite{VanRaamsdonk:2021qgv}. 

Similar black hole microstates have played an important role in recent discussions of the black hole information paradox \cite{Almheiri:2019qdq,Penington:2019kki}. They also provide toy models of the physics of non-unitary dynamics, which have recently received attention in the quantum statistical mechanics community in the context of quantum dynamics in the presence of random measurements~\cite{PhysRevB.98.205136,PhysRevX.9.031009,PhysRevB.99.224307}. Here the non-unitary dynamics is a Euclidean evolution starting at an initial state that can be thought of as the outcome a random measurement. It is therefore of considerable interest to better understand the microscopic underpinnings of such black hole microstates. In particular, we would like to understand under what conditions various bottom-up constructions can be realized in a full microscopic model.

With the cosmological motivation foremost in our thoughts, the goal of the present paper is to use a special case of AdS/CFT duality to relate the near-horizon physics of the braneworld cosmology model proposed in \cite{Antonini:2019qkt}, which involves a near-extremal AdS-Reissner Nordstr\"om black hole (AdS-RN), with specific boundary states in the charged Sachdev-Ye-Kitaev (cSYK) model. In a similar fashion to the Majorana SYK construction by Kourkoulou and Maldacena \cite{Kourkoulou:2017zaj}, the boundary states considered are the fermionic Fock states, evolved for an amount $\tau_0$ of Euclidean time, called the \textit{preparation time}. The boundary states can be prepared by a Euclidean path integral with a boundary condition at Euclidean time $\tau=-\tau_0$ \cite{Takayanagi:2011zk,Fujita:2011fp,Cooper:2018cmb,Antonini:2019qkt,Almheiri:2018ijj,VanRaamsdonk:2020tlr}. Here we continue to work with Einstein-Maxwell gravity as in \cite{Antonini:2019qkt}, although the braneworld states we consider are in a different parameter regime than the one with good gravity localization (near the horizon versus far from the black hole, respectively). The matching we exhibit between braneworld spacetimes and microscopic boundary states is a step towards exhibiting a complete microscopic description of a cosmological spacetime.



We work in the context of the recent but already well-studied example of AdS/CFT duality known as nearly-$AdS_2$/$CFT_1$ duality~\cite{Maldacena:2016hyu,Maldacena:2016upp,Sarosi:2017ykf}. The particular setup we consider involves Jackiw-Teitelboim (JT) gravity \cite{Jackiw:1984je,Teitelboim:1983ux} and the low-energy limit of the Sachdev-Ye-Kitaev (SYK) model \cite{Sachdev:1992fk,kitaev}. JT gravity is a $1+1$-dimensional dilaton-gravity theory whose saddle point solution involves an $AdS_2$ bulk spacetime with a dynamical boundary. The leading order fluctuations above the saddle point are governed by a boundary Schwarzian effective action \cite{Maldacena:2016upp}. JT gravity gives a good approximation to the physics near the horizon of a higher dimensional, near-extremal charged black hole \cite{Maldacena:2016upp,Sarosi:2017ykf,Sachdev:2019bjn,Brown:2018bms,Nayak:2018qej,Moitra:2018jqs,Gaikwad:2018dfc}. In particular, the Schwarzian fluctuations account for the deviation of the near-horizon geometry from $AdS_2\times S^{d-1}$ (for a $(d+1)$-dimensional black hole). For this reason, JT gravity is sometimes referred to as nearly-$AdS_2$ gravity. 


The SYK model is a $(0+1)$-dimensional quantum mechanical model of $N$ interactive Majorana fermions. In the low-energy limit it develops an emergent conformal symmetry under arbitrary time diffeomorphisms Diff($\mathbb{R}$) (or Diff($S^1$) in the thermal case). Such symmetry is spontaneously broken down to $\textrm{SL}(2,\mathbb{R})$ by the saddle point solution, which is exact in the large $N$ limit. This breaking generates an infinite number of degenerate zero modes. The symmetry is also explicitly broken down to $\textrm{SL}(2,\mathbb{R})$ by a Schwarzian effective action at low but non-vanishing temperatures, which lifts the degeneracy of the zero modes. The residual $\textrm{SL}(2,\mathbb{R})$ symmetry can be regarded as a gauge symmetry \cite{Maldacena:2016hyu}. These features mimic exactly the properties of JT gravity \cite{Maldacena:2016upp}, making the duality between low-energy SYK in the large $N$ limit and JT gravity manifest (at least near its saddle point; for a non-perturbative treatment see for example \cite{Saad:2018bqo,Saad:2019lba,Johnson:2019eik,Johnson:2020exp}). The SYK model can be generalized to higher dimensions \cite{stanford} and to include continuous global symmetries \cite{Ferrari:2019ogc,sahoo,Sachdev:2015efa,sachdev,Chaturvedi:2018uov,Liu:2019niv,Davison:2016ngz}. A model of particular interest for our purposes is the SYK model with complex fermions (cSYK) \cite{Sachdev:2015efa,sachdev,Chaturvedi:2018uov,Liu:2019niv,Davison:2016ngz}, which possesses a global $U(1)$ symmetry. This is the correct model to describe the near-horizon physics of a dimensionally reduced near-extremal black hole. Indeed, unlike Majorana SYK, it can capture also charge fluctuations and describe charged bulk fermions \cite{Sachdev:2015efa,sachdev}.


The main advantage of this model is that the duality can be realized explicitly in a simple setup. In particular, the cSYK boundary states can be constructed and studied. In analogy with the Majorana SYK case \cite{Kourkoulou:2017zaj}, we show that cSYK boundary states look thermal with inverse temperature $\beta=2\tau_0$ and with properties analogous to those described by the canonical ensemble for a fixed charge superselection sector (for a treatment of the nearly-$AdS_2$/$CFT_1$ duality in the grand-canonical ensemble, including charge fluctuations, see for instance \cite{Sachdev:2019bjn}). In particular, at large $N$ and when the preparation time is sufficiently large, the dynamics is governed by a Schwarzian effective action for the time reparametrization mode. Differently from the thermal case, the mode must satisfy appropriate boundary conditions, which are dictated by the total occupation number of the Fock state chosen (i.e. by the charge subsector considered). Such boundary conditions break two of the three generators of the residual $\textrm{SL}(2,\mathbb{R})$ symmetry. 

We also study the other side of the duality by performing a dimensional reduction of the braneworld cosmology model \cite{Antonini:2019qkt}\footnote{Note that such model is also formulated in the canonical fixed-charge ensemble.}. The near-horizon region is well approximated by JT gravity with a regularized boundary. We then show that configurations where the brane is tangent to the regularized boundary and the Schwarzian fluctuations of the boundary preserve such intersection point exhibit exactly the same pattern of symmetry breaking that we observe in the cSYK boundary states. 

These results provide strong evidence for the duality between asymptotically-AdS black holes containing an ETW brane and pure boundary states in a dual quantum mechanical theory, thereby realizing explicitly a simplified version of the holographic braneworld cosmology proposals \cite{Cooper:2018cmb,Antonini:2019qkt}. However, we remark that this simple example is not directly relevant for the description of cosmology. Indeed, it describes a setup where the brane is in the near-horizon region of a large AdS-RN black hole, and therefore gravity is not localized on the brane \cite{Antonini:2019qkt}. The holographic description of spacetimes with branes that are far from the black hole horizon and therefore can localize gravity likely requires more complicated boundary states of higher dimensional CFTs such as $\mathcal{N}=4$ SYM \cite{VanRaamsdonk:2021qgv}.

The paper is organized as follows. In Section \ref{section2} the relevant features of the cSYK model are quickly reviewed. In Section \ref{section3} the boundary states are defined in analogy with \cite{Kourkoulou:2017zaj}. We study their properties, emphasizing their similarities with the thermal state using both analytic arguments and numerical evidence. Using the collective field formulation of the path integral, the Schwarzian effective action governing the dynamics of the time reparametrization mode is derived, and we find the two boundary conditions that the mode needs to satisfy, commenting on the resulting symmetry breaking pattern. In Section \ref{section4} we briefly review the bottom-up holographic braneworld cosmology model proposed in \cite{Antonini:2019qkt}. Section \ref{section5} is devoted to the dimensional reduction of such model. The Schwarzian effective action is derived along with an equation of motion describing the trajectory of the brane. We show that when the brane is tangent to the regularized boundary and such property is preserved under Schwarzian fluctuations of the boundary, the system displays a symmetry breaking pattern analogous to that found for the cSYK boundary states. In Section \ref{section6} we discuss our findings and give final remarks. Our choice of units is $\hbar=c=\varepsilon_0=k_B=1$.

\section{Review of the cSYK model}
\label{section2}

We consider the SYK model \cite{sachdev,Ferrari:2019ogc,Liu:2019niv,Davison:2016ngz} with $N$ complex fermions, with Hamiltonian
\small
\begin{equation}
    \hat{H}=\sum_{\substack{j_1<...<j_{q/2}=1\\k_1<...<k_{q/2}=1}}^NJ_{j_1...j_{q/2},k_1...k_{q/2}}\mathcal{A}\left(\hat{\psi}_{j_1}^\dag...\hat{\psi}_{j_{q/2}}^\dag\hat{\psi}_{k_1}...\hat{\psi}_{k_{q/2}}\right),
\label{hamiltonian}
\end{equation}
\normalsize
where $\mathcal{A}$ denotes the completely anti-symmetrized product\footnote{The completely anti-symmetric definition of the Hamiltonian implies a particle-hole symmetry \cite{sachdev}, which is not important for our purposes but simplifies the analysis.} and the fermions satisfy the anti-commutation relations $\{\hat{\psi}_i,\hat{\psi}^\dag_j\}=\delta_{ij}$, $\{\hat{\psi}_i,\hat{\psi}_j\}=0=\{\hat{\psi}^\dag_i,\hat{\psi}^\dag_j\}$. The couplings $J_{j_1j_2,k_1k_2}$ are independently drawn from a complex Gaussian distribution with zero mean and variance
\begin{equation}
\overline{\Big |J_{j_1...j_{q/2},k_1...k_{q/2}}\Big |^2}=J^2\frac{(q/2)!(q/2-1)!}{N^{q-1}}
\end{equation}
and satisfy the symmetry properties $J_{ij,kl}=-J_{ji,kl}=-J_{ij,lk}=J^*_{kl,ij}$.

The cSYK model possesses a globally conserved $U(1)$ charge $\hat{Q}=\sum_j\mathcal{A}\left(\hat{\psi}^\dag_j\hat{\psi}_j\right)=\sum_{j}\hat{\psi}^\dag_j\hat{\psi}_j-\frac{N}{2}\hat{\mathbbm{1}}$, with $\hat{\mathbbm{1}}$ the identity matrix. Since $[\hat{Q},\hat{H}]=0$, it is possible to find a basis of energy eigenstates that diagonalizes the charge. The Hilbert space splits into superselection sectors labeled by their charge $\mathcal{Q}$. For even $N$, $\mathcal{Q}$ is an integer number satisfying $-N/2<\mathcal{Q}<N/2$. Each charge subsector admits a basis of energy eigenstates, and the ground state is in the $\mathcal{Q}=0$ subsector \cite{sahoo}. 

For $q=4$, every energy eigenstate with $\mathcal{Q}\neq 0$ is twofold degenerate due to the particle-hole symmetry: for every energy eigenstate in the $\mathcal{Q}=\bar{\mathcal{Q}}$ charge subsector, there is an energy eigenstate with the same energy eigenvalue in the $\mathcal{Q}=-\bar{\mathcal{Q}}$ charge subsector. Energy eigenstates with $\mathcal{Q}=0$ are non-degenerate. 

The two-point function over a generic state $\ket{\psi}$ is defined as
\begin{equation}
    G(\tau_1,\tau_2)=-\frac{1}{N}\sum_{i=1}^N\braket{\psi|T\left[\hat{\psi}_i(\tau_1)\hat{\psi}_i^\dag (\tau_2)\right]|\psi}
    \label{2pfdef}
\end{equation}
where $T$ is the fermionic time-ordered product, and therefore $G(\tau,\tau\pm 0^+)=\braket{\hat{Q}}/N\pm 1/2$, where $\braket{\hat{Q}}$ is the expectation value of the charge in the state $|\psi\rangle$. In a given charge subsector, this will be just the charge $\mathcal{Q}$ of the subsector. Note that the two-point function presents an ultraviolet (UV) asymmetry. The ``off-diagonal'' two-point correlators involving the product of two different fermions are suppressed by a power of $1/N$ \cite{Maldacena:2016hyu,stanford}, as follows from a diagrammatic calculation of the variance. 

In particular, we can consider a thermal state in the canonical ensemble at fixed charge\footnote{Since the braneworld cosmologies we consider are saddle point solutions that dominate in the gravitational canonical ensemble at fixed charge, the canonical partition function is the one relevant for our simple model. But in general charge fluctuations can be included in the analysis by considering the grand-canonical partition function for a fixed chemical potential $\mu$ \cite{sachdev}.}. Then after averaging over the random couplings, the partition function is encoded in an effective action for two bi-local collective fields $(G,\Sigma)$ \cite{Maldacena:2016hyu,sachdev}. On-shell, the $G$ field is the Green's function defined in equation (\ref{2pfdef}) with the state taken to be thermal, while $\Sigma$ is the self-energy. The Euclidean action is given by 
\vspace{0.2cm}
\begin{widetext}
\begin{equation}
    \frac{S_E}{N}=-\log \det \left[\delta(\tau_1-\tau_2)\partial_{\tau_1}-\Sigma(\tau_1,\tau_2)\right]-\int d\tau_1 d\tau_2\left[\Sigma(\tau_1,\tau_2)G(\tau_2,\tau_1)+\frac{J^2}{q}\left(-G(\tau_1,\tau_2)G(\tau_2,\tau_1)\right)^{\frac{q}{2}}\right].
        \label{gsigmaaction}
    \end{equation}
\end{widetext}
Note that in the large $N$ limit the saddle point solution fully determines the partition function with fluctuations suppressed by $1/N$. In the infrared (IR), the first term in the determinant can be dropped and the action develops a symmetry under arbitrary time reparametrizations $\tau\to f(\tau)$, analogously to the Majorana SYK model. In the complex SYK case, there is an additional emergent local $U(1)$ symmetry. The IR saddle point solutions for the collective fields transform conformally under time reparametrization. The general transformation law also includes a local phase shift:
\small
\begin{equation}
    \begin{split}
        &G_{f,\lambda}(\tau_1,\tau_2)=\left[f'(\tau_1)f'(\tau_2)\right]^\Delta e^{i\left[\lambda(\tau_1)-\lambda(\tau_2)\right]} G(f(\tau_1),f(\tau_2)),\\[5pt]
        &\Sigma_{f,\lambda}(\tau_1,\tau_2)= \left[f'(\tau_1)f'(\tau_2)\right]^{1-\Delta} e^{i\left[\lambda(\tau_1)-\lambda(\tau_2)\right]} \Sigma(f(\tau_1),f(\tau_2)).
        \end{split}
        \label{conftransf}
\end{equation}
\normalsize
The conformal IR (zero temperature) solution for the Green's function $G$ is given by
\begin{equation}
    G_{\beta=\infty}(\pm \tau)=\mp \textrm{e}^{\pm\pi\mathcal{E}}b^\Delta\tau^{-2\Delta}, \hspace{1cm} N\gg J\tau\gg 1
    \label{longtime2pf}
\end{equation}
where $\tau=\tau_1-\tau_2$, $\beta$ is the inverse temperature and $\mathcal{E}$ is the asymmetry parameter, which is related to the average charge density \cite{sachdev}. The finite temperature solution in the $N\gg \beta J>\tau J\gg 1$, $(\beta-\tau)J\gg 1$ limit can be found by a time reparametrization $\tau\to \tan(\pi\tau/\beta)$ and an appropriate local phase shift $i\lambda(\tau)=-2\pi\mathcal{E}\tau/\beta$:
\begin{equation}
    G_c(\pm\tau)=\mp b^{\Delta}\frac{\textrm{e}^{\pm 2\pi\mathcal{E}\left(\frac{1}{2}-\frac{\tau}{\beta}\right)}}{\left[\frac{\beta}{\pi}\sin\left(\frac{\pi\tau}{\beta}\right)\right]^{2\Delta}}, \hspace{1cm} 0<\tau<\beta.
    \label{conformal2pf}
\end{equation}
Note that this solution is correctly anti-periodic with period $\beta$.

The saddle point solutions are not invariant under the transformation (\ref{conftransf}), meaning that both the symmetries are spontaneously broken. In particular, the $\textrm{Diff}(\mathbb{R})$ (or $\textrm{Diff}(S^1)$ at finite temperature) time reparametrization symmetry is spontaneously broken down to $\textrm{SL}(2,\mathbb{R})$\footnote{Considering only bosonic fields (like $G$), we are not able to tell whether the residual symmetry group is $\textrm{SL}(2,\mathbb{R})$ or $\textrm{PSL}(2,\mathbb{R})=\textrm{SL}(2,\mathbb{R})/\{\pm \hat{\mathbbm{1}}\}$. Since we are interested in general in a bulk theory that contains also fermions, we will consider the residual gauge group to be $\textrm{SL}(2,\mathbb{R})$, which is the appropriate one for a theory with fermions \cite{Stanford:2019vob}, see Appendix \ref{appendixa2}.}, and the local $U(1)$ symmetry is completely broken to global $U(1)$, which is an exact symmetry of the theory (see Appendix \ref{appendixa2} and \cite{Chaturvedi:2018uov} for additional details).

The symmetries are also explicitly broken when including corrections away from the IR action. This means that, given the true saddle point solution $(G_*,\Sigma_*)$, the solutions obtained by the transformation (\ref{conftransf}) have now an action which is lifted by a term of order $1/(\beta J)$. They are ``quasi-solutions'' that dominate the fluctuations around the saddle point at leading order in the temperature. Such dynamics is encoded at large $N$ in a Schwarzian effective action for the time reparametrization mode and an additional action for the phase field $\lambda(\tau)$ \cite{sachdev}, which is conjugate to charge density fluctuations in the grand-canonical ensemble \cite{Davison:2016ngz}. 

The two modes can be decoupled by a change of variables 
\begin{equation}
    \lambda(\tau)\to \lambda(\tau)+i2\pi\mathcal{E}\frac{\tau-\phi(\tau)}{\beta}.
    \label{decoupling}
\end{equation}
In the fixed charge ensemble we are interested in, the dynamics at low temperature and large $N$ is then governed only by the Schwarzian action for the reparametrization mode:
\begin{equation}
    I_{eff}[f]=-\frac{N\gamma}{4\pi^2}\int_0^\beta d\tau\left\{\tan\left(\frac{\pi f(\tau)}{\beta}\right),\tau\right\}
    \label{effactionsyk}
\end{equation}
where now $f(\tau)$ is a monotonic function obeying $f(\tau+\beta)=f(\tau)+\beta$, $\gamma$ is the coefficient of the linear term of the specific heat at fixed charge, and $\{f(\tau),\tau\}$ is the Schwarzian derivative
\begin{equation}
    \left\{f(t),t\right\}\equiv \frac{f'''(t)}{f'(t)}-\frac{3}{2}\left(\frac{f''(t)}{f'(t)}\right)^2.
    \label{schwarziander}
\end{equation}
The Schwarzian action is $\textrm{SL}(2,\mathbb{R})$ invariant. Therefore, the diffeomorphism invariance is both spontaneously and explicitly broken down to $\textrm{SL}(2,\mathbb{R})$. The residual symmetry must be regarded as an unphysical gauge symmetry \cite{Maldacena:2016hyu}: modes that are $\textrm{SL}(2,\mathbb{R})$ equivalent correspond to the same physical configuration and must be accounted for only once in the path integral. In other words, the path integral for the reparametrization mode is carried over modes $f(\tau)$ in the left quotient of $\textrm{Diff}(S^1)$ by $\textrm{SL}(2,\mathbb{R})$. We will show that when the path integral computes amplitudes in a given boundary state, the reparametrization mode must satisfy specific boundary conditions that further break the residual $\textrm{SL}(2,\mathbb{R})$ symmetry, leaving only one unbroken generator. In the next section we will take $q=4$ in the Hamiltonian (\ref{hamiltonian}) for simplicity. Our results should however generalize immediately to any $q$ integer multiple of 4. The generalization to other values of $q$ might present some differences due to the different spectral structure \cite{sahoo}.

\section{cSYK boundary states and symmetry breaking}
\label{section3}

\subsection{Definition of the boundary states}
\label{coeff}

Consider the Fock basis of the cSYK Hilbert space. A Jordan-Wigner transformation allows us to map the fermions $\hat{\psi}_i$ to spin operators. We can then denote a generic state of the Fock basis as
\begin{equation}
    \ket{S_i}=\ket{\uparrow_1\downarrow_2...\uparrow_{N-1}\uparrow_N}.
\end{equation}
These states are evidently eigenstates of both the occupation number operator for a single fermion $\hat{n}_k=\hat{\psi}^\dag_k\hat{\psi}_k$ with eigenvalue 0 or 1, and of the total charge operator $\hat{Q}$ with eigenvalue $\mathcal{Q}=n_{\uparrow,\mathcal{Q}}-N/2$, where $n_{\uparrow,\mathcal{Q}}$ is the total number of fermions in the $\ket{\uparrow}$ state. However they are clearly not energy eigenstates, and therefore their dynamics is non-trivial.

The Euclidean time evolution under the interaction Hamiltonian (\ref{hamiltonian}) maps any given Fock state $\ket{S_i}$ with charge $\mathcal{Q}$ to a superposition of all the Fock states in the same charge subsector, whose dimension is $D_{\mathcal{Q}}=N!/[n_{\uparrow,\mathcal{Q}}!(N-n_{\uparrow,\mathcal{Q}})!]$. Since the Hamiltonian can be diagonalized in each charge subsector, we can write a given Fock state with charge $\mathcal{Q}$ as a linear superposition of all the energy eigenstates in the same charge subsector:
\begin{equation}
    \ket{S_i}=\sum_{k=1}^{D_{\mathcal{Q}}}c^i_k\ket{E_k}.
    \label{superpos}
\end{equation}

From now on, we will assume to be working within a charge subsector with charge $\mathcal{Q}$. The complex coefficients $c_k^i$ must satisfy the normalization condition $\sum_{k=1}^{D_{\mathcal{Q}}}|c^i_k|^2=1$. But given the structure of the Hamiltonian (\ref{hamiltonian}), we can say something more. As we reviewed in the previous section, the couplings $J_{j_1j_2,k_1k_2}$ governing the all-to-all interactions between the fermions are all drawn from the same distribution with zero mean and fixed $1/N^3$-suppressed variance. Given an energy eigenstate $\ket{E_k}=\sum_{i=1}^{D_{\mathcal{Q}}}c_i^k\ket{S_i}$, we then expect the coefficients $c_i^k=(c_k^i)^*$ (where ${}^*$ indicates the complex conjugate) weighting the contribution of each Fock state $\ket{S_i}$ in the superposition to be randomly distributed. In particular, we expect\footnote{Physical intuitions and numerical evidence motivating these claims are reported in Appendix \ref{appendixb1}.}: 
\begin{itemize}
    \item Their phases to be uniformly distributed in $[-\pi,\pi]$;
    \item After averaging over disorder, their squared norm to have mean $\overline{|c_i^k|^2}=1/D_{\mathcal{Q}}$ and variance $\textrm{var}(|c_i^k|^2)=a/D_{\mathcal{Q}}^2+\mathcal{O}(D_{\mathcal{Q}}^{-3})$, where $a>1$ is an order $1$ c-number;
    \item The covariance $\textrm{cov}(|c_i^k|^2,|c_j^k|^2)$ to be suppressed at least by a power $D_{\mathcal{Q}}^{-3}$. Note that it cannot vanish exactly due to the normalization constraint;
    \item The correlation between the coefficients and the energy eigenvalues to be also strongly suppressed.
\end{itemize} 
We emphasize that whenever the charge per particle differs by a non-vanishing amount from $\pm 1/2$ at large $N$, the quantity $D_{\mathcal{Q}}$ is exponential in $N$ and much greater than $N$ at large $N$.

These assumptions are reasonable in part because the model enjoys an enhanced $U(N)$ symmetry at large $N$ arising from the suppressed variance of the random couplings. In particular, such symmetry permutes the various Fock states in a given charge subsector. Note that this is the analog of the emergent $O(N)$ symmetry in the $N$-fermion Majorana SYK model \cite{Kourkoulou:2017zaj}. The presence of the $U(N)$ symmetry simplifies the analysis of the physics in the boundary states, especially when considering collective operators (see Appendices \ref{appendixb1} and \ref{appendixb3}). 

Given the relation (\ref{superpos}) and the properties of the coefficients $c_k^i$ that we have described, we expect Fock states to have energy of order zero. Since the ground state energy of the cSYK model satisfies $E_0\propto -N$ \cite{sachdev}, these are relatively high-energy states in the middle of the spectrum (for a given charge). In order to study the properties of boundary states in the well-understood and analytically treatable IR limit of the cSYK model, we can define our boundary states by evolving Fock states for an amount $\tau_0$ of Euclidean time, called the preparation time:
\begin{equation}
    \ket{B_i}=\frac{\textrm{e}^{-\tau_0 \hat{H}}\ket{S_i}}{\sqrt{\braket{S_i|\textrm{e}^{-2\tau_0 \hat{H}}|S_i}}}=\frac{\sum_{k=1}^{D_{\mathcal{Q}}}c^i_k\textrm{e}^{-\tau_0 E_k}\ket{E_k}}{\sqrt{\sum_{k=1}^{D_{\mathcal{Q}}}|c^i_k|^2\textrm{e}^{-2\tau_0 E_k}}}.
    \label{bdystate}
\end{equation}
If $\tau_0$ is sufficiently large, higher energy eigenstates in the superposition are strongly suppressed and $\ket{B_i}$ is a low-energy state. We remark that, in the large $N$ limit we are interested in, $N$ is the largest parameter in the theory and the spacing between energy eigenvalues is exponentially suppressed in $N$. The typical level spacing at a given energy density and charge is of order $\textrm{e}^{-\text{entropy}}$ which is $1/D_{\mathcal{Q}}$ in the middle of the spectrum. Therefore, the dynamics of the boundary state $\ket{B_i}$ remains substantially different from the one of the ground state, even for large preparation times $\tau_0$ (provided $\tau_0$ does not scale with $N$).

\subsection{Single-fermion two-point function}
\label{singlefermionsection}

We can now study the behavior of the ``diagonal'' two-point correlator of one specific fermion over a boundary state:
\begin{equation}
    \tilde{G}_{i,k}(\tau_1,\tau_2)=-\braket{B_i|T\left[\hat{\psi}_k (\tau_1)\hat{\psi}_k^\dag (\tau_2)\right]|B_i}
\end{equation}
with $\tau_{1,2}\in [-\tau_0,\tau_0]$. Using the definition (\ref{bdystate}), this can be rewritten as
\begin{equation}
    \tilde{G}_{i,k}(\tau_1,\tau_2)=-\frac{\braket{S_i|\textrm{e}^{-\tau_0 \hat{H}}T\left[\hat{\psi}_k (\tau_1)\hat{\psi}_k^\dag (\tau_2)\right]\textrm{e}^{-\tau_0 \hat{H}}|S_i}}{\braket{S_i|\textrm{e}^{-2\tau_0 \hat{H}}|S_i}}.
    \label{2pf1fermion}
\end{equation}

Let us analyze the denominator first. Since the Fock states $\ket{S_i}$ form a basis of the charge subsector, summing over all the Fock states in the subsector we obtain the partition function at fixed charge $\mathcal{Q}$ and temperature $\beta=2\tau_0$:
\begin{equation}
    \sum_{i=1}^{D_{\mathcal{Q}}}\braket{S_i|\textrm{e}^{-2\tau_0 \hat{H}}|S_i}=\Tr_{\mathcal{Q}}\left(\textrm{e}^{-2\tau_0 \hat{H}}\right)=Z_\mathcal{Q}[2\tau_0].
\end{equation}
But given the properties of the coefficients outlined in the previous section, after averaging over the couplings we expect the value of the denominator to be independent of the specific Fock state chosen. Additionally, since the distribution of the norms of the coefficients is peaked around their mean value, we expect this property to be approximately self-averaging. Therefore we get
\begin{equation}
    \braket{S_i|\textrm{e}^{-2\tau_0 \hat{H}}|S_i}=\frac{1}{D_{\mathcal{Q}}}Z_\mathcal{Q}[2\tau_0]+\mathcal{O}(D_{\mathcal{Q}}^{-1}).
    \label{denom}
\end{equation}
Note that the first term is of order $D_{\mathcal{Q}}^0$.
The analysis of the numerator of equation (\ref{2pf1fermion}) is similar in fashion but slightly subtler, and is reported in Appendix \ref{appendixb2}. We can expect that if one of the two fermion operators is inserted at time $\tau\approx\pm \tau_0$, or if the preparation time $\tau_0$ is not long enough, the correlator appearing in the numerator is strongly dependent on the occupation number of the $k$-th fermion (but not on the occupation number of all the other fermions). Conversely, if the preparation time is long enough ($\tau_0 J\gg 1$) and the fermions are both inserted far from the Fock states ($(\tau_0-|\tau_{1,2}|)J\gg 1$), the correlator is approximately independent of all the occupation numbers, up to corrections of order $\exp(-(\tau_0-|\tau_{1,2}|) J)$. A self-averaging property analogous to the one described for the denominator holds also for the numerator.

In general, the correlation function (\ref{2pf1fermion}) takes the form (up to $1/D_{\mathcal{Q}}$-suppressed corrections)
\begin{equation}
\begin{split}
    &\tilde{G}_{i,k}^\uparrow(\tau_1,\tau_2)=-\frac{2N}{2\mathcal{Q}+N}C_{\mathcal{Q},2\tau_0,k}(\tau_1,-\tau_0;-\tau_0+0^+,\tau_2)\\
    &\tilde{G}_{i,k}^\downarrow(\tau_1,\tau_2)=-\frac{2N}{N-2\mathcal{Q}}C_{\mathcal{Q},2\tau_0,k}(\tau_1,-\tau_0+0^+;-\tau_0,\tau_2)
    \end{split}
    \label{2pf4pf}
\end{equation}
where the superscripts $\uparrow$ and $\downarrow$ indicate whether the $k$-th fermion is in the $\ket{\uparrow}$ or $\ket{\downarrow}$ state in the initial Fock state chosen and $C_{\mathcal{Q},\beta,k}(\tau_1,\tau_2;\tau_3,\tau_4)$ is the thermal four-point function at inverse temperature $\beta$, defined as \cite{coleman_2015}
\begin{equation}
    \frac{1}{Z_\mathcal{Q}[\beta]}\Tr_{\mathcal{Q}}\left(\textrm{e}^{-\beta\hat{H}}T\left[\hat{\psi}_k(\tau_1)\hat{\psi}_k(\tau_2)\hat{\psi}_k^\dag(\tau_3)\hat{\psi}_k^\dag(\tau_4)\right]\right).
    \label{therm4pf}
\end{equation}
The result (\ref{2pf4pf}) is valid at leading order in $1/D_\mathcal{Q}$ for any value of $\tau_0$, $\tau_1$ and $\tau_2$.

\begin{figure*}
    \centering
    \subfloat[]{
        \includegraphics[width=0.33\linewidth]{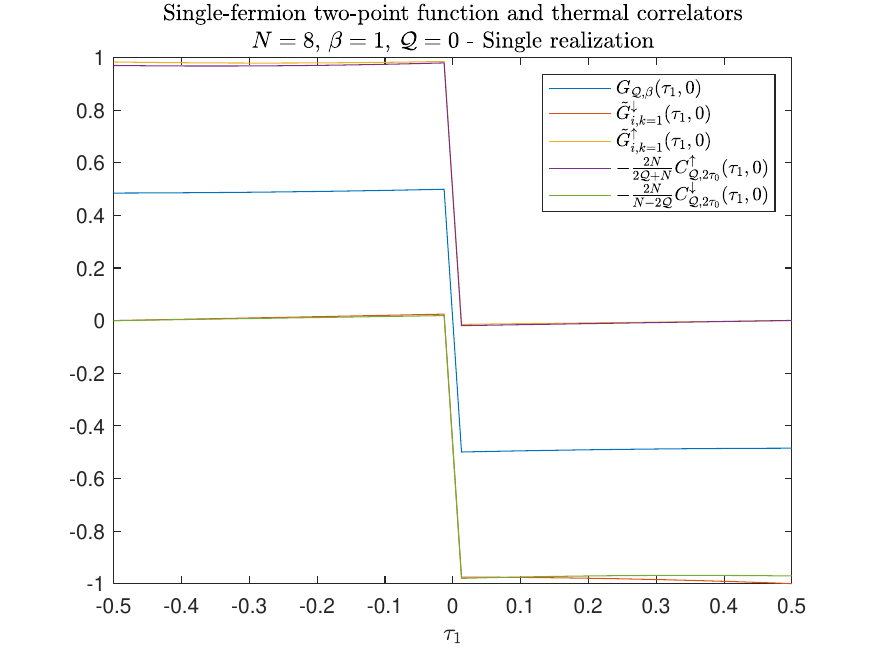}}
    \subfloat[]{
        \includegraphics[width=0.33\linewidth]{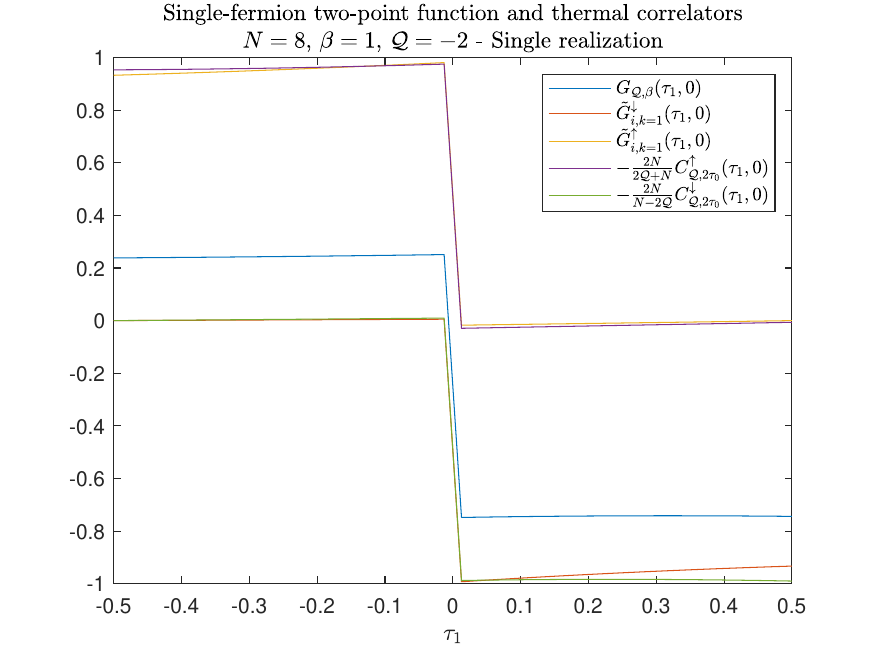}}
    \subfloat[]{
        \includegraphics[width=0.33\linewidth]{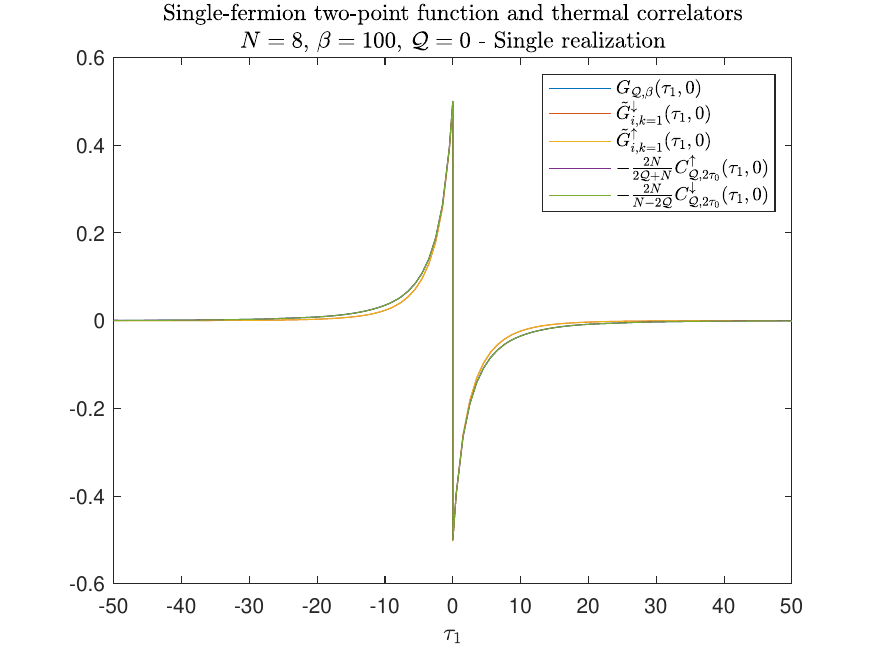}}
    \caption{\textbf{Single fermion two-point functions.} The left hand sides and right hand sides of equation (\ref{2pf4pf}) are plotted along with the thermal two-point function in the corresponding charge subsector for $N=8$, $k=1$ and a single realization of the Hamiltonian. The expectations outlined in Section \ref{singlefermionsection} are matched to good accuracy. Note that for $\tau_1 \approx \tau_2=0$ the thermal two-point function correctly takes the value $G(0^\pm,0)\approx \mathcal{Q}/N\mp 1/2$. (a) $\mathcal{Q}=0$, $\beta=2\tau_0=1$. (b) $\mathcal{Q}=-2$, $\beta=2\tau_0=1$. (c) $\mathcal{Q}=0$, $\beta=2\tau_0=100$.}
    \label{n8single}
\end{figure*}

The properties of the coefficients $c_i^k$ outlined in the previous subsection guarantee that the value of the thermal correlator (\ref{therm4pf}) is independent of $k$ up to corrections suppressed by an inverse power of $D_{\mathcal{Q}}$. Therefore, we can substitute $C_{\mathcal{Q},2\tau_0,k}\to C_{\mathcal{Q},2\tau_0}$, where $C_{\mathcal{Q},2\tau_0}$ is averaged over all the fermions. Applying Wick's theorem at finite temperature \cite{coleman_2015} to the four-point function, and in the limit $\tau_0 J\gg 1$, $(\tau_0-|\tau_{1,2}|)J\gg 1$, we find
\begin{equation}
    \tilde{G}_{i,k}^\uparrow(\tau_1,\tau_2)\approx \tilde{G}_{i,k}^\downarrow(\tau_1,\tau_2)\approx  G_{\mathcal{Q},\beta}(\tau_1,\tau_2)
    \label{2pf1feequivalence}
\end{equation}
where $G_{\mathcal{Q},\beta}(\tau_1,\tau_2)$ is the thermal two-point function at temperature $\beta=2\tau_0$ averaged over all the fermions (which reduces to equation (\ref{conformal2pf}) in the large $N$ and $\tau J\gg 1$ limit). As expected, in the appropriate limit the two-point correlator on the boundary state is independent of the specific Fock state chosen. Note that this result holds for any fermion $k$. The exact diagonalization numerical results reported in Figure \ref{n8single} confirm the predicted behavior\footnote{\label{footnote}Note that for the $N=8$ case we consider in our exact diagonalization results, evolving for a large amount of Euclidean time (for instance in the $\beta=100$ case of Figure \ref{n8single}(c)) means to effectively end up in the ground state of the corresponding charge subsector. Additional details on this point are reported in Appendix \ref{appendixb2}.}. 
Equation (\ref{2pf1feequivalence}) suggests that if the preparation time $\tau_0$ is large and if we measure observables not involving the insertion of a single fermion near the original Fock state, the dynamics in a generic boundary state with charge $\mathcal{Q}$ can be well approximated using the canonical ensemble at fixed charge $\mathcal{Q}$ and temperature $\beta=2\tau_0$. In the next section we will show how this property becomes even stronger by considering a collective two-point correlator, and in Section \ref{thermal} we will give a more general argument to link the dynamics in a boundary state to that one in the canonical ensemble. 

\subsection{Collective two-point function}
\label{collectivesection}

\begin{figure*}
    \centering
        \centering
    \subfloat[]{
        \includegraphics[width=0.33\linewidth]{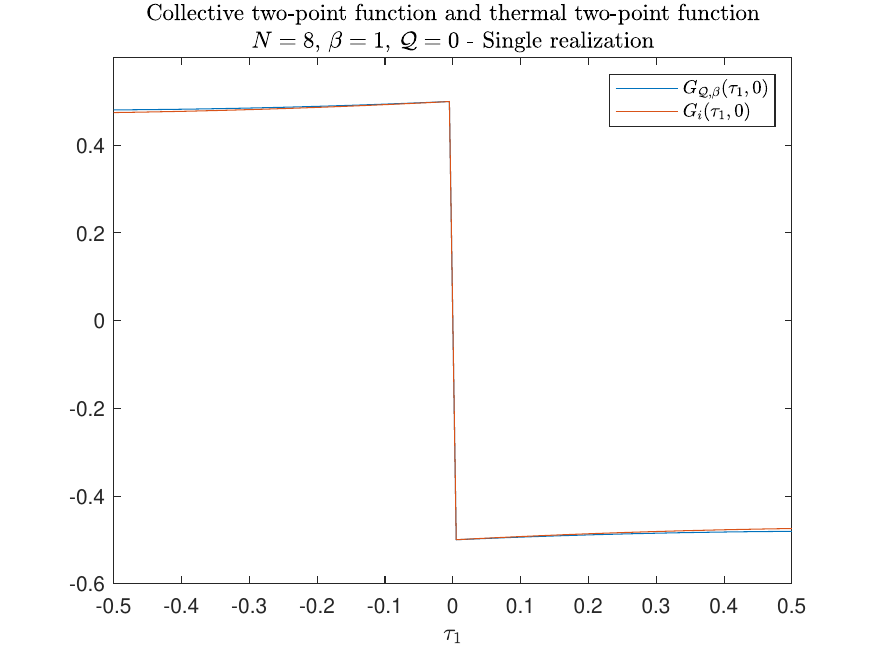}}
    \subfloat[]{
        \includegraphics[width=0.33\linewidth]{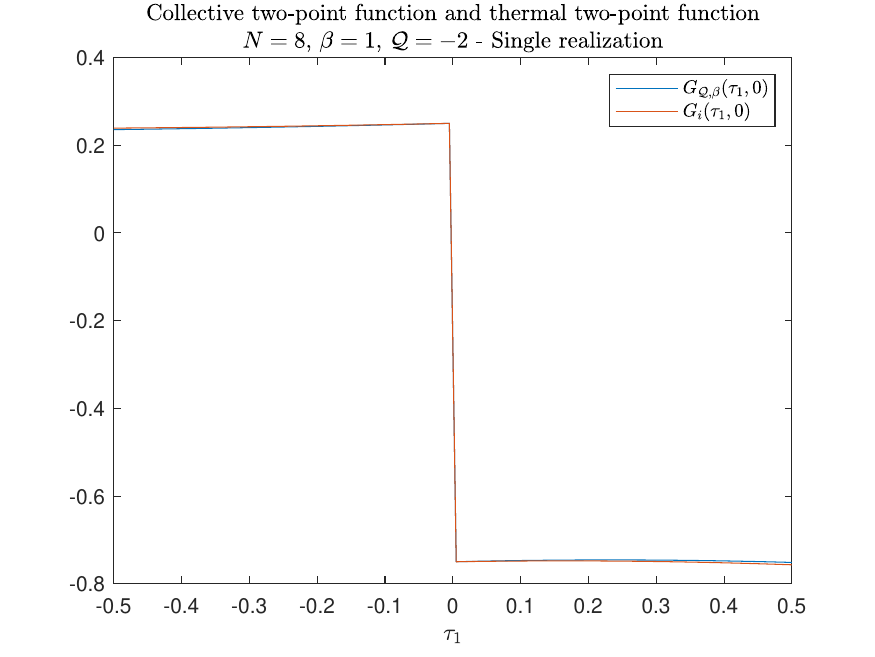}}
    \subfloat[]{
        \includegraphics[width=0.33\linewidth]{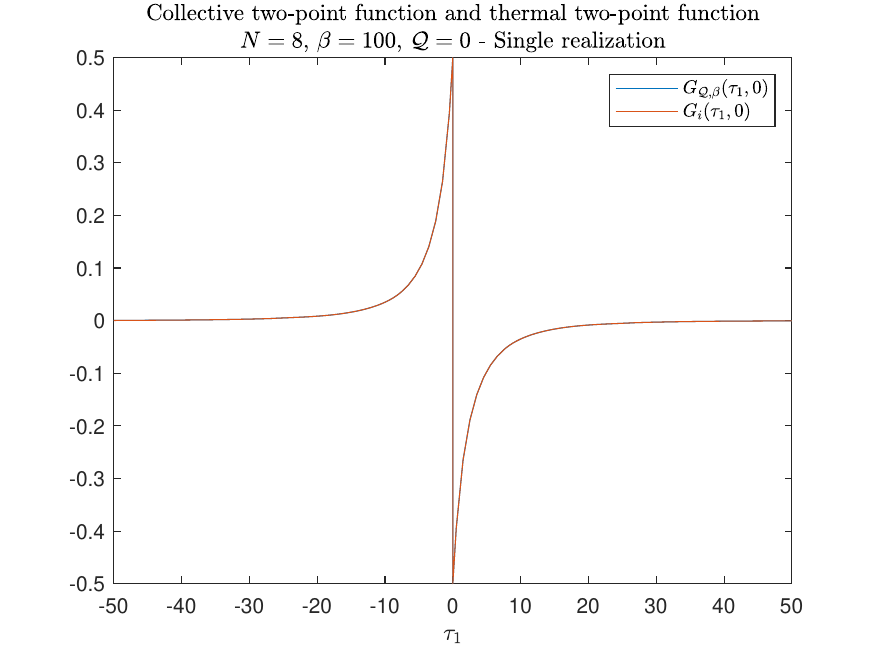}}
    \caption{\textbf{Collective two-point functions.} The left hand side and right hand side of equation (\ref{coll2pf}) are plotted for $N=8$, $k=1$ and a single realization of the Hamiltonian. The expectations outlined in Section \ref{collectivesection} are matched to good accuracy. Note that for $\tau_1 \approx \tau_2=0$ the collective and the thermal two-point functions correctly take the value $G(0^\pm,0)\approx \mathcal{Q}/N\mp 1/2$. (a) $\mathcal{Q}=0$, $\beta=2\tau_0=1$. (b) $\mathcal{Q}=-2$, $\beta=2\tau_0=1$. (c) $\mathcal{Q}=0$, $\beta=2\tau_0=100$.}
    \label{collective8}
\end{figure*}

We are ultimately interested in formulating the cSYK dynamics in a boundary state $\ket{B_i}$ in terms of the collective fields introduced in Section \ref{section2}. Indeed, in the large $N$ and large $\beta J$ limit, they are suitable to make the connection to the dual JT gravity picture. The first natural step towards a collective field formulation is to study the behavior of the collective two-point function over a boundary state:
\begin{equation}
    G_{i}(\tau_1,\tau_2)=-\braket{B_i|T\left[\frac{1}{N}\sum_{k=1}^N\hat{\psi}_k (\tau_1)\hat{\psi}_k^\dag (\tau_2)\right]|B_i}.
    \label{2pfcoll}
\end{equation}
We have already pointed out that, up to corrections suppressed by powers of $1/D_{\mathcal{Q}}$, the value of the single-fermion correlators is independent of the specific fermion $k$ chosen, as long as all the fermions considered have the same occupation number in the initial Fock state (or $\tau_0 J\gg 1$). Therefore at large $N$ we can drop the subscript $k$ from the single-fermion correlator $\tilde{G}_{i}$. Since a Fock state with a given charge $\mathcal{Q}$ has $N/2+\mathcal{Q}$ fermions in the $\ket{\uparrow}$ state and $N/2-\mathcal{Q}$ fermions in the $\ket{\downarrow}$ state, the collective two-point function (\ref{2pfcoll}) then takes the form
\begin{equation}
    G_{i}(\tau_1,\tau_2)\approx\left(\frac{1}{2}+\frac{\mathcal{Q}}{N}\right)\tilde{G}_{i}^\uparrow(\tau_1,\tau_2)+\left(\frac{1}{2}-\frac{\mathcal{Q}}{N}\right)\tilde{G}_{i}^\downarrow(\tau_1,\tau_2).
\end{equation}
Using equation (\ref{2pf4pf}) and Wick's theorem at finite temperature, we immediately obtain
\begin{equation}
    G_{i}(\tau_1,\tau_2)\approx G_{\mathcal{Q},\beta}(\tau_1,\tau_2).
    \label{coll2pf}
\end{equation}
for $\tau_{1,2}\in [-\tau_0,\tau_0]$. Note that, unlike the single-fermion correlator case, for the collective two-point function this result holds regardless of the value of the preparation time $\tau_0$ and of whether the fermions are inserted next to the Fock state. The numerical evidence reported in Figure \ref{collective8} confirms the result (\ref{coll2pf}). 
When both fermions are inserted right next to the bra or the ket, the boundary value of the collective two-point functions is completely determined by its charge:
\begin{equation}
\begin{split}
    &G_i(-\tau_0,-\tau_0^-)=G_i(\tau_0^-,\tau_0)=\frac{N+2\mathcal{Q}}{2N}\\
    &G_i(-\tau_0^-,-\tau_0)=G_i(\tau_0,\tau_0^-)=-\frac{N-2\mathcal{Q}}{2N}
\end{split}
\label{Gconditions}
\end{equation}
where $\tau_0^-=\tau_0-0^+$. In general, if the two fermions are inserted at the same time $\tau$, we obtain $G_i(\tau,\tau+0^+)=(N+2\mathcal{Q})/(2N)$ and $G_i(\tau+0^+,\tau)=-(N-2\mathcal{Q})/(2N)$. When the two fermions are inserted one next to the bra and one next to the ket, at leading order the collective field takes the form
\begin{equation}
\begin{split}
    &G_i(-\tau_0,\tau_0)=\frac{N+2\mathcal{Q}}{2N}\frac{D_\mathcal{Q}}{Z_\mathcal{Q}[2\tau_0]}\frac{Z_{\mathcal{Q}-1}[2\tau_0]}{D_{\mathcal{Q}-1}}\equiv C_\mathcal{Q}^{(1)}\\
    &G_i(\tau_0,-\tau_0)=-\frac{N-2\mathcal{Q}}{2N}\frac{D_\mathcal{Q}}{Z_\mathcal{Q}[2\tau_0]}\frac{Z_{\mathcal{Q}+1}[2\tau_0]}{D_{\mathcal{Q}+1}}\equiv C_\mathcal{Q}^{(2)}.
\end{split}
\label{Gconditions2}
\end{equation}

An important remark is in order here. Equations (\ref{coll2pf}), (\ref{Gconditions}) and (\ref{Gconditions2}) are independent of the specific boundary state $\ket{B_i}$. Therefore at large $N$ the collective two-point function provides information about the charge subsector, but is not able to distinguish between two boundary states within the same charge subsector. In principle, at leading order in $1/N$ it also cannot distinguish a boundary state of the form described above from a boundary state obtained by evolving in Euclidean time a generic superposition of Fock states. Conversely, knowledge about the boundary value of all the single-fermion correlators introduced in the previous subsection is enough to identify the boundary state uniquely. The Schwarzian effective action will be derived in the collective field formulation. Therefore, the Schwarzian sector of our theory describes the physics in any boundary state in a given charge subsector, or in a boundary state built from a generic superposition of Fock states. This implies that the Schwarzian sector of our gravitational bulk theory will be dual to any boundary state of such class in a given charge subsector. From a bulk perspective, in order to tell apart two different boundary states in the same charge subsector we would need to take into account bulk fermions and provide a set of boundary conditions for them on the dimensionally reduced ETW brane. Note that $N$ such fields are naturally present in a bulk model dual to the low-energy limit of the cSYK model, and their limiting value on the regularized boundary represents the source for the cSYK fermions, according to the HKLL dictionary \cite{Lensky:2020ubw}. Additional remarks on this point will be made at the end of Section \ref{section5}. Although interesting, the question of how to implement such boundary conditions is beyond the goals of the present paper.

\subsection{Boundary states look thermal}
\label{thermal}

In this subsection we generalize the results obtained for two-point functions to generic operators $\hat{O}$. In particular, we show that in the large $N$ limit the expectation value of generic operators in a boundary state approaches the thermal expectation value computed by the canonical ensemble at inverse temperature $\beta=2\tau_0$. We will make use of the following assumptions:
\begin{enumerate}
    \item The distribution of the coefficients $c_i^k$ satisfies the properties outlined in Section \ref{coeff}, i.e. the phases are uniformly distributed in $[-\pi,\pi]$, the norms of the coefficients satisfy $\overline{|c_i^k|^2}=1/D_{\mathcal{Q}}$, $\textrm{var}(|c_i^k|^2)=a/D_{\mathcal{Q}}^2+\mathcal{O}\left(D_{\mathcal{Q}}^{-3}\right)$, $\textrm{cov}(|c_i^k|^2,|c_j^k|^2)=\mathcal{O}\left(D_{\mathcal{Q}}^{-3}\right)$ and the correlation between the coefficients and the energy eigenvalues is strongly suppressed.
    \item The matrix elements $\braket{E_i|\hat{O}|E_j}$ of the operator $\hat{O}$, which in general is a non-local operator in time, are weakly correlated with the coefficients $c_i^k$. This implies that the matrix elements $\braket{B_i|\hat{O}|B_i}$ depend weakly on the choice of the initial Fock state $\ket{S_i}$.
    \item The off-diagonal matrix elements are suppressed in the large $N$ limit: $\braket{E_i|\hat{O}|E_j}=\mathcal{O}(\textrm{e}^{-\text{entropy}/2})$, which is $\mathcal{O}(1/\sqrt{D_{\mathcal{Q}}})$ in the center of the spectrum. 
\end{enumerate}
Assumption 1 has been already motivated, and is confirmed by the numerical results reported in Appendix \ref{appendixb1}. Assumption 2  holds up to $1/N$ corrections for any operator $\hat{O}$ which is invariant under permutations of the fermions, or is inserted far from the Fock state at $\tau=\pm\tau_0$. This excludes operators involving the insertion of single fermions at time $\pm\tau_0$. In fact, that would cause $\braket{B_i|\hat{O}|B_i}$ to vanish identically for some boundary states. Such operators can clearly distinguish a boundary state from the thermal state (see Appendix \ref{appendixb3} for further discussion on this point). Assumption 3 is a standard form of the eigenstate thermalization hypothesis (ETH)~\cite{eth1,eth2,eth3}.

Let us now study the behavior of the expectation value of an operator $\hat{O}$ over a boundary state with charge $\mathcal{Q}$: $\braket{\hat{O}}_i\equiv\braket{B_i|\hat{O}|B_i}$. We outline here the reasoning and results of our analysis, while the details of the proof are reported in Appendix \ref{appendixb3}. All the results are true in the large $N$ limit at leading order in $1/D_{\mathcal{Q}}$. Using the definition of the boundary states (\ref{bdystate}), the average over disorder of the expectation value $\braket{\hat{O}}_i$ is equal to the expectation value of the operator $\hat{O}$ in the canonical ensemble at fixed charge $\mathcal{Q}$ and inverse temperature $\beta=2\tau_0$:
\begin{equation}
    \overline{\overline{\braket{O}_i}}= \frac{1}{Z_{\mathcal{Q}}[2\tau_0]}\Tr_{\mathcal{Q}}\left(\textrm{e}^{-2\tau_0\hat{H}}\hat{O}\right)+\mathcal{O}\left(\frac{1}{D_\mathcal{Q}}\right)
    \label{oaverage}
\end{equation}
where we used a double overline to indicate that we averaged over both the phases and the norms of the coefficients $c_k^i$ (see Appendix \ref{appendixb3}).
Therefore, after averaging over the random couplings, every boundary state $\ket{B_i}$ ``looks'' thermal in the large $N$ limit. It is then natural to ask to what extent this property remains valid for a single realization of the Hamiltonian (\ref{hamiltonian}). The answer to this question can be found in the size of the fluctuations around the average value (\ref{oaverage}):
\begin{equation}
    \overline{\overline{\braket{O}_i^2}}-\overline{\overline{\braket{O}_i}}^2=\mathcal{O}\left(\frac{1}{D_\mathcal{Q}}\right).
    \label{fluctuations}
\end{equation}
Equation (\ref{fluctuations}) shows that fluctuations around the average value (\ref{oaverage}) are suppressed in the large $N$ limit. This implies that even for a single realization of the Hamiltonian the expectation value of a generic operator $\hat{O}$ over a boundary state approaches the thermal expectation value computed in the canonical ensemble, up to $1/N$-suppressed corrections. 

The takeaway is clear: unless measurements of very specific operators involving the insertion of single fermions at time $\tau\approx\pm \tau_0$ are performed, at leading order in the $1/N$ expansion it is impossible to discriminate between a boundary state $\ket{B_i}$ and the thermal state. In other words, boundary states look thermal.

\subsection{Path integral, Schwarzian action and symmetry breaking}

Euclidean correlators in a boundary state $\ket{B_i}$ can be computed using the Euclidean path integral with boundary conditions at Euclidean time $\tau=\pm \tau_0$:
\begin{equation}
   \braket{B_i|\hat{O}|B_i}= \int_{\ket{\phi(\pm\tau_0)}=\ket{S_i}}\mathcal{D}\psi\mathcal{D}\bar{\psi}O\textrm{e}^{-S_{SYK}[\psi,\bar{\psi}]}
\end{equation}
where we denoted with $\bar{\psi}$ the Grassmann variables associated with the fermion creation operators $\psi^\dag$. The boundary condition $\ket{\phi(\pm\tau_0)}=\ket{S_i}$ means that the occupation number of each fermion at time $\tau=\pm\tau_0$ is fixed.

It is then possible to average over the random couplings, introduce the collective fields $(G,\Sigma)$ and integrate out the fermions in complete analogy with the usual collective field formulation of the SYK model we reviewed in Section \ref{section2}. We arrive then at a path integral over the collective fields, where the action is given by equation (\ref{gsigmaaction}) and now the boundary conditions $\ket{\phi(\pm \tau_0)}=\ket{S_i}$ become a set of boundary conditions on the collective field $G$, which is completely analogous to the one in equations (\ref{Gconditions}) and (\ref{Gconditions2}).

When the preparation time is large $\tau_0 J\gg 1$ all the IR analysis carried out in Section \ref{section2} is still valid. In particular, the action develops an emergent time reparametrization symmetry which is spontaneously and explicitly broken down to $\textrm{SL}(2,\mathbb{R})$. The leading order fluctuations around the saddle point are still dominated at large $N$ by Schwarzian fluctuations of the reparametrization mode, governed by the effective action (\ref{effactionsyk}) with $\beta=2\tau_0$. Under time reparametrization $\tau\to f(\tau)$ (with $f(\tau)$ monotonic) the $(G,\Sigma)$ fields transform according to equation (\ref{conftransf}), where we can ignore the phase field\footnote{We ignore the phase field because it is conjugate to charge density fluctuations in the grand-canonical ensemble, and we are working at fixed charge $\mathcal{Q}$.} after decoupling it from the reparametrization mode using the transformation (\ref{decoupling}).

There are two differences from the thermal case. The first one is that now the collective fields are defined only for $\tau_{1,2}\in [-\tau_0,\tau_0]$. Therefore, given the transformation law (\ref{conftransf}), every time diffeomorphism $\tau\to f(\tau)$ should map the interval $[-\tau_0,\tau_0]$ to itself. The second one is that the collective field $G$ must satisfy the boundary conditions (\ref{Gconditions}) and (\ref{Gconditions2}). Given the true saddle point solution $G_*$, which satisfies by definition equations (\ref{Gconditions}) and (\ref{Gconditions2}), we can use the transformation law (\ref{conftransf}) to derive two boundary conditions for the time reparametrization mode $f(\tau)$. Time reparametrizations satisfying such boundary conditions will map the true saddle to quasi-solutions $G_f$ that also satisfy equations (\ref{Gconditions}) and (\ref{Gconditions2}).

We will use only the first lines of equations (\ref{Gconditions}) and (\ref{Gconditions2}). It is immediate to check that the boundary conditions for the reparametrization mode we find guarantee that the second lines of equations (\ref{Gconditions}) and (\ref{Gconditions2}) are satisfied as well. First, we can impose the boundary condition
\begin{equation}
    \frac{N+2\mathcal{Q}}{2N}=G_f(\tau_0^-,\tau_0)=\left[f'(\tau_0)\right]^{2\Delta}G_*(f(\tau_0^-),f(\tau_0)).
\end{equation}
Using the monotonicity of $f(\tau)$ and the fact that the true saddle satisfies $G_*(\tau,\tau+0^+)=(N+2\mathcal{Q})/(2N)$ for any $\tau\in [-\tau_0,\tau_0]$, we get $G_*(f(\tau_0^-),f(\tau_0))=(N+2\mathcal{Q})/(2N)$ and therefore $f'(\tau_0)=1$. Repeating the same argument for $G_f(-\tau_0,-\tau_0^-)$ we also obtain $f'(-\tau_0)=1$. Let us now impose the third boundary condition:
%
\begin{equation}
    C_\mathcal{Q}^{(1)}=G_f(-\tau_0,\tau_0)=\left[f'(-\tau_0)f'(\tau_0)\right]^{\Delta}G_*(f(-\tau_0),f(\tau_0)).
\end{equation}
%
Using the first boundary condition $f'(\pm\tau_0)=1$, this reduces to $G_*(f(-\tau_0),f(\tau_0))=C_\mathcal{Q}^{(1)}$. Since $f(\tau_0)\neq f(-\tau_0)$ and we know that $G_*(-\tau_0,\tau_0)=C_\mathcal{Q}^{(1)}$, the only way to satisfy this condition is to impose $f(\pm \tau_0)=\pm \tau_0$. Therefore we arrived at a set of boundary conditions for the time reparametrization mode:
\begin{equation}
    f(\pm\tau_0)=\pm\tau_0, \hspace{1cm} f'(\pm\tau_0)=1.
    \label{repbc}
\end{equation}
The same result was achieved by Kourkoulou and Maldacena in the context of the Majorana SYK model.

Note that, given a time diffeomorphism $f(\tau)$ that does not satisfy the boundary conditions (\ref{repbc}), it is always possible to perform a $\textrm{SL}(2,\mathbb{R})$ transformation to a $\tilde{f}(\tau)$ that satisfies (\ref{repbc}) (see Appendix \ref{appendixa2}). In Section \ref{section2} we emphasized that $\textrm{SL}(2,\mathbb{R})$-equivalent modes must be accounted for only once in the path integral. Therefore, the boundary conditions (\ref{repbc}) do not exclude any reparametrization mode from the path integral. They rather further break two out of the three generators of the residual $\textrm{SL}(2,\mathbb{R})$ symmetry. Note that the same pattern of symmetry breaking is obtained, as we have pointed out, by starting with a generic superposition of Fock states instead of a single one.

A simple calculation yields the only unbroken generator\footnote{We thank Rodrigo A. Silva for pointing this out.}. Consider a $\textrm{SL}(2,\mathbb{R})$ transformation mapping the circle of length $2\tau_0$ to itself:
\begin{equation}
    \textrm{e}^{i\frac{\pi}{\tau_0}f(\tau)}=\textrm{e}^{i\theta}\frac{\textrm{e}^{i\frac{\pi}{\tau_0}\tau}+\alpha}{\alpha^*\textrm{e}^{i\frac{\pi}{\tau_0}\tau}+1}, \hspace{0.5cm} |\alpha|\leq 1
\end{equation}
where ${}^*$ denotes the complex conjugate, $\alpha$ is a complex number and $\theta \in [0,2\pi]$. The infinitesimal transformation ($|\alpha|\ll 1$) is
\begin{equation}
    f(\tau)\approx \tau+ a + b\cos\left(\frac{\pi\tau}{\tau_0}\right)+c\sin\left(\frac{\pi\tau}{\tau_0}\right)
    \label{infinitesimal}
\end{equation}
with $a=\tau_0\theta/\pi$, $b=2\tau_0 \Im(\alpha)/\pi$, $c=-2\tau_0\Re(\alpha)/\pi$. The three generators of the transformation are then $(1,\cos(\pi\tau/\tau_0),\sin(\pi\tau/\tau_0))$. Imposing the boundary conditions (\ref{repbc}) in equation (\ref{infinitesimal}) we find $a=b$, $c=0$: the only $\textrm{SL}(2,\mathbb{R})$ generator unbroken by the boundary conditions is the combination $1+\cos(\pi\tau/\tau_0)$. Note that this generator correctly maps the points $\pm\tau_0$ to themselves. 

Finally, an intuitive picture that will be useful to guide us in the analysis of the gravity dual is that the physics in a boundary state $\ket{B_i}$ is analogous to the physics in the thermal state, but we have a fixed, special point on the thermal circle (which has length $\beta=2\tau_0$) at the identified Euclidean times $\tau=\pm\tau_0$, where boundary conditions for the collective fields in the path integral must be imposed.
\begin{figure}[H]
    \centering
    \includegraphics[width=0.45\textwidth]{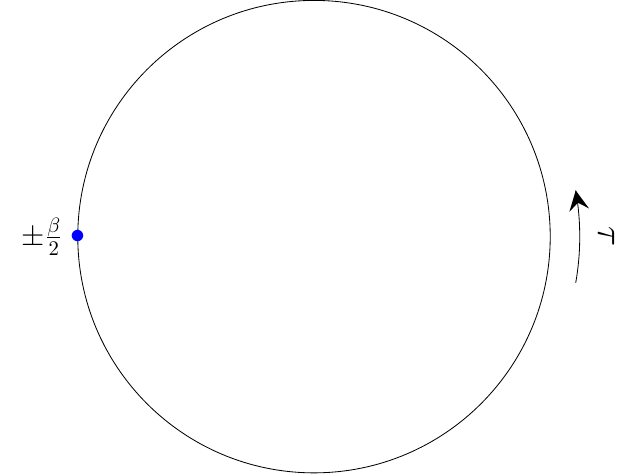}
    \caption{\textbf{Boundary states and canonical ensemble.} The symmetry breaking pattern we encountered when evaluating correlators over a boundary state can be visualized by considering the thermal circle with length $\beta=2\tau_0$ with a fixed, special point at $\tau=\pm\tau_0$. The physics in a boundary state will then be equivalent to the one described by the canonical ensemble at fixed charge in the same charge subsector, except the collective fields must satisfy the appropriate boundary conditions at the special point.}
    \label{sykcartoon}
\end{figure}

\section{Review of holographic braneworld cosmologies}
\label{section4}

In this section we briefly review the main features of the bottom-up holographic braneworld cosmology model built in \cite{Antonini:2019qkt}. The model is a generalization to the AdS-Reissner Nordstr\"om (AdS-RN) case of the AdS-Schwarzschild model originally proposed in \cite{Cooper:2018cmb}. 

The full $(d+1)$-dimensional gravitational Euclidean action (reported in Appendix \ref{appendixc}) is given by $S_g=S_{bulk}+S_{etw}$. $S_{bulk}$ is the Einstein-Maxwell action with a negative cosmological constant, and includes a Gibbons-Hawking-York (GHY) term and an electromagnetic boundary term for the asymptotic boundary, needed when considering the gravitational ensemble at fixed charge \cite{Hawking:1995ap,Chamblin:1999tk}. $S_{etw}$ is the $d$-dimensional action for the constant tension end-of-the-world (ETW) brane, also involving an electromagnetic boundary term. According to the AdS/BCFT prescription \cite{Takayanagi:2011zk,Fujita:2011fp}, all the bulk fields satisfy Neumann boundary conditions on the brane, which possesses a dynamical metric and cuts off the Euclidean geometry. 

The saddle point solution of the bulk action is given by the Euclidean AdS-RN wormhole, whose metric is:
\begin{equation}
    ds^2=F(r)d\tau^2+\frac{dr^2}{F(r)}+r^2d\Omega_{d-1}^2
\end{equation}
where $d\Omega_{d-1}^2$ is the line element of the $(d-1)$-dimensional unit sphere and 
\begin{equation}
    F(r)=1+\frac{r^2}{L_{AdS}^2}-\frac{2\mu}{r^{d-2}}+\frac{Q^2}{r^{2(d-2)}}
\end{equation}
with $L_{AdS}$ AdS radius and $\mu$ and $Q$ mass and charge parameters of the black hole respectively. The metric has two horizons: an outer event horizon at $r=r_+$, and an inner Cauchy horizon at $r=r_-$. Varying the brane action (or, equivalently, imposing Israel junction conditions) yields an equation of motion for the brane
\begin{equation}
    \frac{dr}{d\tau}=\pm\frac{F(r)}{T_{etw}r}\sqrt{F(r)-T_{etw}^2r^2}
\end{equation}
where $T_{etw}\in (0,1/L_{AdS})$ is the tension of the brane and the $\pm$ signs correspond to the expanding and contracting phases of the brane, which reaches its minimum radius $r=r_0$ at $\tau=\pm\beta/2$, where $\beta$ is the inverse temperature of the black hole. The preparation time $\tau_0$ of the dual CFT state can then be computed in terms of bulk parameters by subtracting from the total Euclidean periodicity $\beta$ the total time $2\Delta\tau$ needed for the brane to complete its trajectory from and back to the asymptotic boundary: $\tau_0=(\beta-2\Delta\tau)/2$.

\begin{figure}
    \centering
    \includegraphics[scale=0.7]{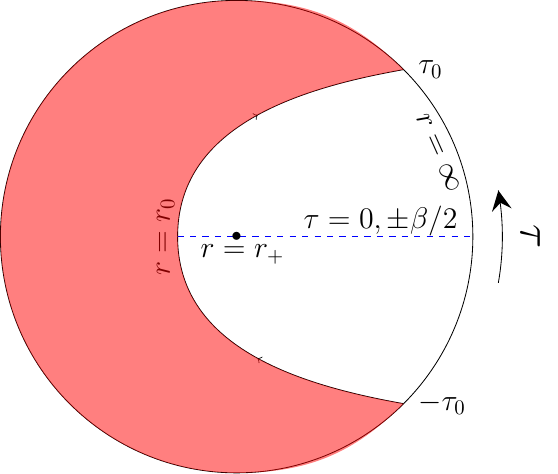}
    \caption{Trajectory of the brane in the Euclidean wormhole. The angular coordinate is the Euclidean time $\tau$ and the radial coordinate is the radius $r$. The central point represents the outer horizon $r=r_+$ and the circumference is the asymptotic AdS boundary. Each point in the diagram is a $(d-1)$-dimensional sphere. The ETW brane contracts from the boundary to a minimum radius $r_0$ at $\tau=\pm\beta/2$, and then expands back to the boundary. The red region is outside the ETW brane and therefore is not part of the geometry.}
    \label{eucltraj}
\end{figure}
\begin{figure}
    \centering
    \includegraphics[scale=0.5]{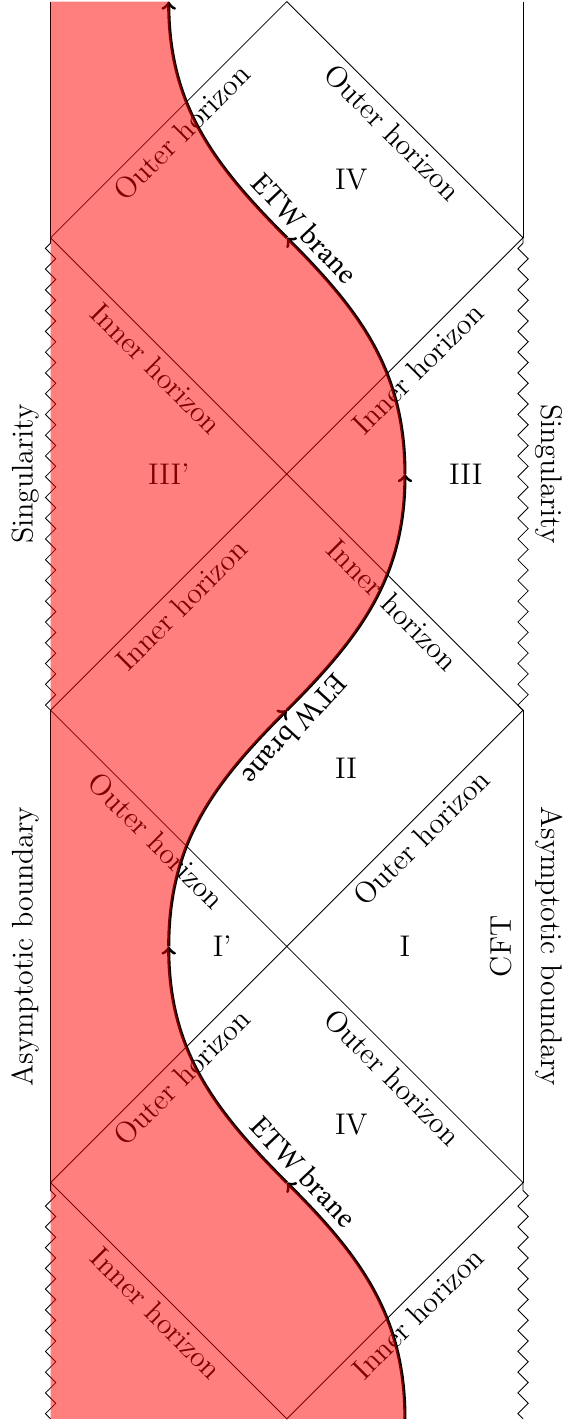}
    \caption{Trajectory of the brane in Lorentzian signature. The brane emerges from the past horizon in the left asymptotic region and collapses into the future horizon. The red region, including the left asymptotic boundary, is cut off by the ETW brane and is not part of the geometry. We glued here more patches of the AdS-RN spacetime, extending the trajectory of the brane. However, the trajectory is reliable only between the intersection points of the brane with the (inner) Cauchy horizon.}
    \label{penrose}
\end{figure}

The $\tau=0,\pm\beta/2$ slice is taken as initial condition for the evolution in Lorentzian time. In other words, the state on the $\tau=0,\pm\beta/2$ slice is prepared by the gravitational Euclidean path integral with appropriate boundary conditions at $\tau=-\tau_0$, and the resulting state is further evolved in Lorentzian time. The corresponding Lorentzian geometry is given by the maximally extended AdS-RN black hole, where the left asymptotic region is cut off by the ETW brane. The minimum radius $r_0$ in Euclidean signature becomes a maximum radius in Lorentzian signature, and the brane emerges from the past event horizon in the left asymptotic region and collapses into the future event horizon. In principle, we can glue multiple patches of the AdS-RN spacetime, and extend farther the brane trajectory. It would then cross the inner horizon, reach a minimum value radius $r_0^-$ and then start expanding again, emerging in a new patch of the universe (see Figure \ref{penrose}). However, the evolution of the brane trajectory after it crosses the Cauchy horizon is not reliable due to the instability of the latter \cite{hovdebo}. Since gravity localization is also efficient only when the brane is far from the black hole horizon \cite{Antonini:2019qkt}, we will focus our attention on the portion of trajectory that resides in the exterior region of the black hole.
In this picture, the whole geometry is dual to a single BCFT with a $U(1)$ global symmetry living on the right asymptotic boundary. The Euclidean version of the CFT lives on $S^{d-1}\times [-\tau_0,\tau_0]$. From the point of view of an observer comoving with the brane, the Lorentzian trajectory of the brane in the black hole spacetime looks like the expansion and contraction of a closed FLRW universe, where the brane radius plays the role of the scale factor and satisfies the Friedmann equation:
\begin{equation}
    \left(\frac{\dot{r}}{r}\right)^2=-\frac{1}{r^2}+\frac{2\mu}{r^d}-\frac{Q^2}{r^{2d-2}}+\left(T_{etw}^2-\frac{1}{L^2_{AdS}}\right)
    \label{friedmann}
\end{equation}
where $\dot{r}$ denotes a derivative with respect to the brane proper time. Although for $d=4$ the motion of the brane resembles the evolution of a cosmological universe, we obtain an effective four-dimensional braneworld cosmology only if an observer living on the brane perceives gravity as effectively four-dimensional and localized on the brane. Gravity localization can be achieved via a Randall-Sundrum II mechanism \cite{RS1,RS2} (see \cite{Karch:2000ct,resonances2} for generalizations relevant to our discussion). In the presence of a bulk black hole, it is localized only locally and when the brane is far from the black hole horizon \cite{resonances2,Antonini:2019qkt}. Therefore, in order to have a portion of the brane trajectory where gravity is localized, we need the maximum radius of the brane $r_0$ (i.e. the minimum radius in Euclidean signature) to be much larger than the black hole event horizon $r_0\gg r_+$. This can be obtained by considering a near-critical brane ($T_{etw}\lesssim 1/L_{AdS}$) and a large black hole ($r_+\gg L_{AdS}$).

In order for this gravitational model to describe a braneworld cosmology and have a BCFT dual, we need two conditions to be satisfied at the same time:
\begin{enumerate}
    \item The preparation time computed in terms of bulk parameters must be positive: $\tau_0>0$;
    \item For part of its trajectory the brane must sit far from the black hole horizon, i.e. $r_0\gg r_+$.
\end{enumerate}
As we have already pointed out, condition 2 guarantees that gravity can be localized on the brane and that we can properly consider the evolution of the brane as the one of a cosmological universe. Condition 1 on the other hand guarantees the existence of a portion $S^{d-1}\times [-\tau_0,\tau_0]$ of the asymptotic Euclidean AdS boundary where to define the dual CFT, and therefore the existence of a CFT dual to our bulk construction. Intuitively, if $\tau_0<0$, the brane in Figure \ref{eucltraj} would overlap itself, leaving no portion of the asymptotic boundary in the geometry. 

When both the conditions are satisfied, the properties of the resulting braneworld cosmology are encoded in the dual BCFT state, and can be extracted in principle by measuring appropriate observables. For example, at early times the Ryu-Takayanagi surface \cite{rt1,rt2} associated to large regions of the CFT ends on the ETW brane, and therefore the entanglement entropy of such regions has a time dependence that probes the brane evolution \cite{Cooper:2018cmb}. This is analogous to what happens in doubly holographic setups involving an evaporating black hole on the brane \cite{Almheiri:2019hni}, and suggests a connection between the braneworld models studied here and the physics of entanglement islands \cite{Almheiri:2019yqk,Penington:2019npb,Penington:2019kki,Almheiri:2019qdq,VanRaamsdonk:2021qgv}. 

The models reviewed in this section provide a framework to describe cosmology in AdS/CFT correspondence and therefore study quantum gravity in a cosmological universe. However, it is worth noting that in the simplest possible realization of this bottom-up proposal, involving an AdS-Schwarzschild black hole and no matter in the bulk, conditions 1 and 2 cannot be simultaneously satisfied: when the brane is far enough to allow gravity localization, it also overlaps itself, violating condition 1 \cite{Cooper:2018cmb}. Including a bulk gauge field, and therefore considering an AdS-RN black hole, the two conditions can be satisfied at the same time, provided that the black hole is large and near-extremal ($r_-\lesssim r_+$). The corresponding saddle point solution is also the dominant one in the thermodynamical ensemble \cite{Antonini:2019qkt}. An alternative possible solution to the ``overlap problem'' was recently proposed in \cite{VanRaamsdonk:2021qgv}, where the bulk theory involves additional fields associated to the dual CFT, which is constructed by coupling two 3-dimensional CFTs (which are joined in the bulk IR by a connected brane) using a 4-dimensional CFT with many less degrees of freedom ($c_{4D}\ll c_{3D}$, with $c$ central charge).

An explicit realization of a model similar to the one described in this section was provided in \cite{Almheiri:2018ijj} in the context of the $AdS_3$/$CFT_2$ duality. A first attempt to build a top-bottom model of the present braneworld construction appeared recently in \cite{VanRaamsdonk:2021qgv}. The CFT is taken to be $\mathcal{N}=4$ SYM theory on a 3-dimensional manifold times an interval, coupled to two superconformal 3-dimensional CFTs with many more degrees of freedom, living at the boundaries of the interval. The bulk dual picture is described by a specific configuration involving brane/antibrane pairs in type IIB string theory.

In the present paper, we are interested in a specific region of the parameter space of the AdS-RN bottom-up braneworld cosmology model \cite{Antonini:2019qkt}. In particular, we are interested in studying configurations where the minimum brane radius $r_0$ (in Euclidean signature) is just outside the horizon of a large, near-extremal black hole $r_0\gtrsim r_+\gtrsim r_-\gg L_{AdS}$. This setup, which we analyze in detail in the next section, is clearly not relevant for cosmology, since local gravity localization can be achieved only when $r_0\gg r_+$. However, it is well known \cite{Maldacena:2016upp,Sarosi:2017ykf,Sachdev:2019bjn,Brown:2018bms,Nayak:2018qej,Moitra:2018jqs} that the physics of the near-horizon region of a near-extremal AdS-RN black hole is well captured by two-dimensional JT gravity, whose semiclassical description is in turn dual to the low-energy limit of the SYK (or cSYK) model. Therefore, when the brane enters the near-horizon region of the higher dimensional black hole, we will be able to obtain by dimensional reduction an effective 2-dimensional description involving an ``end-of-the-world particle'' (a one-dimensional ETW brane) cutting off part of the hyperbolic disk (i.e. Euclidean $AdS_2$). The dynamics in such geometry is governed by JT gravity, and the dual description of specific configurations is given by the cSYK boundary states described in Section \ref{section3}. 

Although this construction does not provide an example of braneworld cosmology in AdS/CFT correspondence, the simplicity of our setup allows us to gain insight about the structure of the duality between black hole spacetimes involving an ETW brane and specific boundary states in a dual holographic quantum mechanical theory.

\section{Bulk dual of cSYK boundary states}
\label{section5}

The geometry of the near-horizon region of a $(d+1)$-dimensional near-extremal AdS-RN black hole approaches $AdS_2\times S^{d-1}$ \cite{Maldacena:2016upp,Sarosi:2017ykf,Sachdev:2019bjn,Brown:2018bms,Nayak:2018qej,Moitra:2018jqs}. At the low-energy scales we are interested in in our analysis, spherical perturbations are dominant \cite{Sachdev:2019bjn} and the dynamics is well captured by an effective two-dimensional theory obtained by dimensional reduction. As usual when performing dimensional reduction, a dilaton field $\phi(\tau,r)$ is introduced, which plays the role of the radius of the $(d-1)$-dimensional sphere $S^{d-1}$ on-shell. Away from the black hole horizon $r=r_+$, there are deviations from the product geometry $AdS_2\times S^{d-1}$, and the dimensionally-reduced model describes nearly-$AdS_2$ gravity. At leading order (which is near to the black hole horizon) such deviations are encoded in the fluctuations of a regularized boundary, and the physics is well approximated by JT gravity \cite{Maldacena:2016upp,Sarosi:2017ykf,Sachdev:2019bjn,Moitra:2019bub}.

In this section, we report the results\footnote{Detailed calculations and additional considerations are reported in Appendix \ref{appendixc}} of the dimensional reduction of the bottom-up holographic braneworld cosmology model reviewed in Section \ref{section4}. We will find that, under appropriate conditions, the resulting theory mimics the properties of the cSYK boundary states we described in Section \ref{section3}, providing evidence for the duality between such boundary states and JT gravity setups involving an ETW particle\footnote{In the late development stages of this work reference \cite{Gao:2021uro} appeared, where the authors explore the semiclassical and quantum properties of JT gravity setups involving ETW branes in Lorentzian signature. The results obtained there are complementary to the ones achieved in the present paper.}.

\subsection{Dimensional reduction of braneworld cosmologies}

We focus our attention on a fixed charge canonical ensemble with charge $\mathcal{Q}$ \cite{Antonini:2019qkt}. Before dimensionally reducing the corresponding braneworld cosmology action, it is useful to perform a change of coordinates:
\begin{equation}
    r\to r_e+\frac{R_2^2}{z}
    \label{varchange}
\end{equation}
where $r_e$ is the horizon radius of the extremal black hole with charge $\mathcal{Q}$ \cite{Antonini:2019qkt} and we conveniently introduced the quantity
\begin{equation}
    R_2=\frac{r_eL_{AdS}}{\sqrt{d(d-1)r_e^2+(d-2)^2L_{AdS}^2}}
    \label{ads2radius}
\end{equation}
\vspace{0.3cm}

\noindent which will be the $AdS_2$ radius of the dimensionally reduced theory. The near-horizon region is then identified by the condition $z\gg R_2^2/r_e$. For the near-extremal case of our interest $r_+\gtrsim r_e$, after a Weyl rescaling the metric of the near-horizon region reads
\begin{equation}
     ds^2\approx \frac{R_2^2}{z^2}\left[\left(1-4\pi T^2 z^2\right)dt^2+\frac{dz^2}{\left(1-4\pi T^2 z^2\right)}\right]+r_e^2d\Omega^2_{d-1}
    \label{nearmetric}
\end{equation}
where $T=1/\beta$ is the Hawking temperature of the black hole and the horizon is at $z_H=1/(2\pi T)$. In the extremal limit $T\to 0$ the two-dimensional metric reduces to Poincar\'e AdS\footnote{Note that in Euclidean signature Poincar\'e coordinates cover the whole hyperbolic disk \cite{Sarosi:2017ykf}.}. The two-dimensional metric in the near-extremal case (\ref{nearmetric}) is still $AdS_2$, and can be mapped to Poincar\'e AdS by a change of coordinates \cite{Sachdev:2019bjn}.

As we have already mentioned, the asymptotic $AdS_2$ boundary must be regularized to the near-horizon region to cut off any large deviation from the $AdS_2\times S^{d-1}$ geometry. When the black hole is large $r_+\gtrsim r_e\gg L_{AdS}$, it is possible to place the regularized boundary in the near-horizon region $z\gg R_2^2/r_e$ and at the same time deep in the asymptotic $AdS_2$ region $z\ll R_2\approx L_{AdS}$ \cite{Nayak:2018qej}. In this region, corrections above extremality are also small if the temperature satisfies $2\pi T\lesssim 1/R_2$. The region where the regularized boundary sits is then defined by the conditions
\begin{equation}
     2\pi T R_2^2\lesssim R_2\approx L_{AdS}\ll\frac{R_2^2}{z}\ll r_e.
    \label{allcond}
\end{equation}
We can now perform the dimensional reduction of the holographic braneworld cosmology action (\cite{Antonini:2019qkt}, and reported in Appendix \ref{appendixc}). Working at fixed charge, we can further integrate out the bulk gauge field \cite{Brown:2018bms} to obtain the following dimensionally reduced effective action describing the physics inside the regularized boundary (the expressions of $S_0$ and $S_0^{etw}$ are reported in Appendix \ref{appendixc}):
\begin{widetext}
\begin{equation}
\begin{split}
    S_{tot}=&S_0+S_0^{etw}
    -\frac{V_{d-1}}{16\pi G}\int_\mathcal{M}d^2x\sqrt{g}\Phi_1\left[R+\frac{2}{R_2^2}\right]-\frac{V_{d-1}}{8\pi G}\int_{\partial\mathcal{M}_\infty}du\sqrt{\gamma_{uu}}\Phi_1K^\infty\\
    &-\frac{V_{d-1}}{8\pi G}\int_{etw}dv\sqrt{h_{vv}}\Phi_1\left[K^{etw}-\frac{d}{2}\Phi_0^{-\frac{d-2}{2(d-1)}}T_{etw}\right]+...
    \end{split}
    \label{JTaction}
\end{equation}
\end{widetext}
$S_0$ is a topological term, $S_0^{etw}$ is a term proportional to the proper length of the trajectory of the ETW particle, $V_{d-1}$ is the volume of the $(d-1)$-dimensional unit sphere, $G$ is the $(d+1)$-dimensional Newton's constant, $\mathcal{M}$ is the two-dimensional spacetime manifold, $g_{ij}$ is the two-dimensional Weyl-rescaled metric, $\Phi_1$ is a redefined dynamical dilaton, $R$ is the two-dimensional Ricci scalar. $\partial\mathcal{M}_{\infty}$ is the regularized boundary, $\gamma_{uu}$ the metric induced on it, and $K_{\infty}$ its extrinsic curvature. Finally, $h_{vv}$ is the metric induced on the ETW particle, $K^{etw}$ its extrinsic curvature and $\Phi_0=r_e^{d-1}$. The last two terms of the first line are the JT gravity action, while the second line is the ETW particle action, and the dots account for higher order corrections away from the near-horizon region.

From the action (\ref{JTaction}) it is clear that the bulk geometry is fixed on-shell to be $AdS_2$, with AdS radius $R_2$. Euclidean $AdS_2$ possesses an asymptotic $\textrm{Diff}(S^1)$ boundary time reparametrization symmetry. Being an emergent asymptotic symmetry, it is spontaneously broken by $AdS_2$, and only a $\textrm{SL}(2,\mathbb{R})$ subgroup remains unbroken \cite{Maldacena:2016upp}. As we will see in the next section, all the dynamics is encoded in the fluctuations of the boundary $\partial \mathcal{M}_\infty$. The leading order fluctuations are described by a Schwarzian action that explicitly breaks the $\textrm{Diff}(S^1)$ symmetry down to $\textrm{SL}(2,\mathbb{R})$.

\subsection{Boundary conditions and Schwarzian action}

The action (\ref{JTaction}) must be supported by a set of boundary conditions for the metric and the dilaton at the regularized boundary and at the location of the ETW particle. At the regularized boundary, we impose the Dirichlet-Dirichlet (DD) boundary conditions usually implemented in JT gravity (see \cite{Goel:2020yxl} for an analysis of the possible boundary conditions in JT gravity):
\begin{equation}
    \gamma_{uu}=\frac{1}{\varepsilon^2}, \hspace{2cm} \Phi_1|_{\partial\mathcal{M}_\infty}=\frac{\phi_b}{\varepsilon}
    \label{ddbc}
\end{equation}
where $\varepsilon$ is a small cutoff parameter determining the location of the regularized boundary (whose total proper length is set to be $L=\beta/\varepsilon$) and $\phi_b$ is a constant. On the other hand, according to the AdS/BCFT prescription \cite{Takayanagi:2011zk,Fujita:2011fp} and to the higher dimensional braneworld cosmology model reviewed in Section \ref{section4}, we must impose Neumann-Neumann (NN) boundary conditions on the ETW particle. They are given by (a derivation of these boundary conditions is given in Appendix \ref{appendixd})
\begin{equation}
        K^{etw}-\frac{d}{2}r_e^{-\frac{d-2}{2}}T_{etw}=0
        \label{nnbcmet}
        \end{equation}
        \begin{equation}
        n_{etw}^i\partial_i\Phi_1-\Phi_1 K^{etw}-(d-1)r_e^{\frac{d}{2}}T_{etw}=0.
    \label{NNbcdil}
\end{equation}
We can now obtain an effective action governing the dynamics of the gravitational theory by integrating out the dilaton field $\Phi_1$, evaluating the action on-shell and imposing the boundary conditions (\ref{ddbc}), (\ref{nnbcmet}) and (\ref{NNbcdil}). This fixes the bulk geometry to be $AdS_2$, with the metric given by the two-dimensional part of equation (\ref{nearmetric}). 

\begin{figure}
    \centering
    \includegraphics[width=0.45\textwidth]{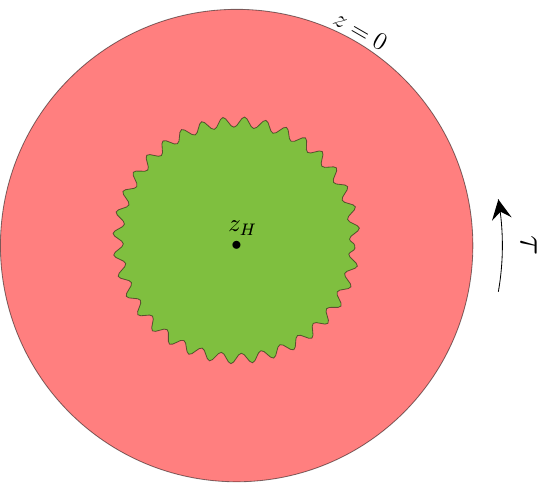}
    \caption{\textbf{Time reparametrization and $AdS_2$ chunks.} Different time reparametrizations $\tau(u)$ correspond to different shapes of the regularized boundary in the rigid $AdS_2$ bulk. The spacetime region described by our effective action is the one enclosed by the regularized boundary, and shaded in green. Chunks of $AdS_2$ of different shapes are weighted differently by the Schwarzian action in the path integral.}
    \label{JT1cartoon}
\end{figure}

The only dynamical term left in the effective action comes from the GHY term for the regularized boundary $\partial\mathcal{M}_\infty$. Parametrizing the boundary with the boundary proper time $u$, the location of the boundary in the fixed Euclidean $AdS_2$ bulk is given by $(\tau(u),z(u))$, with $\tau$ bulk time coordinate. Up to third order in $\varepsilon$, the boundary conditions (\ref{ddbc}) impose
\begin{equation}
    z(u)\approx\varepsilon R_2 \tau'(u)+\varepsilon^3 R_2^3\left[\frac{\left(\tau''(u)\right)^2}{2\left(\tau'(u)\right)^2}-2\pi^2 T^2\left(\tau'(u)\right)^3\right]
    \label{boundaryparam}
\end{equation}
where a prime indicates a derivative with respect to the boundary proper time $u$. The shape of the regularized boundary is then completely determined by the time reparametrization mode $\tau(u)$, and the dynamics is encoded in its fluctuations, which are described at leading order in $1/\varepsilon$ by a Schwarzian action:
\begin{equation}
    I_{eff}\approx -\frac{V_{d-1} R_2\phi_b}{8\pi G}\int du\left\{\tan\left(\pi T\tau(u)\right),u\right\}.
    \label{effactionsch}
\end{equation}
In the absence of the ETW particle, the integral over $u$ runs from $-\beta/2$ to $\beta/2$ and $\tau(u)$ is a monotonic function satisfying $\tau(u+\beta)=\tau(u)+\beta$. The action (\ref{effactionsch}) explicitly breaks the asymptotic $\textrm{Diff}(S^1)$ symmetry down to $\textrm{SL}(2,\mathbb{R})$. This pattern of symmetry breaking is completely analogous to the one found in the SYK model, and makes the duality between low-energy limit of the SYK model and JT gravity manifest. In particular, the gravitational Schwarzian sector can be described either by the Majorana SYK model, or by a fixed charge subsector of the cSYK model. 
The picture emerging from the analysis so far is that one of a rigid $AdS_2$ bulk cutoff by a regularized boundary whose shape is determined by the function $\tau(u)$ (see Figure \ref{JT1cartoon}). The Schwarzian action (\ref{effactionsch}) takes different values on ``chunks'' of $AdS_2$ with different shapes\footnote{Up to an $\textrm{SL}(2,\mathbb{R})$ transformation connecting physically equivalent configurations.}, which are then weighted differently in the gravitational path integral. The saddle point solution is given by $\tau(u)=u$, which represents a circular boundary sitting at $z=\varepsilon R_2\equiv z_b$.

\subsection{ETW particle trajectory}
\label{trajectorysection}

So far we have focused on the dynamics of the regularized boundary. Imposing the boundary conditions (\ref{nnbcmet}) and (\ref{NNbcdil}) the action for the ETW particle clearly vanishes on-shell. Therefore, in the resulting effective description the ETW particle follows a trajectory (fixed by the boundary conditions) in the rigid $AdS_2$ bulk, and its evolution is independent of the regularized boundary fluctuations, i.e. of the dynamics. Parametrizing the ETW particle worldline by the bulk time $\tau$ (i.e. picking $v=\tau$ in the action (\ref{JTaction})), the equation of motion for the brane particle is given by
\begin{widetext}
\begin{equation}
    z'(\tau)=\pm\frac{1-4\pi^2 T^2 z^2}{T_{etw}\left(\frac{dR_2}{2}+\frac{r_e}{R_2}z\right)}\sqrt{1-4\pi^2T^2z^2-\left(\frac{dR_2}{2}+\frac{r_e}{R_2}z\right)^2T_{etw}^2}
    \label{braneeom}
\end{equation}
\end{widetext}
where the $+$ ($-$) sign corresponds to the contraction (expansion) phase of the ETW particle. As described in Section \ref{section4}, we are interested in time-reflection symmetric states. Therefore, the brane reaches its minimum radius (i.e. maximum value of $z$) $z=z_{max}$ for $\tau=\pm\beta/2$. $z=z_{max}$ is the inversion point $z'(\pm\beta/2)=0$, and is determined by imposing the argument of the square root in equation (\ref{braneeom}) to vanish. If we could trust this trajectory all the way to the asymptotic $AdS_2$ boundary, it would describe a particle hitting the asymptotic boundary at times $\tau=\pm \bar{\tau}$, and reaching its minimum radius at the inversion point $z_{max}$. However, we remark that the trajectory described by equation (\ref{braneeom}) is reliable only inside the regularized boundary, where the physics is well-described by the dimensionally reduced action (\ref{JTaction}).

Since our effective two-dimensional description is cut off at the regularized boundary, the ETW particle is present in our model only if it crosses the boundary and enters the near-horizon region. At leading order in $\varepsilon$ this implies
\begin{equation}
    z_{max}\geq \varepsilon R_2 \tau'(u)= z_b \tau'(u).
    \label{entercond}
\end{equation}
For $\tau(u)=u$, condition (\ref{entercond}) imposes a bound on the tension $T_{etw}$:
\begin{equation}
    0<T_{etw}\leq T_{etw}^{max}=\frac{\sqrt{1-4\pi^2 T^2 z_b^2}}{\frac{dR_2}{2}+\frac{r_e}{R_2}z_b}.
    \label{tensioncond}
\end{equation}
Using the conditions (\ref{allcond}) determining the region where the regularized boundary sits, we immediately find that equation (\ref{tensioncond}) implies $0<T_{etw}\ll 1/L_{AdS}$, which means that the tension of the higher dimensional ETW brane must be far from its critical value. This result was to be expected because in the higher dimensional picture we want the ETW brane to enter the near-horizon region.

The near-horizon condition also guarantees that in the portion of trajectory we are interested in (i.e. when the ETW particle is inside the regularized boundary) $z(\tau)\gg R_2^2/r_e$ holds. In this limit, the equation of motion (\ref{braneeom}) can be solved analytically, obtaining
\begin{widetext}
\begin{equation}
    z(\tau)\approx\frac{R_2}{\sqrt{4\pi^2 T^2 R_2^2+r_e^2 T_{etw}}}\sqrt{1-\frac{r_e^2T_{etw}}{4\pi^2 T^2 R_2^2}\tan^2\left[\frac{2\pi}{\beta}\left(\tau\mp \frac{\beta}{2}\right)\right]}
    \label{analyticsol}
\end{equation} 
\end{widetext}
and now $z_{max}\approx R_2/\sqrt{4\pi^2 T^2 R_2^2+r_e^2 T^2_{etw}}$, while the maximum brane tension is $T_{etw}^{max}\approx R_2\sqrt{1-4\pi^2 T^2 z_b^2}/(z_b r_e)$.

\subsection{Tangent trajectory and symmetry breaking}

Let us initially focus on the Schwarzian saddle point $\tau(u)=u$. In this case the regularized boundary is circular and sits at $z=z_b$. When the condition (\ref{entercond}) is satisfied, the brane will in general intersect the boundary in two points, at times $\tau=u=\pm u_0$, with $u_0<\beta/2$. The AdS/BCFT prescription suggests that the quantity $u_0$ represents the preparation time of the state dual to the spacetime geometry. Although this solution is completely meaningful on the gravity side, it is not clear at the moment what the dual cSYK state would look like. Indeed, the preparation time of the boundary states introduced in Section \ref{section3} is by construction $u_0=\beta/2$. 

One possibility is that the geometry with two intersection points is dual to some different kind of cSYK boundary state, one with a different structure at the microscopic scale. This is perfectly reasonable since the brane intersection with the regularized boundary is by definition at the cutoff scale in the dimensionally reduced theory.

It is also conceivable that only holographic boundary states with preparation time $u_0=\beta/2$ can be built in the cSYK model, and that the two-intersection solutions are only an indication that the bulk theory is richer than the boundary one. This would not come as a surprise for two reasons. First, the analogy with the $AdS_3$/$CFT_2$ case, where only holographic boundary states with preparation time $u_0=\beta/4$ (and therefore the brane anchored at antipodal points) can be constructed using conformal boundary conditions \cite{Almheiri:2018ijj}. Second, the fact that the bulk theory described in the present section comes from the dimensional reduction of a higher dimensional theory, and is therefore expected to be richer than the (0+1)-dimensional boundary theory. This is made explicit by the presence of an additional bulk parameter, the tension $T_{etw}$, which has no known counterpart in our cSYK description. In higher dimensional braneworld cosmologies, the bulk tension parameter has a counterpart in the boundary dual theory, with a precise physical meaning: it accounts for additional CFT degrees of freedom living on the $(d-1)$-dimensional boundary of the $d$-dimensional manifold where the BCFT is defined \cite{Cooper:2018cmb,VanRaamsdonk:2021qgv}. But the dual theory considered in the present paper is (0+1)-dimensional quantum mechanics, and therefore there can be no additional CFT boundary degrees of freedom. The tension is then just a bulk free parameter inherited from the higher dimensional description.

To make contact with the boundary states defined in Section \ref{section3}, we therefore impose $u_0=\beta/2$, which implies that the two intersection points reduce to a single one at bulk coordinates $\tau=\pm u_0=\pm\beta/2$ and $z=z_{max}=z_b$. This condition can be met for a brane trajectory sufficiently close to the horizon provided we choose the cutoff surface $z_b$ such that $T_{etw}=T_{etw}^{max}(z_b)$. In other words, the trajectory of the particle is tangent to the regularized boundary. One then has a special point on the regularized boundary where Neumann boundary condition for all the bulk fields are imposed, which is reminiscent of the cSYK thermal circle with a special point. If we took into account the presence of bulk fermions in our description, appropriate boundary conditions for the fermions should also be specified at the special point.
\begin{figure}
    \centering
    \includegraphics[width=0.45\textwidth]{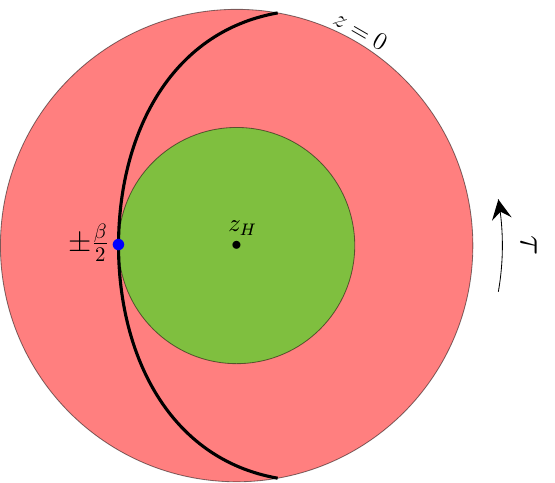}
    \caption{\textbf{Tangent ETW particle at saddle point.} At the Schwarzian saddle point $\tau=u$ the regularized boundary is a circle sitting at $z=z_b$. When the ETW particle trajectory is tangent to the regularized boundary, a single special point at $\tau=\pm\beta/2$ is identified, where Neumann bondary conditions for all the bulk fields must be imposed. This configuration is the one relevant to the description of the cSYK boundary states analyzed in Section \ref{section3}.}
    \label{JT2cartoon}
\end{figure}
Let us now consider dynamical fluctuations of the regularized boundary, governed by the effective action (\ref{effactionsch}). From a bulk point of view, we could consider arbitrary boundary reparametrizations $u\to \tau(u)$, i.e. arbitrary shapes for the regularized boundary in the rigid $AdS_2$ bulk (see Figure \ref{JT1cartoon}). The single intersection point at $\tau=u=\pm\beta/2$ can then be mapped to a different single intersection point at an arbitrary $\tau(\beta/2)=\tau(-\beta/2)+\beta$, and it would not reside on the time-reflection symmetric line $\tau=\pm\beta/2$ anymore. Or there could be more than one intersection point, or no intersection points at all. We know of no principle restricting such configurations from the bulk point of view, and this is again a sign of the additional freedom present in the higher-dimensional bulk theory relative to the lower-dimensional boundary dual. But since our goal is to build a geometry which can be described by the cSYK boundary states defined in Section \ref{section3}, a necessary condition for our construction to work is that both the time-reflection symmetry of the ETW particle trajectory in the bulk and the tangent condition must be preserved under reparametrizations of the boundary. In other words, the intersection of the ETW particle trajectory and the boundary must remain a single point on the time-reflection symmetric line $\tau=\pm\beta/2$.

The time-reflection symmetry condition immediately implies that the intersection point $u=\pm\beta/2$ is mapped by an admissible boundary reparametrization to $\tau(\pm\beta/2)=\pm\beta/2$. Since the trajectory of the ETW particle in the bulk is completely independent of the boundary fluctuations, $z_{etw}(\tau=\pm\beta/2)=z_{max}=z_b$ still holds after the boundary reparametrization. By means of equation (\ref{boundaryparam}), at leading order in $\varepsilon$ the tangent condition implies $z(u=\pm\beta/2)=z_b\tau'(\pm\beta/2)=z_b$, and therefore $\tau'(\pm\beta/2)=1$. Note that, since the Schwarzian fluctuations are $1/N$-suppressed (which is, of Planckian size), the preservation of the tangent condition under time reparametrizations also implies that no additional intersection points appear in perturbation theory: the special point at $\tau=\pm\beta/2$ remains the only intersection point between the ETW particle trajectory and the regularized boundary.  After the identification $u_0=\beta/2$, we obtain two boundary conditions for the boundary reparametrization mode:
\begin{equation}
    \tau(\pm u_0)=\pm u_0, \hspace{1cm} \tau'(\pm u_0)=1
    \label{repbc2}
\end{equation}
which are completely analogous to the ones obtained for the cSYK boundary states (see equation (\ref{repbc})). Therefore we obtained the exact same pattern of symmetry breaking: the effect of the boundary conditions (\ref{repbc2}) is to further break the residual $\textrm{SL}(2,\mathbb{R})$ symmetry, with only one generator left unbroken. This result provides clear evidence of the duality between the cSYK boundary states constructed in Section \ref{section3} and the dimensionally reduced braneworld cosmology model described in the present section. 

As we have already explained at the end of Section \ref{collectivesection}, the boundary states described in Section \ref{section3} are not the only ones to exhibit the properties described in the present analysis: starting with a generic superposition of Fock states (instead of a single Fock state) and evolving it for an amount $\tau_0$ of Euclidean time would lead to analogous results\footnote{Note that this is not true in general for a highly excited state. As an example, consider an energy eigenstate. The eigenstate thermalization hypothesis guarantees that correlators of  generic operators would look thermal in such state. However, it is enough to compute quantum energy fluctuations to tell apart an energy eigenstate (for which they would exactly vanish) from the thermal state. This is different from the expectations for our boundary states and could be a signal that energy eigenstates do not have a smooth interior.}. Note, however, that a complete dual description of the low-energy limit of the cSYK model must necessarily include $N$ bulk fermions, related to sources of the cSYK fermions by the HKLL construction \cite{Lensky:2020ubw}. A set of boundary conditions for such bulk fields must be specified at the intersection point between the regularized boundary and the ETW particle. A given set of boundary conditions will then uniquely identify a specific cSYK boundary state from a bulk perspective. From this point of view, the construction described in the present section, where we ignored the presence of the bulk fermions, can be regarded as the common dual geometric background associated to the whole class of cSYK boundary states that break the $\textrm{SL}(2,\mathbb{R})$ symmetry in the specific way we analyzed. It is then a conceptually straightforward step to specify a set of boundary conditions for the $N$ bulk fields at the intersection point, and therefore restrict to a single cSYK boundary state.


We emphasize that we only used the intuition provided by the cSYK boundary states as a guideline to understand the symmetry properties that a hypothetical bulk dual should satisfy. The fact that requiring such properties to hold in our dimensionally reduced braneworld cosmology model leads to the same symmetry breaking pattern that we observed on the cSYK side is non-trivial, and should be regarded as a key result of the present paper.

\section{Discussion}
\label{section6}

In this paper we built an explicit example of the duality between asymptotically AdS spacetimes with an end-of-the-world brane and boundary states of a dual CFT. In particular, we considered specific boundary states in the cSYK model and showed that, under appropriate conditions, the dimensional reduction of the holographic braneworld cosmology model proposed in \cite{Antonini:2019qkt} is dual to such boundary states. Although the range of bulk parameters where the duality holds does not allow gravity localization on the ETW brane (and therefore the model is not suitable to properly describe a braneworld cosmology), our result confirms the possibility to describe similar spacetimes using BCFTs and sheds light on how to explicitly realize similar constructions. It also has the merit to provide a novel, simple framework to explore the properties of the AdS/BCFT correspondence.

A direction calling for further development is to understand which observables in the cSYK model can probe the bulk trajectory of the ETW particle. In a Lorentzian picture corresponding to our Euclidean model, the ETW particle worldline would sit in the left asymptotic region, cutting off the left asymptotic boundary, analogously to what happens in the higher dimensional model depicted in Figure \ref{penrose} (see \cite{Gao:2021uro} for a recent analysis). The duality between the cSYK boundary states and such geometries implies that the trajectory of the ETW particle in the left asymptotic region can be probed by measuring expectation values of appropriate observables in the boundary state of the cSYK model living on the right asymptotic boundary, which is dual to the whole spacetime. In particular, the expectation values should acquire some time-dependence, which would necessarily be related to the evolution of the ETW particle in the bulk. Natural candidates to probe such behind-the-horizon physics are the entanglement entropy and the holographic complexity \cite{Cooper:2018cmb}. Our construction allows us to further extend this list of possible observables by working within a well-understood, controlled framework.

We also remark one more time that in the present paper we explored only the Schwarzian sector of the dimensionally reduced theory, and in particular we did not explicitly take into account bulk fermions, which are necessarily present in a bulk dual of the low-energy sector of the cSYK model. Therefore, the geometries we obtained can be equivalently described at this level by either the cSYK boundary states in a specific charge subsector constructed in Section \ref{section3} (or analogous states built by starting with a generic superposition of Fock states instead of a single Fock state), or the Majorana SYK boundary states studied by Kourkoulou and Maldacena \cite{Kourkoulou:2017zaj}. Additionally, our purely Schwarzian bulk construction is incapable of distinguishing between two different boundary states in the same cSYK charge subsector (or any two boundary states in the Majorana SYK model). However, when the presence of bulk fermions (which are related to sources of the cSYK fermions by the HKLL construction \cite{Lensky:2020ubw}) is taken into account, the cSYK boundary states are uniquely identified by the specific set of boundary conditions imposed on the bulk fields at the intersection point between the ETW particle and the regularized boundary. Our bulk construction is then dual to a whole class of boundary states, which are distinguished by the boundary conditions imposed on bulk fermions propagating on such geometric background. One advantage of considering cSYK boundary states instead of Majorana SYK ones is that it allows an immediate generalization of our setup to more complicated configurations. For example, it would be possible to study the higher dimensional holographic braneworld cosmologies in the gravitational grand-canonical ensemble, allowing small charge fluctuations. The corresponding cSYK effective action would then also include a term governing the dynamics of the phase field $\lambda(\tau)$ and encoding small charge density fluctuations \cite{Sachdev:2019bjn}. Alternatively, we could include in our description $N$ bulk fermions charged under the $U(1)$ bulk gauge symmetry. These would be naturally described within the cSYK model \cite{sachdev, Lensky:2020ubw} and would determine the specific cSYK boundary state dual to the spacetime geometry, as we have already pointed out.

Another interesting open question is whether or not it is possible to build a dual description of bulk setups where the brane intersects the regularized boundary in two distinct points. If the answer is yes, the corresponding boundary states should look quite different from the ones built in Section \ref{section3}, whose preparation time is by construction equal to half the Euclidean periodicity. It is also possible that such bulk setups are just a manifestation of the additional bulk degrees of freedom inherited from the higher dimensional theory. If this is the case, they might not have a dual representation in the cSYK model.

Finally, a fascinating open question is how the holographic braneworld cosmology models originally introduced in \cite{Cooper:2018cmb,Antonini:2019qkt} relate to the physics of entanglement islands. The dimensionally reduced model presented in this paper could be a useful building block to explore the answer to this question in a simplified setup analogous to the ones where entanglement islands were originally introduced.

\textbf{Acknowledgements}
We would like to thank J. Bringewatt, F. Ferretti, R. A. Silva, and J. L. Villaescusa Nadal for useful discussions. This work was supported in part by the U.S. Department of Energy, Office of Science,
Office of High Energy Physics QuantISED Award DE-SC0019380, by the U.S. Department of
Energy, Office of Science, Office of Advanced Scientific Computing Research, Accelerated Research for Quantum Computing program, and by the Simons Foundation via the It From Qubit collaboration.

\begin{appendices}
\renewcommand\thesubsection{\arabic{subsection}}
\appendixpage

\section{cSYK model and symmetry breaking}
\label{appendixa2}

As we have pointed out in Section \ref{section2}, in general the saddle point solution (\ref{conformal2pf}) for $G$ is not invariant under the transformation (\ref{conftransf}), meaning that the time reparametrization symmetry $\textrm{Diff}(S^1)$ and the local $U(1)$ symmetry are spontaneously broken. However, there is a subgroup of time reparametrizations under which the thermal solution is invariant provided that we also perform an appropriate phase shift \cite{Chaturvedi:2018uov}. Let us start by considering the zero-temperature solution (\ref{longtime2pf}) when the asymmetry parameter vanishes $\mathcal{E}=0$ (i.e. for the $\mathcal{Q}=0$ charge subsector). We can perform a time reparametrization
\begin{equation}
    \tau\to F(\tau)=\tan\left[\frac{\pi f(\tau)}{\beta}\right]
    \label{timerep}
\end{equation}
without performing any local phase shift. Then, from equation (\ref{conftransf}) we obtain
\begin{equation}
     G_f(\tau_1,\tau_2)\propto\left(\frac{F'(\tau_1)F'(\tau_2)}{[F(\tau_1)-F(\tau_2)]^2}\right)^\Delta.
     \label{sl2inv}
\end{equation}
For $f(\tau)=\tau$, this is the saddle point thermal solution (\ref{conformal2pf}) at $\mathcal{E}=0$ (and it is correctly antiperiodic: $G_\beta(-\beta/2)=G_{\beta}(\beta/2)$), while for generic $f(\tau)$ we get a generic reparametrization of the thermal solution. But note that equation (\ref{sl2inv}) is invariant under any reparametrization of the form
\begin{equation}
    \tilde{F}=\frac{aF+b}{cF+d}
    \label{sl2rtransf}
\end{equation}
with $ad-bc=1$, i.e. the thermal solution is invariant under global $\textrm{SL}(2,\mathbb{R})$ transformations.\footnote{It is a global $\textrm{SL}(2,\mathbb{R})$ group because the transformation (\ref{sl2rtransf}) does not depend on $\tau$.} Note that the M\"obius transformation (\ref{sl2rtransf}) does not distinguish elements in the center of $\textrm{SL}(2,\mathbb{R})$. In other words, it does not distinguish the transformation with $a=d=1$, $b=c=0$ (i.e. the identity transformation $\hat{\mathbbm{1}}$) from the one with $a=d=-1$, $b=c=0$ (i.e. the transformation $-\hat{\mathbbm{1}}$). Therefore, looking only at the bosonic sector we are considering, the residual symmetry group is $\textrm{PSL}(2,\mathbb{R})=\textrm{SL}(2,\mathbb{R})/\{\pm\hat{\mathbbm{1}}\}$. However, since in general we are interested in a bulk theory with fermions and fermions are sensitive to the cover of $\textrm{PSL}(2,\mathbb{R})$, we will consider $\textrm{SL}(2,\mathbb{R})$ to be the residual symmetry group \cite{Stanford:2019vob}. Therefore, the $\textrm{Diff}(S^1)$ symmetry is spontaneously broken down to $\textrm{SL}(2,\mathbb{R})$, and the corresponding pseudo-Nambu-Goldstone boson $f(\tau)$, whose dynamics is governed by the Schwarzian action, belongs to the \textit{left} quotient of $\textrm{Diff}(S^1)$ by $\textrm{SL}(2,\mathbb{R})$. Note that the $\textrm{SL}(2,\mathbb{R})$ symmetry is preserved by both the spontaneous and the explicit symmetry breaking, and must be regarded as a gauge symmetry as we explained in Section \ref{section2}. Indeed, the effective action (\ref{effactionsyk}) is invariant under the $\textrm{SL}(2,\mathbb{R})$ transformation (\ref{sl2rtransf}). If we did not exclude the redundant $\textrm{SL}(2,\mathbb{R})$-equivalent configurations when performing the path integral, we would have an infinite degeneracy leading to a diverging Euclidean path integral, as it always happens when we do not deal with gauge symmetries properly. 

We also remark that any local phase shift does not leave either the thermal two-point or the effective action invariant: the emergent local $U(1)$ symmetry is completely broken, both spontaneously and explicitly, leading to a second pseudo-Nambu-Goldstone boson $\lambda(\tau)$, whose effective action governs charge fluctuations in the grand-canonical ensemble and is given by \cite{sachdev}
\begin{equation}
    I_{eff}[\lambda,f]=\frac{N K}{2}\int_0^\beta d\tau\left[\lambda'(\tau)+i\frac{2\pi\mathcal{E}}{\beta}f'(\tau)\right]^2
    \label{lambdaaction}
\end{equation}
where $K$ is the charge compressibility at zero temperature \cite{sachdev}.
On the other hand, a global phase shift is an exact symmetry of the theory.

Let us now generalize our analysis to the case $\mathcal{E}\neq 0$. Now, due to the spectral asymmetry, we cannot obtain the thermal solution (\ref{conformal2pf}) from equation (\ref{longtime2pf}) by simply performing the time reparametrization (\ref{timerep}). Indeed, at the saddle point $f(\tau)=\tau$ we would not obtain an antiperiodic result with period $\beta$, as we would expect since $\tau$ and $\tau+\beta$ are identified. To restore the periodicity, we must also perform a local phase shift as in equation (\ref{conftransf}) with
\begin{equation}
    i\lambda(\tau)=-\frac{2\pi\mathcal{E}}{\beta}\tau, \hspace{1cm} -\beta<\tau<\beta
    \label{localphaseshift}
\end{equation}
which leads to the thermal two-point function (\ref{conformal2pf}). We see immediately that for $\mathcal{E}\neq 0$ such saddle point solution, which transforms as in equation (\ref{conftransf}), is not invariant if we only perform a $\textrm{SL}(2,\mathbb{R})$ transformation as defined by equation (\ref{sl2rtransf}). Indeed, the time reparametrization affects also the local phase shift (\ref{localphaseshift}). However, the thermal two-point function is still invariant under the combined action of a $\textrm{SL}(2,\mathbb{R})$ transformation and a local phase shift given by \cite{Chaturvedi:2018uov}
\begin{equation}
    i\bar{\lambda}(\tau)=\frac{2\pi\mathcal{E}}{\beta}\left(f(\tau)-\tau\right).
\end{equation}
Note that the total effective action (given by the sum of the Schwarzian action (\ref{effactionsyk}) and the phase field action (\ref{lambdaaction})) is invariant under the combined $\textrm{SL}(2,\mathbb{R})\otimes U(1)$ transformation. This is indeed equivalent to performing the change of variables
\begin{equation}
    \lambda(\tau)\to\lambda(\tau)+i\frac{2\pi\mathcal{E}}{\beta}\left(\tau-f(\tau)\right),
    \label{varchange2}
\end{equation}
which decouples the $U(1)$ mode from the time reparametrization mode, and then perform a $\textrm{SL}(2,\mathbb{R})$ time reparametrization, which leaves invariant the actions for both modes. Substituting the change of variable (\ref{varchange2}) into the transformation law (\ref{conftransf}) of the two-point function, we immediately see how a time reparametrization of the conformal thermal solution implies also a (imaginary) phase shift:
\begin{widetext}
\begin{equation}
    G^\beta_{f,\lambda}(\tau_1,\tau_2)=\left[f'(\tau_1)\right]^\Delta\left[f'(\tau_2)\right]^\Delta\textrm{e}^{i[\lambda(\tau_1)-\lambda(\tau_2)]}\textrm{e}^{\frac{2\pi\mathcal{E}}{\beta}\left[f(\tau_1)-f(\tau_2)-(\tau_1-\tau_2)\right]} G_\beta(f(\tau_1),f(\tau_2)).
    \label{thermal2pftransf}
\end{equation}
\end{widetext}
The transformation law (\ref{thermal2pftransf}) leaves the (anti)periodicity of the thermal two-point function unchanged. Moreover, it ensures that a time reparametrization satisfying (\ref{sl2rtransf}) leaves the thermal solution invariant, making the $\textrm{SL}(2,\mathbb{R})$ symmetry explicit again without the need of a local phase shift. Now all the considerations made in the $\mathcal{E}=0$ case apply here too, and the system displays the same pattern of symmetry breaking. In particular, the time reparametrization symmetry is both spontaneously and explicitly broken down to $\textrm{SL}(2,\mathbb{R})$, which is a residual and non-physical ``gauge'' symmetry, while the local $U(1)$ symmetry is completely broken both spontaneously and explicitly to global $U(1)$. Note indeed that a global phase shift still leaves the thermal two-point function invariant, i.e. there is still correctly an unbroken $U(1)$ global symmetry. Such global phase shift clearly also leaves the total effective action invariant: it represents an exact physical symmetry of the theory.

Finally, we would like to remark one more time that the boundary conditions (\ref{repbc}) do not imply the exclusion of any physical configuration from the path integral. Indeed, we have seen that their effect is to further break two out of the three $\textrm{SL}(2,\mathbb{R})$ generators. This implies that in general a $\textrm{SL}(2,\mathbb{R})$ transformation (\ref{sl2rtransf}) maps modes that satisfy the boundary conditions (\ref{repbc}) to modes that do not, and vice versa. In particular, for any given reparametrization mode $\bar{f}(\tau)$ which does not satisfy the boundary conditions, it is always possible to perform an appropriate $\textrm{SL}(2,\mathbb{R})$ transformation to a mode $\tilde{f}(\tau)$ which satisfies (\ref{repbc}). Since time reparametrization modes connected by the $\textrm{SL}(2,\mathbb{R})$ transformation (\ref{sl2rtransf}) are accounted for only once in the path integral, the effect of the boundary conditions (\ref{repbc}) is to select a one-parameter family of admissible reparametrization modes for each class of $\textrm{SL}(2,\mathbb{R})$-equivalent modes.

\section{Boundary states dynamics}
\label{appendixb}

\subsection{Probability distribution of coefficients}
\label{appendixb1}

In Section \ref{section3} we made use of the properties (outlined in Section \ref{coeff}) of the probability distribution of the coefficients $c_k^i$ defined by $\ket{S_i}=\sum_{k=1}^{D_{\mathcal{Q}}}c_k^i\ket{E_k}$. We report here physical motivations and numerical results supporting the features claimed in Section \ref{coeff}.

\begin{figure*}
    \centering
    \subfloat[]{
        \includegraphics[width=0.33\linewidth]{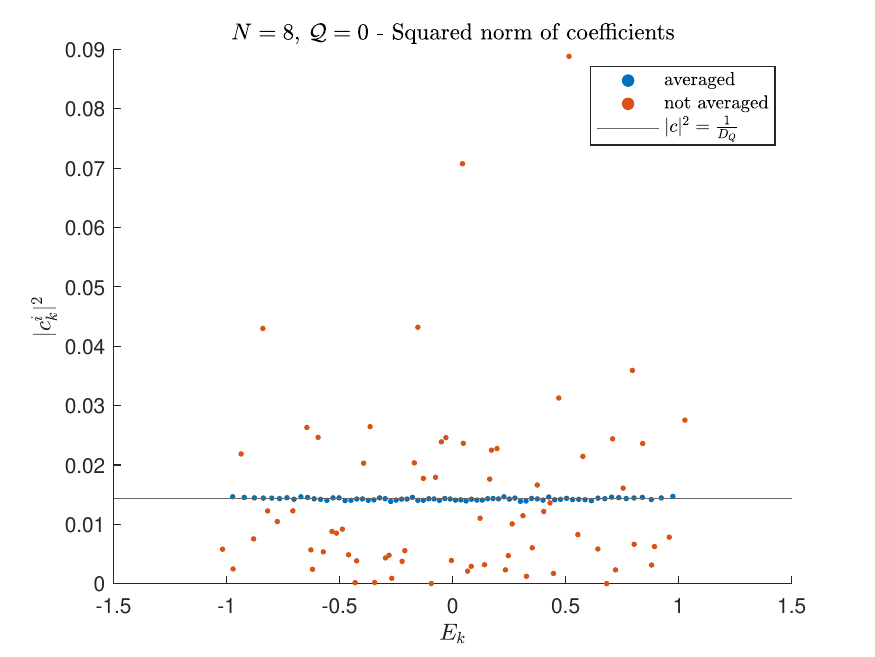}}
    \subfloat[]{
        \includegraphics[width=0.33\linewidth]{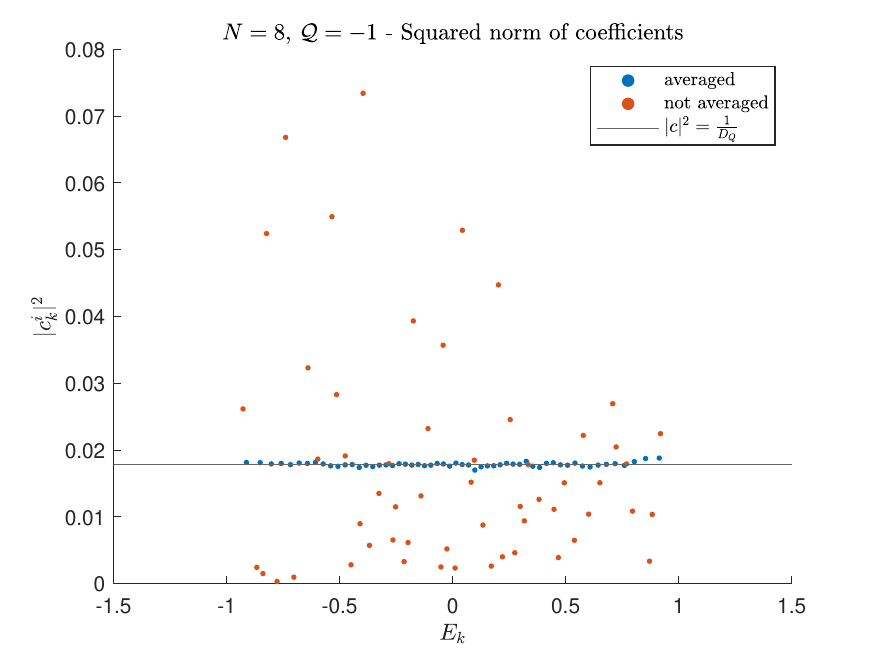}}
    \subfloat[]{
        \includegraphics[width=0.33\linewidth]{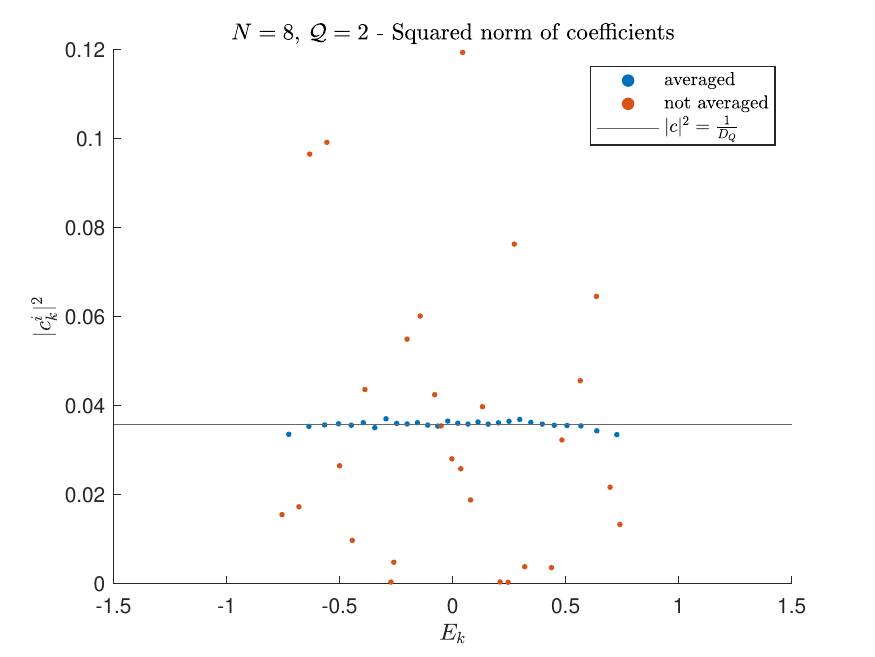}}
    \caption{\textbf{Squared norms of coefficients.} $|c_k^i|^2$ for a given Fock state $\ket{S_i}$ and for every energy eigenstate $\ket{E_k}$ in the same charge subsector. The red points are the squared norms for a single realization of the Hamiltonian, while the blue points are averaged over 5000 realizations. The mean of the squared norm is clearly $1/D_{\mathcal{Q}}$ (horizontal line). (a) $\mathcal{Q}=0$. (b) $\mathcal{Q}=-1$. (c) $\mathcal{Q}=2$. }
    \label{avg}
\end{figure*}

\begin{figure*}
    \centering
    \subfloat[]{
        \includegraphics[width=0.33\linewidth]{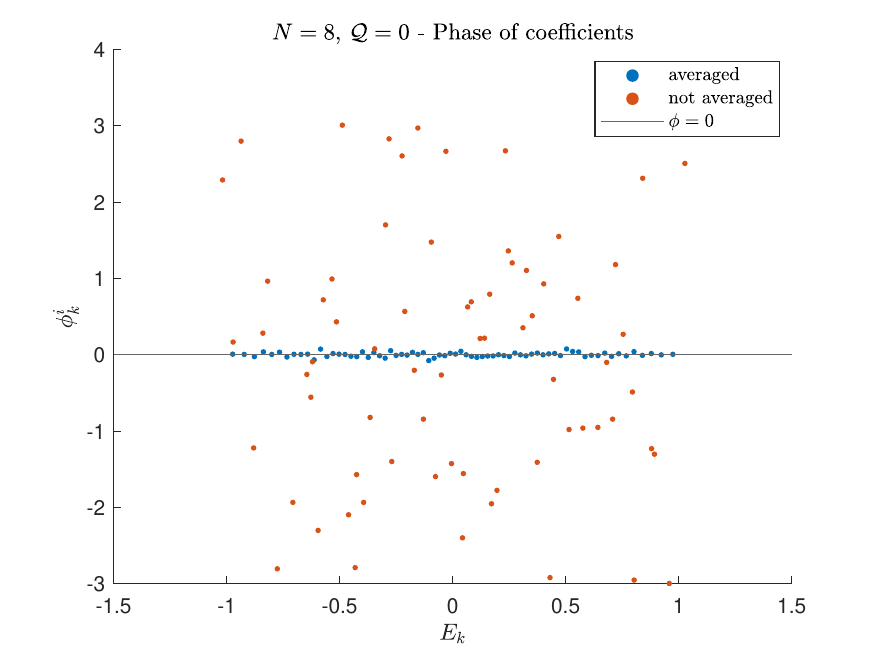}}
    \subfloat[]{
        \includegraphics[width=0.33\linewidth]{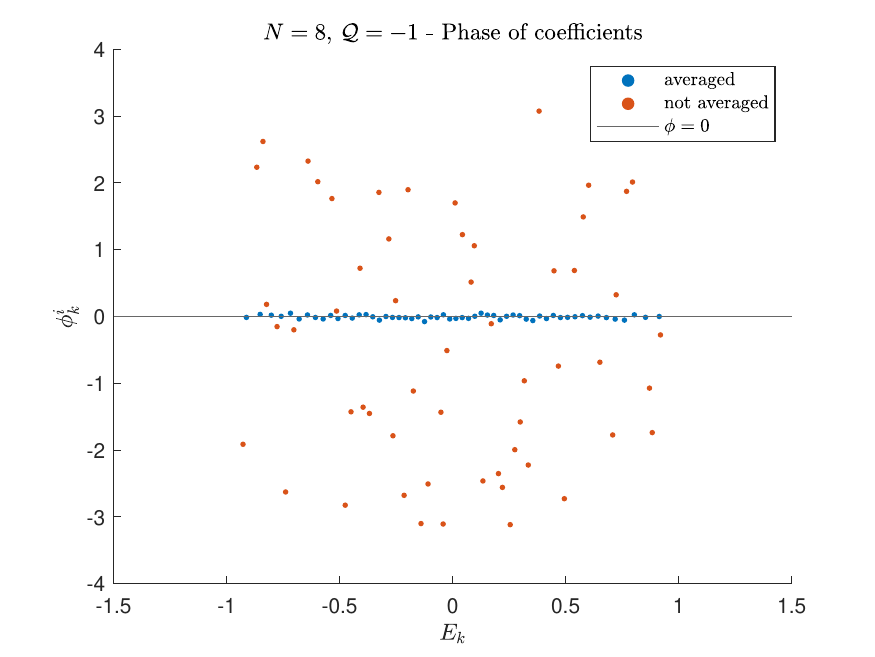}}
    \subfloat[]{
        \includegraphics[width=0.33\linewidth]{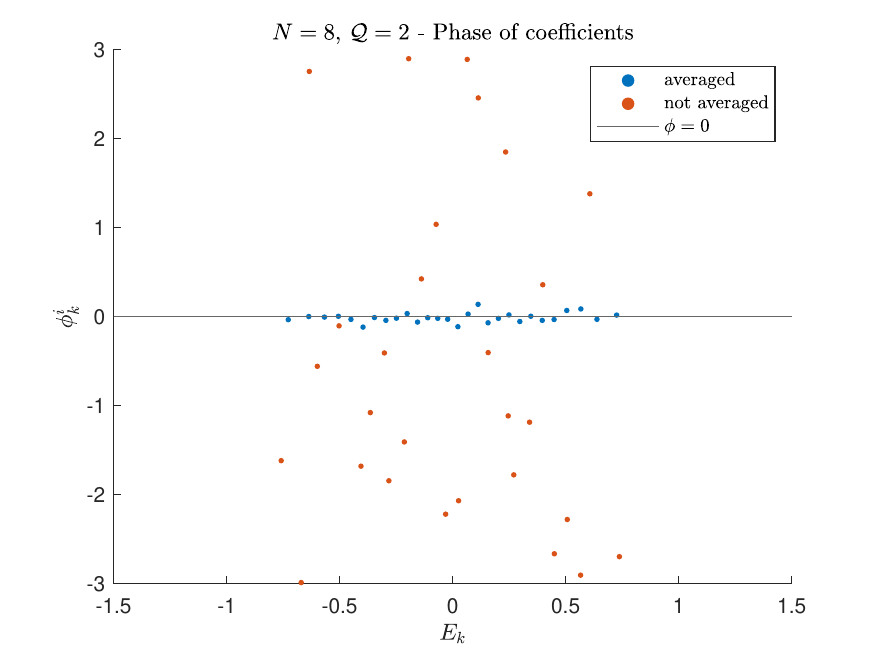}}
    \caption{\textbf{Phases of coefficients.} Phases of $c_k^i$ for a given Fock state $\ket{S_i}$ and for every energy eigenstate $\ket{E_k}$ in the same charge subsector. The red points are the phases for a single realization of the Hamiltonian, while the blue points are averaged over 5000 realizations. The phases are clearly uniformly distributed between $-\pi$ and $\pi$, and therefore their mean vanishes. (a) $\mathcal{Q}=0$. (b) $\mathcal{Q}=-1$. (c) $\mathcal{Q}=2$. }
    \label{phases}
\end{figure*}

An intuitive way to understand this is by focusing on the contribution of each Fock state to the superposition $\ket{E_k}=\sum_{i=1}^{D_{\mathcal{Q}}}c^k_i\ket{S_i}$ (with $c_i^k=(c_k^i)^*$) for multiple realizations of the random Hamiltonian. The emergent $U(N)$ symmetry at large $N$, arising from the properties of the random couplings and mentioned in Section \ref{coeff}, implies that the roles of the different Fock states are ``shuffled'' from one realization to another. Therefore, we expect the coefficient to have, on average, the same magnitude and a uniformly distributed phase. The value of the average magnitude is fixed by the normalization condition to be $1/D_\mathcal{Q}$ (see Figures \ref{avg} and \ref{phases}). 

The simplest model which captures these intuitions takes the coefficients $c_k^i$ as elements of a normalized Haar random vector in a $D_\mathcal{Q}$-dimensional complex vector space. Given such a random state $|\psi\rangle$, we have the standard identity
\begin{equation}
    \int d\psi |\psi \rangle \langle \psi | \otimes |\psi \rangle \langle \psi | = \frac{\mathbbm{1}_2 + F_2}{D_\mathcal{Q}(D_\mathcal{Q}+1)},
\end{equation}
where $\mathbbm{1}_2$ is the identity on two copies and $F_2$ is the swap operator. Based on this formula, a random vector model of the $c^i_k$ predicts that the variance of $|c_k^i|^2$ is $D_\mathcal{Q}^{-2} + \mathcal{O}(D_\mathcal{Q}^{-3})$ and that the covariance of $|c_k^i|^2$ and $|c_l^i|^2$ is $ - D_\mathcal{Q}^{-3} + \mathcal{O}(D_\mathcal{Q}^{-4})$.

\begin{figure*}
    \centering
    \subfloat[]{
        \includegraphics[width=0.33\linewidth]{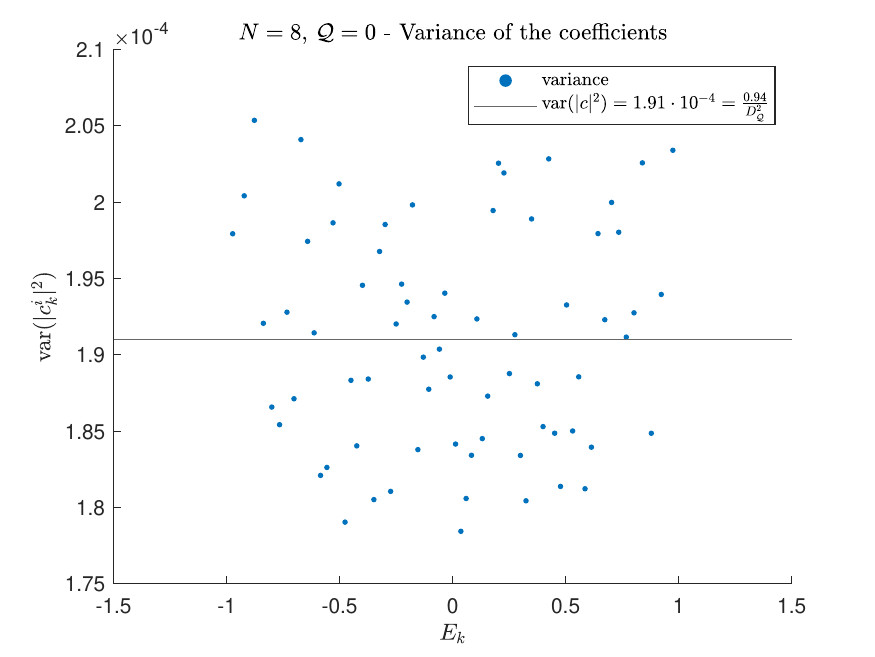}}
    \subfloat[]{
        \includegraphics[width=0.33\linewidth]{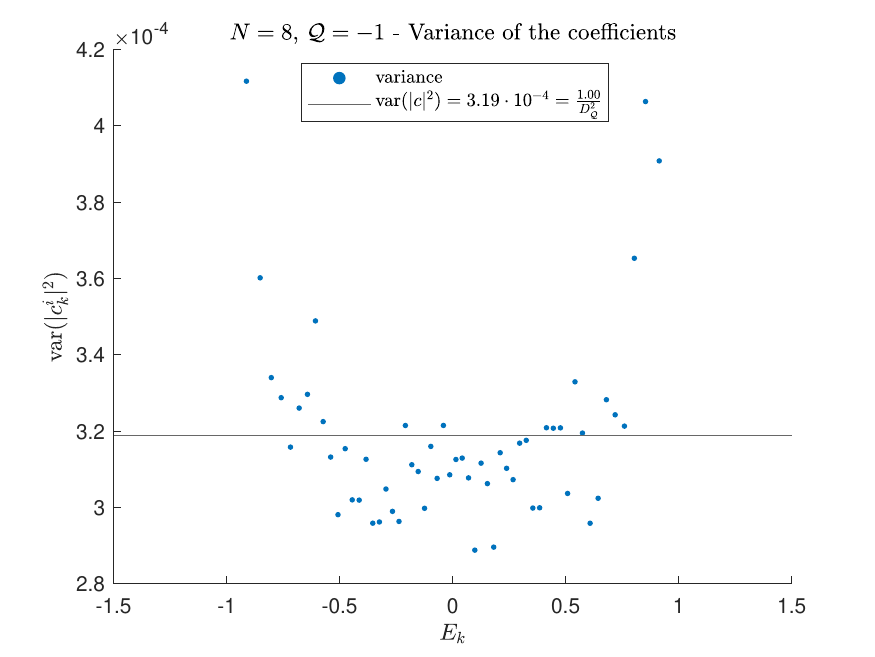}}
    \subfloat[]{
        \includegraphics[width=0.33\linewidth]{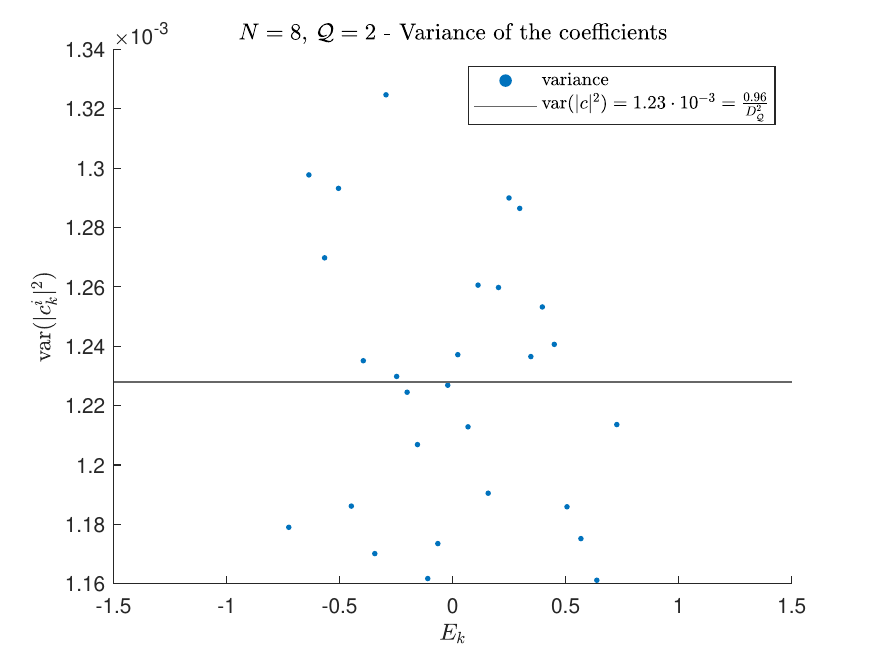}}
    \caption{\textbf{Variance of the squared norms of coefficients.} Variance of $|c_k^i|^2$ for a given Fock state $\ket{S_i}$ and for every energy eigenstate $\ket{E_k}$ in the same charge subsector, averaged over 5000 realizations. The variance is of order $1/D_{\mathcal{Q}}^2$ (horizontal line). (a) $\mathcal{Q}=0$. (b) $\mathcal{Q}=-1$. (c) $\mathcal{Q}=2$. }
    \label{var}
\end{figure*}

\begin{figure*}
    \centering
    \subfloat[]{
        \includegraphics[width=0.33\linewidth]{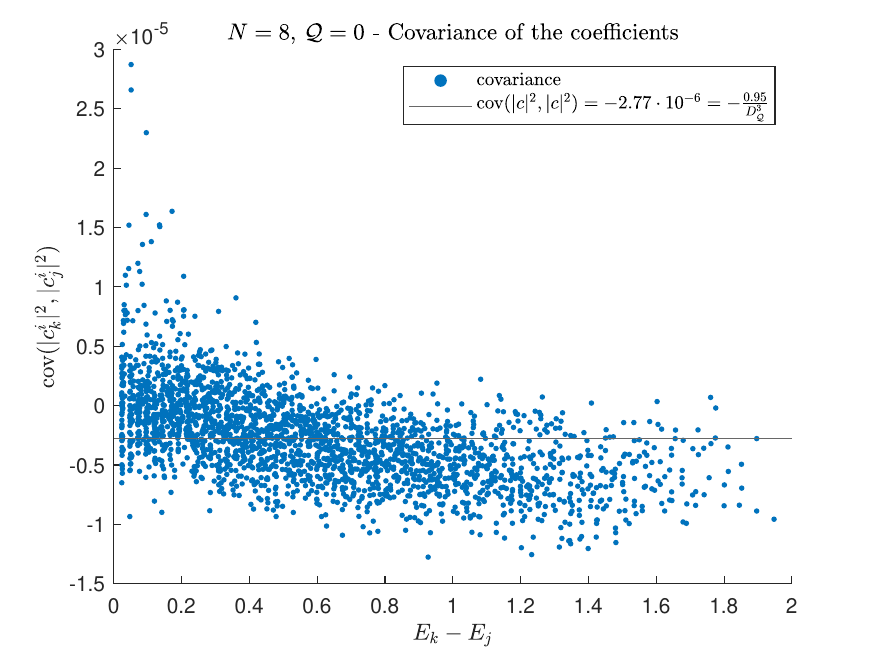}}
    \subfloat[]{
        \includegraphics[width=0.33\linewidth]{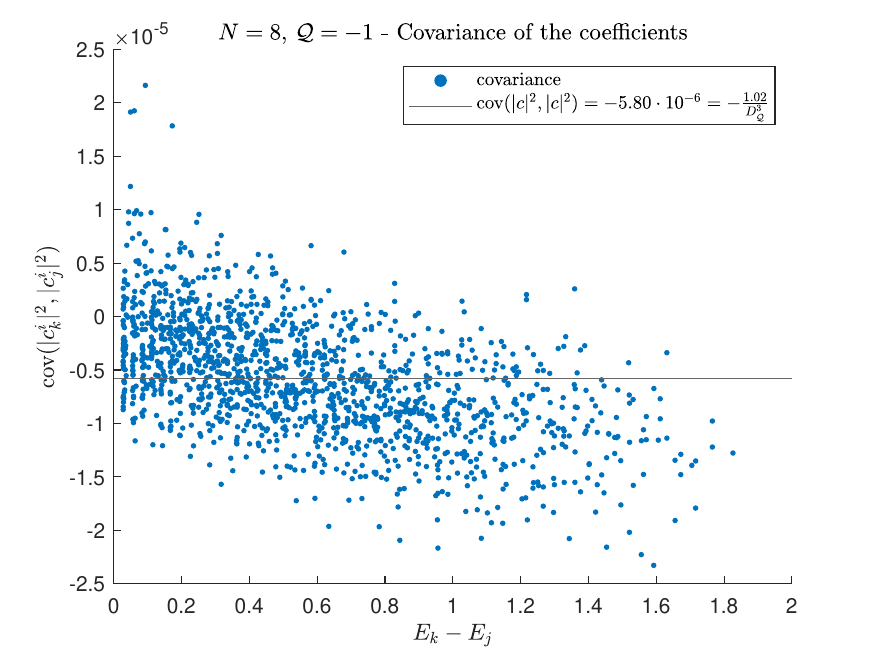}}
    \subfloat[]{
        \includegraphics[width=0.33\linewidth]{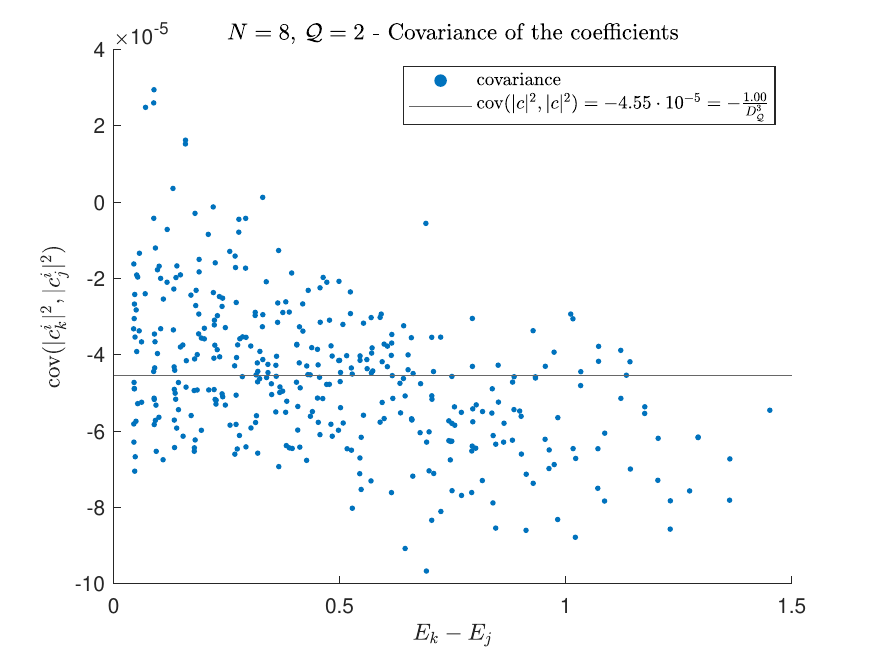}}
    \caption{\textbf{Covariance of the squared norms of coefficients.} Covariance of $|c_k^i|^2$ and $|c_j^i|^2$ for a given Fock state $\ket{S_i}$ and for all energy eigenstates $\ket{E_k}$ and $\ket{E_j}$ in the same charge subsector, averaged over 5000 realizations. The covariance is of order $1/D_{\mathcal{Q}}^3$ (horizontal line). (a) $\mathcal{Q}=0$. (b) $\mathcal{Q}=-1$. (c) $\mathcal{Q}=2$. }
    \label{covar}
\end{figure*}

\begin{figure*}
    \centering
    \subfloat[]{
        \includegraphics[width=0.33\linewidth]{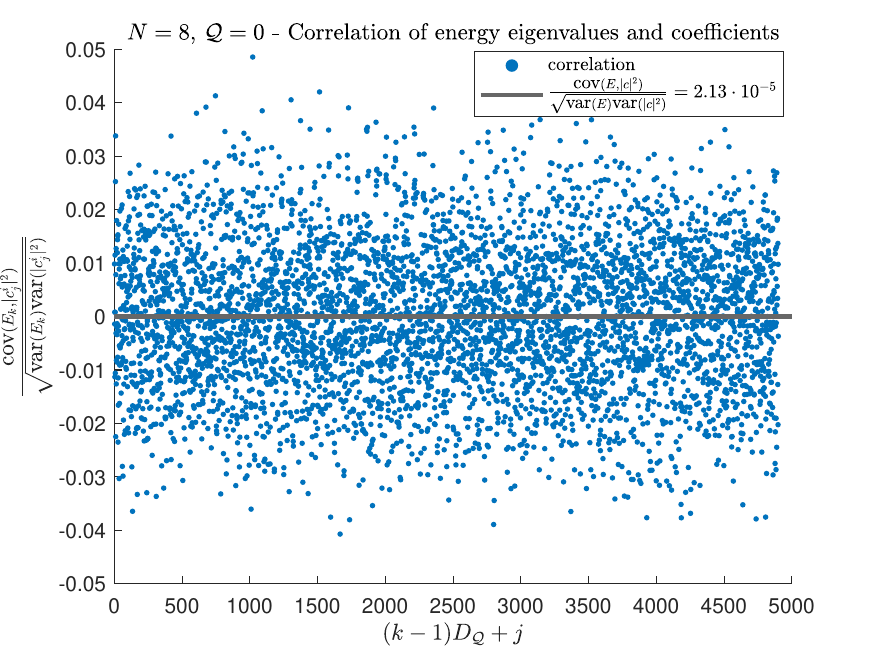}}
    \subfloat[]{
        \includegraphics[width=0.33\linewidth]{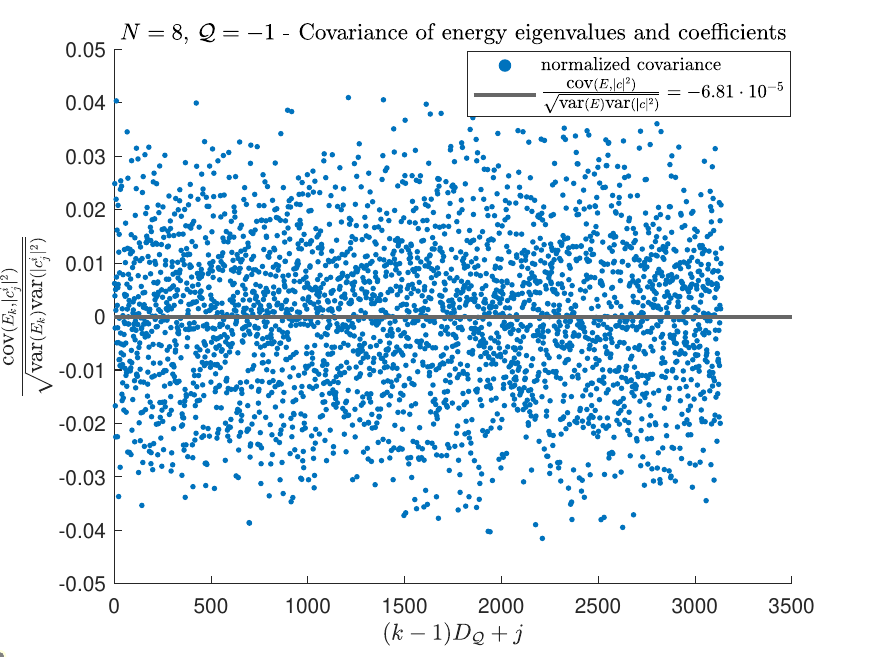}}
    \subfloat[]{
        \includegraphics[width=0.33\linewidth]{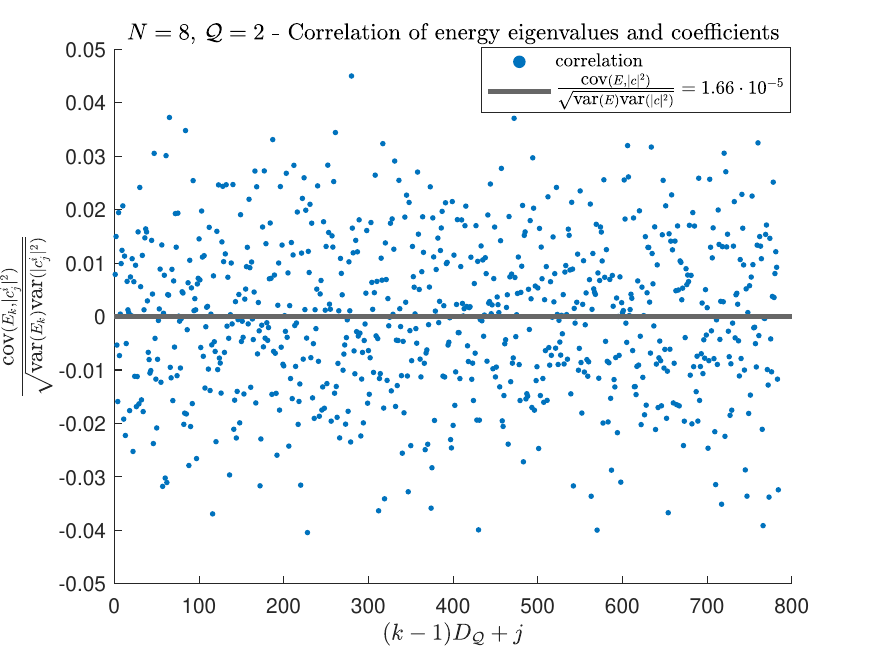}}
    \caption{\textbf{Correlation between squared norms of coefficients and energy eigenvalues.} Correlation between $|c_j^i|^2$ and energy eigenvalues $E_k$ for a given Fock state $\ket{S_i}$ and for all energy eigenstates $\ket{E_j}$ and $\ket{E_k}$ in the same charge subsector, averaged over 5000 realizations. The correlation is clearly strongly suppressed. (a) $\mathcal{Q}=0$. (b) $\mathcal{Q}=-1$. (c) $\mathcal{Q}=2$. }
    \label{enccovar}
\end{figure*}

Our numerical results, reported in Figure \ref{var}, show that the variance of the coefficients is indeed close to $D^{-2}_{\mathcal{Q}}$. Moreover, the numerical data reported in Figure \ref{covar} show a covariance close to $- D_{\mathcal{Q}}^{-3}$. Finally, since all the reasoning above can be applied indistinctly to any energy eigenstate in the charge subsector, the correlation between the energy eigenvalues and the coefficients is also strongly suppressed (see Figure \ref{enccovar}). 


Our numerical results have been obtained using exact diagonalization for $N=8$, and averaging over 5000 realizations of the Hamiltonian. Increasing the number of realizations the points converge to the average values indicated by the horizontal lines, which are therefore very close to the true probability distribution's moments considered. We report here the results for the $\mathcal{Q}=0,-1,2$ charge subsectors.

\subsection{Two-point functions}
\label{appendixb2}

In Section \ref{singlefermionsection} we showed that the denominator of equation (\ref{2pf1fermion}) is given by equation (\ref{denom}). Let us now focus on the numerator of equation (\ref{2pf1fermion}). Since we are inserting a specific fermion $\hat{\psi}_k$, if $\tau_0-|\tau_{1,2}|$ is not large enough to end up in a generic superposition of all the Fock states in the charge subsector before the insertion of the fermion, the numerator of equation (\ref{2pf1fermion}) depends on the occupation number of the $k$-th fermion. However, given the properties of the coefficients $c^i_j$ outlined in Section \ref{coeff} and Appendix \ref{appendixb1}, it is still independent of the occupation number of all the other $N-1$ fermions after averaging over disorder. 

Given a charge subsector with charge $\mathcal{Q}$, there are $n_{\uparrow,\mathcal{Q}}=\mathcal{Q}+N/2$ fermions in the $\ket{\uparrow}$ state. Then the $D_\mathcal{Q}={{N}\choose{n_{\uparrow,\mathcal{Q}}}}$ states in the Fock basis of the charge subsector are split up in $ D_{\mathcal{Q},k=\uparrow}={{N-1}\choose{(n_{\uparrow,\mathcal{Q}})-1}}$ states with the $k$-th fermion in the $\ket{\uparrow}$ state and $ D_{\mathcal{Q},k=\downarrow}={{N-1}\choose{n_{\uparrow,\mathcal{Q}}}}$ states with the $k$-th fermion in the $\ket{\downarrow}$ state.

We can therefore consider the numerator of equation (\ref{2pf1fermion}) and sum over all the $ D_{\mathcal{Q},k=\uparrow}$ (correspondingly, $ D_{\mathcal{Q},k=\downarrow}$) Fock states with the $k$-th fermion in the $\ket{\uparrow}$ (correspondingly, $\ket{\downarrow}$) state. This is equivalent to summing over all the $D_{\mathcal{Q}}$ Fock states in the charge subsector after inserting the projector $P_{k=\uparrow}=\hat{\psi}_k^\dag(-\tau_0)\hat{\psi}_k(-\tau_0)$ (correspondingly, $P_{k=\downarrow}=\hat{\psi}_k(-\tau_0)\hat{\psi}_k^\dag(-\tau_0)$). Using the cyclic property of the trace we obtain
\begin{widetext}
\begin{equation}
    \sum_{\{\ket{S_{i,k=\uparrow}}\}}\braket{S_i|\textrm{e}^{-\tau_0 \hat{H}}T\left[\hat{\psi}_k (\tau_1)\hat{\psi}_k^\dag (\tau_2)\right]\textrm{e}^{-\tau_0 \hat{H}}|S_i}=\Tr_\mathcal{Q}\left(\textrm{e}^{-2\tau_0\hat{H}}T\left[\hat{\psi}_k (\tau_1)\hat{\psi}_k^\dag (\tau_2)\hat{\psi}_k^\dag (-\tau_0+0^+)\hat{\psi}_k (-\tau_0)\right]\right)
\end{equation}
\begin{equation}
    \sum_{\{\ket{S_{i,k=\downarrow}}\}}\braket{S_i|\textrm{e}^{-\tau_0 \hat{H}}T\left[\hat{\psi}_k (\tau_1)\hat{\psi}_k^\dag (\tau_2)\right]\textrm{e}^{-\tau_0 \hat{H}}|S_i}=\Tr_\mathcal{Q}\left(\textrm{e}^{-2\tau_0\hat{H}}T\left[\hat{\psi}_k (\tau_1)\hat{\psi}_k^\dag (\tau_2)\hat{\psi}_k (-\tau_0+0^+)\hat{\psi}_k^\dag (-\tau_0)\right]\right).
\end{equation}
\end{widetext}
After averaging over disorder, each element in the sum on the left hand sides does not depend on the occupation number of the remaining $N-1$ fermions, and has the same occupation number for the $k$-th fermion: all the elements of the sum take the same value. Since the distribution of the coefficients is sharply peaked, at leading order in $1/D_{\mathcal{Q}}$ the latter property is still valid for a single drawing of the couplings. This is the same self-averaging property invoked in Section \ref{singlefermionsection} for the denominator of equation (\ref{2pf1fermion}). Therefore we obtain
\begin{widetext}
\begin{equation}
    \braket{S_{i,k=\uparrow}|\textrm{e}^{-\tau_0 \hat{H}}T\left[\hat{\psi}_k (\tau_1)\hat{\psi}_k^\dag (\tau_2)\right]\textrm{e}^{-\tau_0 \hat{H}}|S_{i,k=\uparrow}}=\frac{1}{D_{\mathcal{Q},k=\uparrow}}\Tr_\mathcal{Q}\left(\textrm{e}^{-2\tau_0\hat{H}}T\left[\hat{\psi}_k (\tau_1)\hat{\psi}_k^\dag (\tau_2)\hat{\psi}_k^\dag (-\tau_0+0^+)\hat{\psi}_k (-\tau_0)\right]\right)
    \label{uparrownum}
\end{equation}
\begin{equation}
    \braket{S_{i,k=\downarrow}|\textrm{e}^{-\tau_0 \hat{H}}T\left[\hat{\psi}_k (\tau_1)\hat{\psi}_k^\dag (\tau_2)\right]\textrm{e}^{-\tau_0 \hat{H}}|S_{i,k=\downarrow}}=\frac{1}{D_{\mathcal{Q},k=\downarrow}}\Tr_\mathcal{Q}\left(\textrm{e}^{-2\tau_0\hat{H}}T\left[\hat{\psi}_k (\tau_1)\hat{\psi}_k^\dag (\tau_2)\hat{\psi}_k (-\tau_0+0^+)\hat{\psi}_k^\dag (-\tau_0)\right]\right).
    \label{downarrownum}
\end{equation}
\end{widetext}
Note that these expectation values are of order $D_{\mathcal{Q}}^0$. Finally, dividing equations (\ref{uparrownum}) and (\ref{downarrownum}) by equation (\ref{denom}), at leading order in $1/D_{\mathcal{Q}}$ we find the result reported in equation (\ref{2pf4pf}). 

\subsubsection*{Numerical results}

\begin{figure*}
    \centering
    \subfloat[]{
        \includegraphics[width=0.4\linewidth]{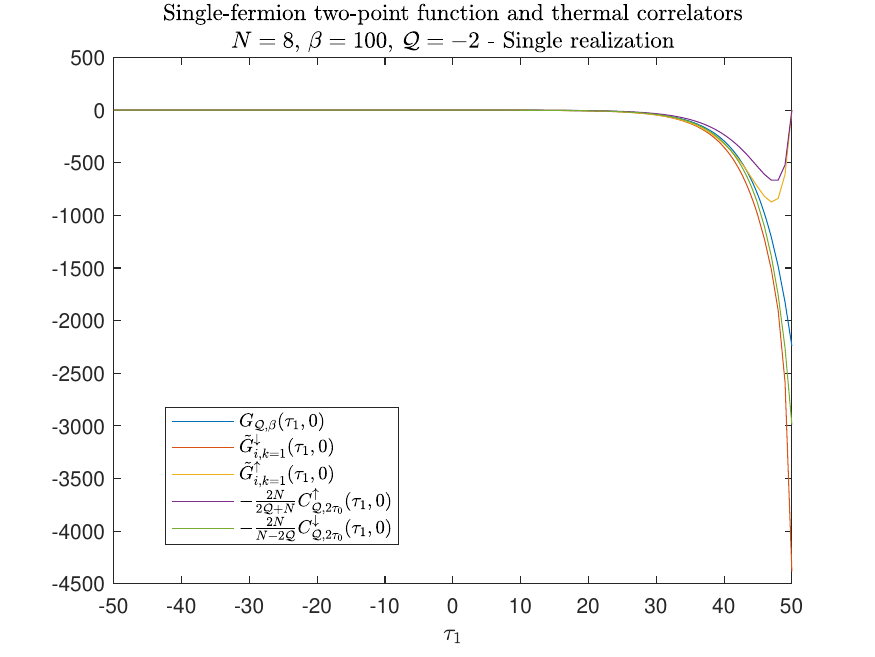}}
    \subfloat[]{
        \includegraphics[width=0.4\linewidth]{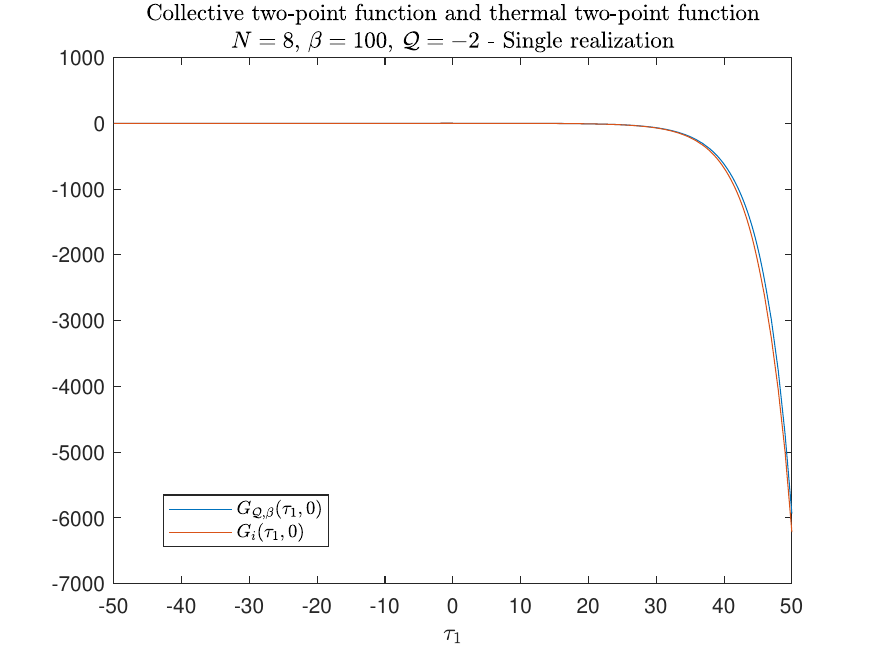}}
   \caption{\textbf{Single fermion and collective two-point functions.} $\mathcal{Q}=-2$, $N=8$, $k=1$, $\beta=2\tau_0=100$, single realization of the Hamiltonian. The expectations outlined in Sections \ref{singlefermionsection} and \ref{collectivesection} are matched to good accuracy. The evident asymmetry is expected and explained in Appendix \ref{appendixb2}. (a) Single fermion two-point functions. The left hand sides and right hand sides of equation (\ref{2pf4pf}) are plotted along with the thermal two-point function in the corresponding charge subsector. (b) Collective two-point functions. The left hand side and right hand side of equation (\ref{coll2pf}) are plotted.}
    \label{singleqm2}
\end{figure*}

\begin{figure*}
    \centering
    \subfloat[]{
        \includegraphics[width=0.33\linewidth]{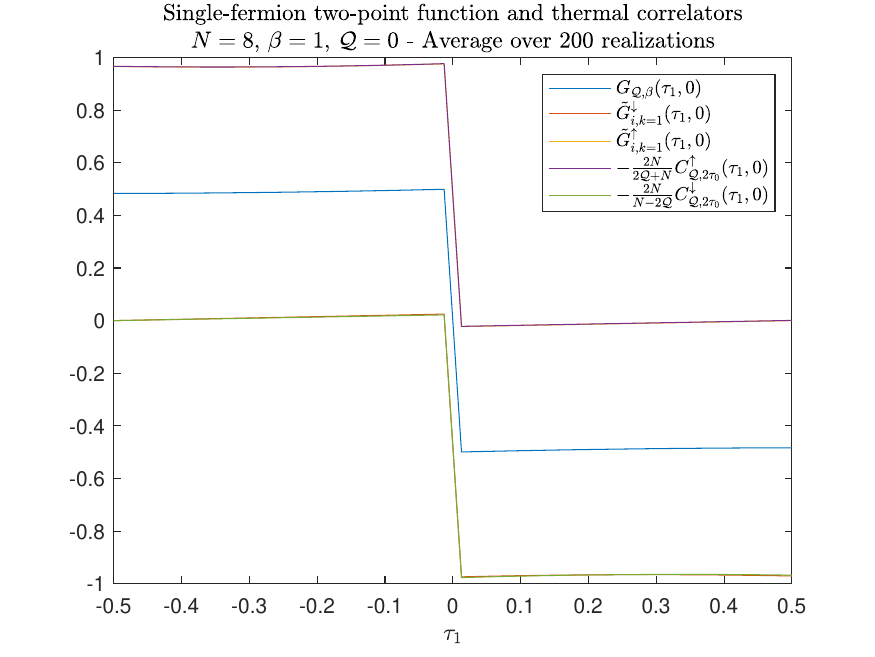}}
    \subfloat[]{
        \includegraphics[width=0.33\linewidth]{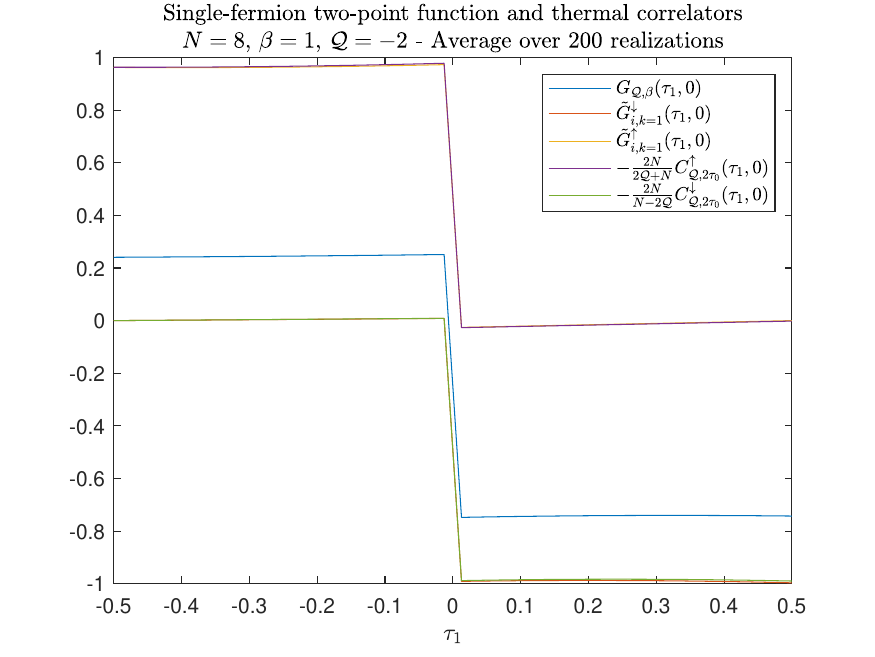}}
    \subfloat[]{
        \includegraphics[width=0.33\linewidth]{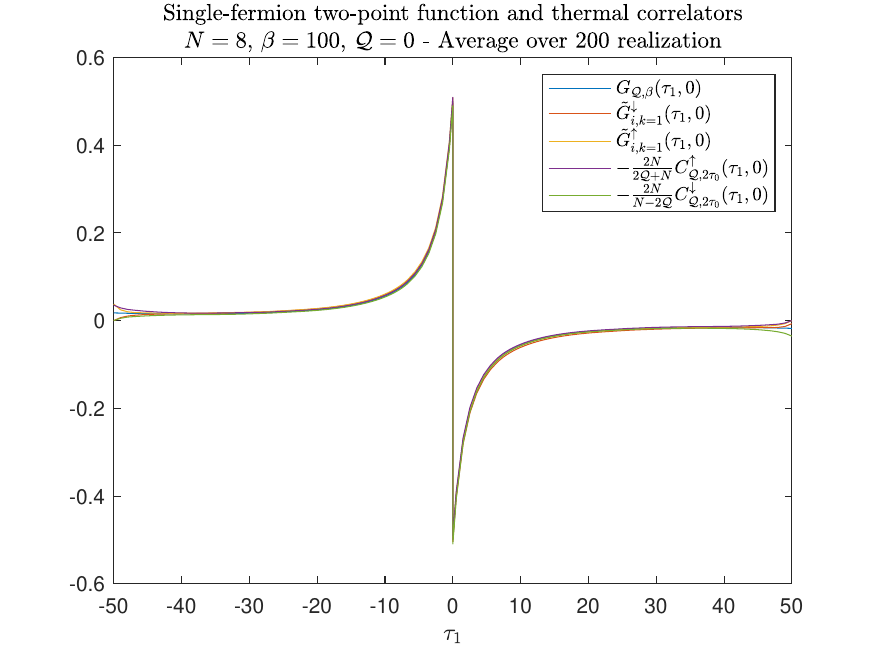}}
   \caption{\textbf{Single fermion two-point functions: average over disorder.} The left hand sides and right hand sides of equation (\ref{2pf4pf}) are plotted along with the thermal two-point function in the corresponding charge subsector for $N=8$ and $k=1$, averaged over 200 realizations of the Hamiltonian. The expectations outlined in Section \ref{singlefermionsection} are matched to good accuracy, and more precisely than for a single realization of the Hamiltonian, as expected. (a) $\mathcal{Q}=0$, $\beta=2\tau_0=1$. (b) $\mathcal{Q}=-2$, $\beta=2\tau_0=1$. (c) $\mathcal{Q}=0$, $\beta=2\tau_0=100$.}
    \label{200single}
\end{figure*}

\begin{figure*}
    \centering
    \subfloat[]{
        \includegraphics[width=0.33\linewidth]{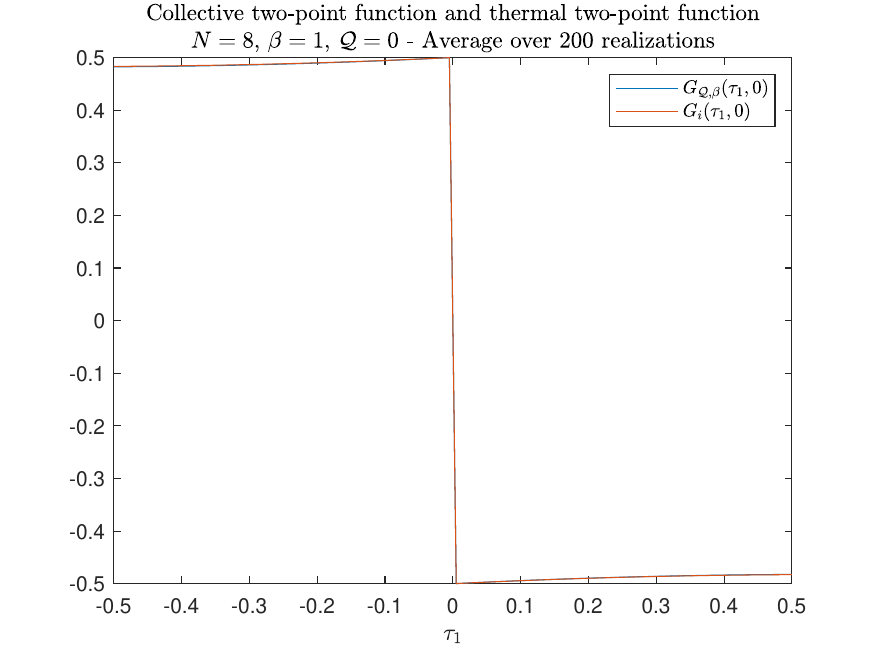}}
    \subfloat[]{
        \includegraphics[width=0.33\linewidth]{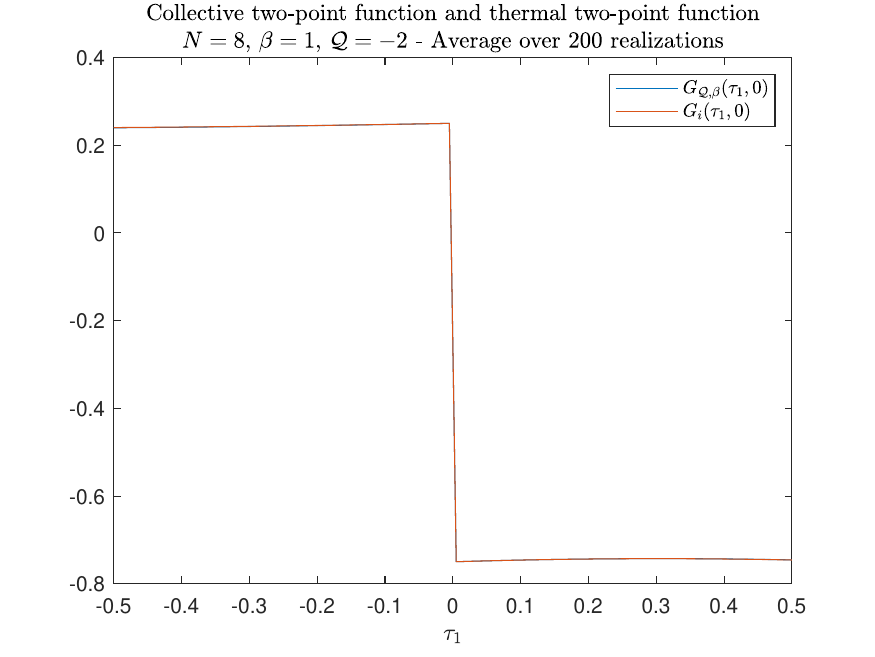}}
     \subfloat[]{
        \includegraphics[width=0.33\linewidth]{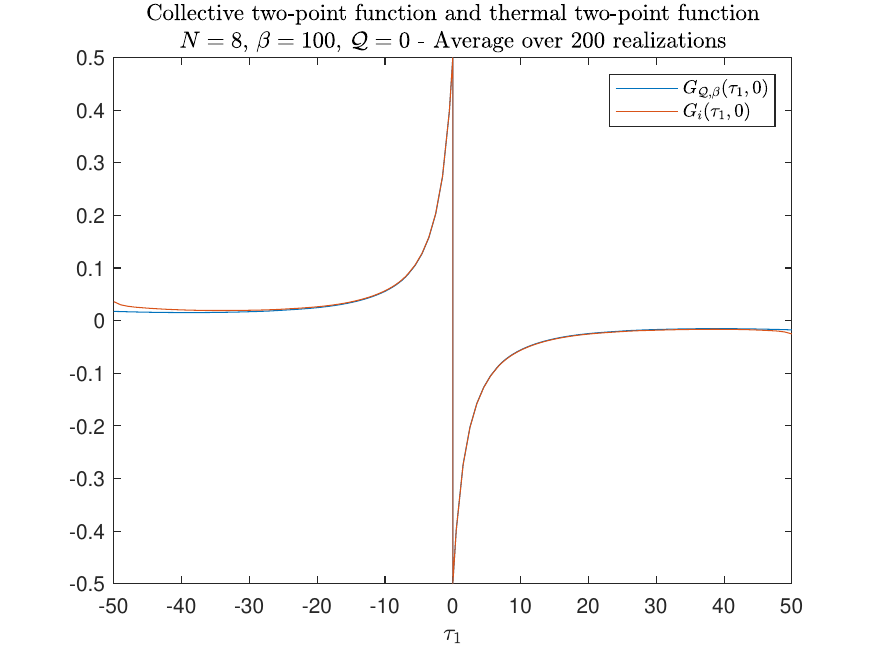}}
   \caption{\textbf{Collective two-point functions: average over disorder.} The left hand side and right hand side of equation (\ref{coll2pf}) are plotted for $N=8$ and $k=1$, averaged over 200 realizations of the Hamiltonian. The expectations outlined in Section \ref{collectivesection} are matched to good accuracy, and more precisely than for a single realization of the Hamiltonian, as expected. (a) $\mathcal{Q}=0$, $\beta=2\tau_0=1$. (b) $\mathcal{Q}=-2$, $\beta=2\tau_0=1$. (c) $\mathcal{Q}=0$, $\beta=2\tau_0=100$.}
    \label{200collective}
\end{figure*}

\begin{figure*}
    \centering
    \subfloat[]{
        \includegraphics[width=0.4\linewidth]{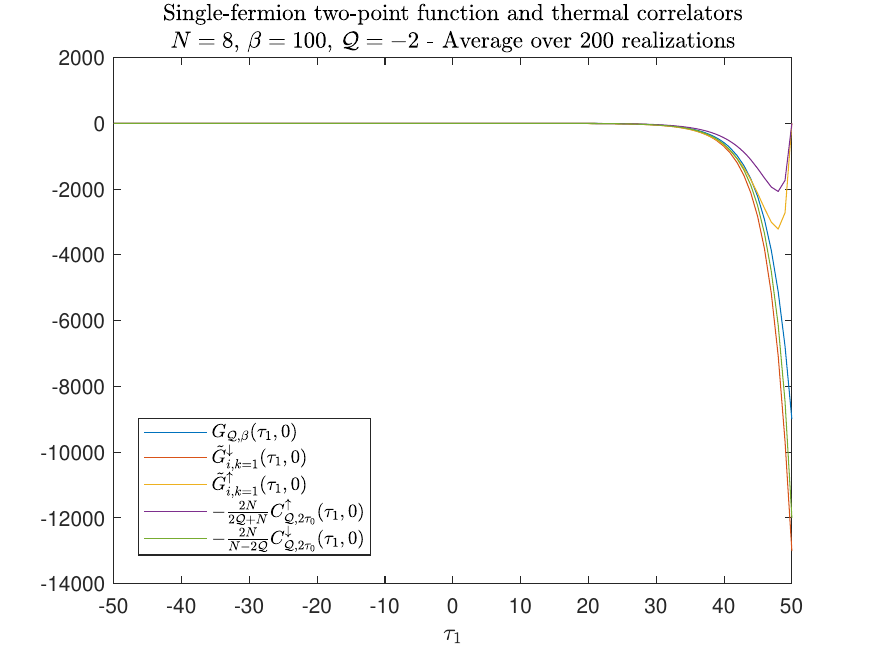}}
    \subfloat[]{
        \includegraphics[width=0.4\linewidth]{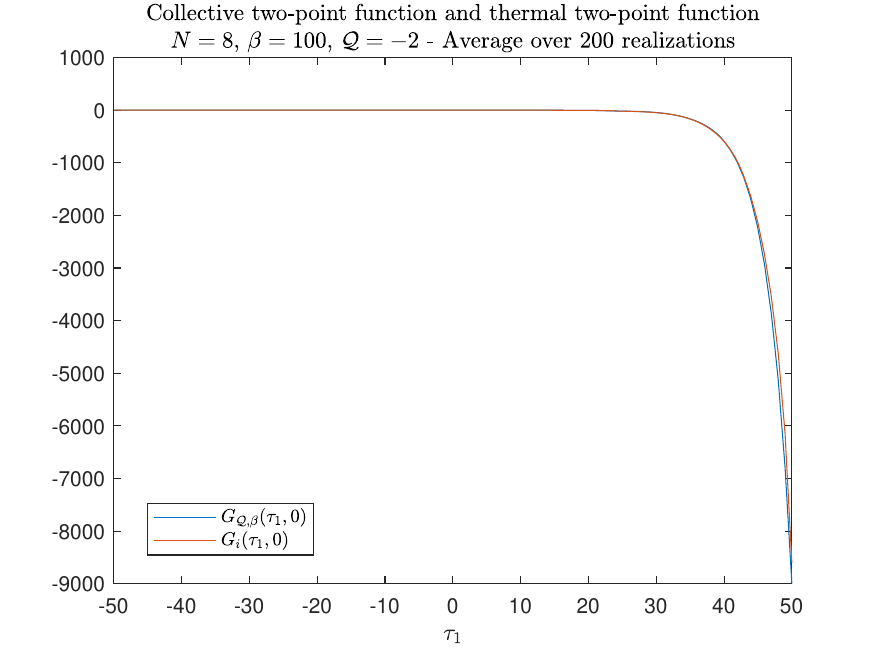}}
   \caption{\textbf{Single fermion and collective two-point functions: average over disorder.} $\mathcal{Q}=-2$, $N=8$, $k=1$, $\beta=2\tau_0=100$, average over 200 realizations of the Hamiltonian. The expectations outlined in Sections \ref{singlefermionsection} and \ref{collectivesection} are matched to good accuracy, and more precisely than for a single realization of the Hamiltonian. The evident asymmetry is expected and explained in Appendix \ref{appendixb2}. (a) Single fermion two-point functions. The left hand sides and right hand sides of equation (\ref{2pf4pf}) are plotted along with the thermal two-point function in the corresponding charge subsector. (b) Collective two-point functions. The left hand side and right hand side of equation (\ref{coll2pf}) are plotted.}
    \label{200qm2}
\end{figure*}

We report here for completeness additional numerical results for the two point functions. In particular, the single fermion and collective two-point functions in the $\mathcal{Q}=-2$ charge subsector for $N=8$, $\tau_0=50$ are plotted in Figure \ref{singleqm2} for a single realization of the Hamiltonian and in Figure \ref{200qm2} averaged over 200 realizations of the Hamiltonian. Note that, as we have pointed out in footnote \ref{footnote}, when evolving for a long amount of Euclidean time we end up in the ground state of the corresponding charge subsector if $N<\beta J$ like in this case. In particular, in Figure \ref{singleqm2} and \ref{200qm2} we insert $\hat{\psi}^\dag$ at time $\tau_2=0$ and $\hat{\psi}$ at time $\tau_1$. Therefore, when $\tau_1\lesssim \tau_0$, at time $\tau=0$ the bra is roughly the ground state of the $\mathcal{Q}=-1$ charge subsector. We then insert $\hat{\psi}^\dag$ and take the scalar product with the ket (which is roughly the ground state of the $\mathcal{Q}=-2$ charge subsector), obtaining in the numerator a result of order $\exp[-\tau_0(E_{-1}+E_{-2})]$, where $E_{-1}$ and $E_{-2}$ are the energy eigenvalues of the ground states of the $\mathcal{Q}=-1$ and $\mathcal{Q}=-2$ charge subsectors, respectively. The normalization factor will clearly be of order $\exp[-2\tau_0E_{-2}]$. Therefore we obtain a result of order $\exp[-\tau_0(E_{-1}-E_{-2})]$. The plots in Figures \ref{singleqm2} and \ref{200qm2} suggest a spacing $|E_{-2}-E_{-1}|=\mathcal{O}(10^{-1})$ and $E_{-1}<E_{-2}$, in accordance with our numerical result and \cite{sachdev}. On the other hand, for $\tau_1\gtrsim -\tau_0$, we expect a result of order $\exp[-\tau_0(E_{-3}-E_{-2})]$, which is exponentially suppressed since $E_{-3}>E_{-2}$, hence the asymmetry observed in Figures \ref{singleqm2} and \ref{200qm2}. For the $\mathcal{Q}=0$ charge subsector (see Figures \ref{n8single}, \ref{collective8}, \ref{200single} and \ref{200collective}), this effect is much smaller because $|E_{0}-E_{\pm 1}|$ is about one order of magnitude smaller than $|E_{-2}-E_{-1}|$. Note also that the behavior of the single fermion two-point function (and the corresponding thermal four-point function) in Figures \ref{singleqm2} and \ref{200qm2} when inserting the fermion very close to $\tau=\tau_0$ depends on the specific boundary state chosen. In particular, when $\hat{\psi}_1$ is inserted right next to the bra Fock state (i.e. at $\tau=\tau_0$) with the first fermion in the $\ket{\uparrow}$ state, the state is annihilated and the correlator correctly vanishes.

Finally, we also report in Figures \ref{200single} and \ref{200collective} the same plots shown in Figures \ref{n8single} and \ref{collective8}, but averaged over 200 realizations of the Hamiltonian. As expected, the accuracy of the result (\ref{2pf4pf}), which is already consistent for a single realization of the Hamiltonian, is increased when averaging over disorder.

\subsection{Expectation values of generic operators}
\label{appendixb3}

In Section \ref{thermal} we claimed that, under appropriate assumptions, the expectation value $\braket{B_i|\hat{O}|B_i}$ of a generic operator $\hat{O}$ over a boundary state is equivalent to the thermal expectation value computed in the same charge subsector. In particular, given the properties of the probability distribution of the coefficients $c_i^k$, such statement is valid at leading order in $1/D_{\mathcal{Q}}$ both after averaging over disorder and for a single realization of the Hamiltonian. Let us know prove this result, working under the assumptions listed in Section \ref{thermal}.

First, let us clarify assumption 2, concerning the set of operators for which the claim holds. As a simple example, consider an operator of the form
\begin{equation}
   \hat{O} = T\left[\prod_{i \in M^-} \hat{\psi}_{i}(\tau^-_i) \prod_{j \in M^+} \hat{\psi}^\dagger_{j}(\tau^+_j) \right]
\end{equation}
where $M^+$ and $M^-$ are index sets of equal size (the case of non-equal size is trivial since operators with unequal numbers of creation and annihilation operators have zero expected value in a fixed charge subsector). Provided the number of insertions is much less than $N$ and provided all the times $\tau^{\pm}_i$ are well separated from $\pm \tau_0$, we expect such operators to have thermal-looking expectation values over a boundary state, up to corrections suppressed by $1/N$ and $\exp(-(\tau_0-|\tau_i^\pm|)J)$. Naturally, this can be the case only if the preparation time is large enough, i.e. $\tau_0 J\gg 1$.

A simple example where assumption 2 fails is when $\hat{O} = \textrm{e}^{\tau_0 \hat{H}} \hat{\psi}_i^\dagger \hat{\psi}_i \textrm{e}^{-\tau_0 \hat{H}}$. In this case, the fermion occupation number operator is inserted right next to the Fock state and hence is sensitive to the precise occupation number of the $i$-th fermion. However, collective fields like
\begin{equation}
   \hat{O}_{\text{coll}} = T\left[\frac{1}{N}\sum_{i=1}^N \hat{\psi}_i^\dagger(\tau_1) \hat{\psi}_i(\tau_2)\right]
   \label{collop}
\end{equation}
are better-behaved. Note that the operator (\ref{collop}) is invariant under permutations of the fermions. The Hamiltonian evolution is also effectively permutation-invariant at large $N$ due to the emergent $U(N)$ symmetry. It follows that the expectation value of permutation-invariant collective operators such as (\ref{collop}) is independent of the specific boundary state up to $1/N$-suppressed corrections, regardless of the value of the preparation time $\tau_0$ and of the location of the fermion insertion points. The collective two-point functions studied in Section \ref{collectivesection} provide a clear example of this behavior.

Given a generic operator $\hat{O}$ satisfying assumptions 2 and 3, we define $\alpha_{ij}\equiv \braket{E_i|\hat{O}|E_j}\exp(-\tau_0(E_i+E_j))$ and $\beta_{i}\equiv \exp(-2\tau_0E_i)$. Using equation (\ref{superpos}), we can then write the expectation value over a boundary state $\ket{B_i}$ as
\begin{equation}
    \braket{O}_i\equiv\braket{B_i|\hat{O}|B_i}=\frac{\sum_{k,l=1}^{D_\mathcal{Q}}|c_k^i||c_l^i|\alpha_{lk}\textrm{e}^{i(\theta_k-\theta_l)}}{\sum_{k=1}^{D_\mathcal{Q}}|c_k^i|^2\beta_k}
\label{expvalueo}
\end{equation}
where $\theta_k$ are the phases of the coefficients $c_k^i$. Averaging over the phases, which are uniformly distributed, we get 
\begin{equation}
    \overline{\braket{O}}_i=\frac{\sum_{k=1}^{D_\mathcal{Q}}|c_k^i|^2\alpha_{kk}}{\sum_{k=1}^{D_\mathcal{Q}}|c_k^i|^2\beta_k}.
    \label{avgtheta1}
\end{equation}
(This statement fails for $\hat{O} = \textrm{e}^{\tau_0 \hat{H}} \hat{\psi}_i^\dagger \hat{\psi}_i \textrm{e}^{-\tau_0 \hat{H}}$ because the $\alpha_{lk}$ can be correlated with the phases.) We will show at the end of the present section that the average over the squared norms of the coefficients can be taken separately in the numerator and denominator, and the average of the ratio is equal to the ratio of the averages up to corrections of order $1/D_{\mathcal{Q}}$. Therefore, we obtain immediately\footnote{A similar result for the expectation value of the Hamiltonian in analogous cSYK pure states was found in \cite{eth3}.}
\begin{equation}
    \overline{\overline{\braket{O}}}_i=\frac{\sum_{k=1}^{D_\mathcal{Q}}\alpha_{kk}}{\sum_{k=1}^{D_\mathcal{Q}}\beta_k}+\mathcal{O}\left(\frac{1}{D_\mathcal{Q}}\right)
\label{oaverageappendix}
\end{equation}
which is the result (\ref{oaverage}) (a single overline indicates an average over the phases and a double overline an average over both the phases and the squared norms of the coefficients). Therefore, averaging over disorder, generic correlators over a boundary state are equal, up to corrections suppressed at large $N$, to the thermal correlators in the same charge subsector. Note that when averaging over the coefficients we used assumption 1 to ignore the (small) correlation between the coefficients and the energy eigenvalues contained in $\alpha_{ij}$ and $\beta_i$.

Let us now analyze the size of fluctuations around the average value (\ref{oaverageappendix}). Taking the square of equation (\ref{expvalueo}) and averaging over the phases we obtain (all the sums are from $1$ to $D_\mathcal{Q}$)
\begin{equation}
    \overline{\braket{O}_i^2}=\frac{\sum_{k\neq l}|c_k^i|^2|c_l^i|^2\left(\alpha_{kk}\alpha_{ll}+\alpha_{lk}\alpha_{kl}\right)+\sum_{k}|c_k^i|^4\alpha_{kk}^2}{\sum_{k,l}|c_k^i|^2|c_l^i|^2\beta_k\beta_l}.
     \label{avgtheta2}
\end{equation}
Again, at leading order in $1/D_{\mathcal{Q}}$ the average over the squared norms can be taken separately in the numerator and denominator. Subtracting the square of equation (\ref{oaverageappendix}) and keeping only the leading orders we obtain the following expression for the variance of the expectation value:
\begin{widetext}
\begin{equation}
    \overline{\overline{\braket{O}_i^2}}-\overline{\overline{\braket{O}}}_i^2=\frac{\sum_{\substack{k\neq l\\ m,n}}\alpha_{lk}\alpha_{kl}\beta_m\beta_n+a\sum_{k,l,m}\left(\alpha_{kk}^2\beta_l\beta_m-\alpha_{kk}\alpha_{ll}\beta_m^2\right)}{\left(\sum_{k,l}\beta_k\beta_l\right)^2}+\mathcal{O}\left(\frac{1}{D_\mathcal{Q}}\right)
    \label{varianceo}
\end{equation}
\end{widetext}
where $a\approx 1$ is the variance of the norm of the coefficients. Let us now evaluate the order of each term in equation (\ref{varianceo}). For a generic $\tau_0$, $\beta_k$ and $\alpha_{kk}$ are $\mathcal{O}(1)$, while $\alpha_{kl}=\mathcal{O}(1/\sqrt{D_\mathcal{Q}})$ for $k\neq l$ by assumption 3. The first sum in the numerator contains $D_{\mathcal{Q}}^3(D_{\mathcal{Q}}-1)\approx D_{\mathcal{Q}}^4$ terms, the second sum in the numerator contains $D_{\mathcal{Q}}^3$ terms and the sum in the denominator contains $D_{\mathcal{Q}}^4$ terms. Therefore we obtain
\begin{equation}
     \overline{\overline{\braket{O}_i^2}}-\overline{\overline{\braket{O}}}_i^2=\mathcal{O}\left(\frac{1}{D_{\mathcal{Q}}}\right)
     \label{varfinal}
\end{equation}
which is equation (\ref{fluctuations}). Note that if we exclude the subsectors with maximal and minimal charge, $D_\mathcal{Q}\geq N$. Since the size of the fluctuations of a generic correlator around its disorder-averaged value is controlled by the square root of the variance (\ref{varfinal}), it is suppressed in the large $N$ limit. This implies that also for a single realization of the Hamiltonian the expectation value of a generic operator $\hat{O}$ over a boundary state approaches in the large $N$ limit the thermal expectation value in the same charge subsector computed by the canonical ensemble at fixed charge. In other words, boundary states ``look'' thermal. This result generalizes the analytic and numerical results for the two-point functions analyzed in Sections \ref{singlefermionsection} and \ref{collectivesection} and in Appendix \ref{appendixb2}.

In order to obtain the results (\ref{oaverageappendix}) and (\ref{varfinal}) we averaged the numerator and denominator separately over the norms of the coefficients, and then took the ratio of the averages. To complete our proof we need to show that, at leading order in $1/D_\mathcal{Q}$, this procedure is equivalent to taking the average of the ratio\footnote{Note that this property holds exactly for the average over the phases.}. Let us denote by $\overline{\mathcal{N}}$ and $\overline{\mathcal{D}}$ the numerator and denominator of equations (\ref{avgtheta1}) or (\ref{avgtheta2}). Then we can write $\overline{\mathcal{N}}=\overline{\overline{\mathcal{N}}}+\overline{\delta \mathcal{N}}$ and $\overline{\mathcal{D}}=\overline{\overline{\mathcal{D}}}+\overline{\delta\mathcal{D}}$. Both $\overline{\overline{\mathcal{N}}}$ and $\overline{\overline{\mathcal{D}}}$ are $\mathcal{O}(1)$ numbers. By construction $\overline{\overline{\delta\mathcal{N}}}=\overline{\overline{\delta\mathcal{D}}}=0$. Using the properties of the coefficients $c_k^i$, the size of the fluctuations of $\overline{\delta\mathcal{N}}$ and $\overline{\delta\mathcal{D}}$ is controlled by $(\overline{\overline{\mathcal{N}^2}}-\overline{\overline{\mathcal{N}}}^2)^{\frac{1}{2}}=\mathcal{O}(D_{\mathcal{Q}}^{-\frac{1}{2}})$ and $(\overline{\overline{\mathcal{D}^2}}-\overline{\overline{\mathcal{D}}}^2)^{\frac{1}{2}}=\mathcal{O}(D_{\mathcal{Q}}^{-\frac{1}{2}})$ respectively. Therefore, in the large $N$ limit $\delta\overline{\mathcal{N}}/\overline{\overline{\mathcal{N}}}\ll 1$ and $\delta\overline{\mathcal{D}}/\overline{\overline{\mathcal{D}}}\ll 1$ with high probability. In the same way, it is possible to show that $\overline{\delta\overline{\mathcal{N}}\delta\overline{\mathcal{D}}}=\overline{\overline{(\mathcal{ND})}}-(\overline{\overline{\mathcal{N}}})(\overline{\overline{\mathcal{D}}})=\mathcal{O}(D_{\mathcal{Q}}^{-1})$. We can then write
\begin{equation}
    \overline{\left(\frac{\mathcal{N}}{\mathcal{D}}\right)}=\frac{\overline{\mathcal{N}}}{\overline{\mathcal{D}}}=\frac{\overline{\overline{\mathcal{N}}}}{\overline{\overline{\mathcal{D}}}}\left(1+\frac{\delta\overline{\mathcal{N}}}{\overline{\overline{\mathcal{N}}}}-\frac{\delta\overline{\mathcal{D}}}{\overline{\overline{\mathcal{D}}}}+\frac{\delta\overline{\mathcal{N}}\delta\overline{\mathcal{D}}}{(\overline{\overline{\mathcal{N}}})(\overline{\overline{\mathcal{D}}})}\right)+\mathcal{O}\left(\frac{1}{D_\mathcal{Q}}\right)
\end{equation}
which immediately implies
\begin{equation}
    \overline{\overline{\left(\frac{\mathcal{N}}{\mathcal{D}}\right)}}=\frac{\overline{\overline{N}}}{\overline{\overline{D}}}+\mathcal{O}\left(\frac{1}{D_\mathcal{Q}}\right)
\end{equation}
which is the desired result.

\section{Dimensional reduction of braneworld cosmology}
\label{appendixc}

We will give here additional details regarding the dimensional reduction of the braneworld cosmology action $S=S_{bulk}+S_{etw}$ proposed in \cite{Antonini:2019qkt}. We will follow mainly the conventions of \cite{Nayak:2018qej,Sachdev:2019bjn} (see also \cite{Maldacena:2016upp,Sarosi:2017ykf,Brown:2018bms,Moitra:2019bub}), to which we refer for additional details and discussion. The action is the one for a $(d+1)$-dimensional Einstein-Maxwell theory at fixed charge with a negative cosmological constant and a dynamical brane with constant tension $T_{etw}$:
\begin{widetext}
\begin{equation}
    \begin{split}
&S_{bulk}=-\frac{1}{16\pi G}\int_{\mathcal{M}^{(d+1)}}d^{d+1}x\sqrt{g_{(d+1)}}(R_{(d+1)}-2\Lambda-4\pi G F^{(d+1)}_{\mu\nu}F_{(d+1)}^{\mu\nu})+S_{GHY}+S_{\infty}^{em}\\
&S_{etw}=-\frac{1}{8\pi G}\int_{{etw}^{(d)}}d^dy\sqrt{h_d}\left[K_{d}^{etw}-(d-1)T_{etw}\right]+S^{em}_{etw}
\end{split}
\end{equation}
\end{widetext}
where $\mathcal{M}^{(d+1)}$ indicates the spacetime manifold, $g_{(d+1)}$ is the determinant of the metric, $R_{(d+1)}$ is the Ricci scalar, $\Lambda$ is the negative cosmological constant (given by $\Lambda=-d(d-1)/(2L^2_{AdS})$), $F^{(d+1)}_{\mu\nu}$ is the electromagnetic tensor, $h_{d}$ is the determinant of the metric induced on the brane and $K_d^{etw}$ is the trace of the extrinsic curvature of the brane. We used $y^A$ (with $A=0,...,d-1$) to denote the coordinates on the brane, and we will use the same notation for the asymptotic boundary. Greek indices $\mu,\nu$ run instead from 0 to $d$. $S_{GHY}$ is the Gibbons-Hawking-York boundary term, given by
\begin{equation}
    S_{GHY}=-\frac{1}{8\pi G}\int_{\partial\mathcal{M}^{(d)}_{\infty}}d^dy\sqrt{\gamma_{d}}K_{d}^{\infty}
\end{equation}
with $\partial\mathcal{M}_{\infty}^{(d)}$ denoting the asymptotic boundary, $\gamma_{d}$ the induced metric on it and $K^\infty_{d}$ the trace of its extrinsic curvature. The electromagnetic boundary terms $S_{\infty}^{em}$ and $S^{em}_{etw}$ are needed to have a well posed variational problem when we hold the charge fixed \cite{Hawking:1995ap}. Their expression is (we report here only the asymptotic one, but the ETW term is completely analogous):
\begin{equation}
    S^{em}_\infty=-\int_{\partial\mathcal{M}^{(d)}_{\infty}}d^dy\sqrt{\gamma_{d}}F_{(d+1)}^{\mu\nu}n_{\mu}^\infty A_\nu
    \label{embdy}
\end{equation}
where $n_\mu^\infty$ is the one-form dual to the unit vector normal to the asymptotic boundary and $A_\mu$ is the electromagnetic potential.

For the sake of dimensional reduction, it is useful to introduce the following ansatz for the spacetime metric, including a Weyl rescaling of the two-dimensional metric:
\begin{equation}
    ds^2=\frac{r_e^{d-2}}{\phi^{d-2}}g_{ij}dx^idx^j+\phi^2(x^i)d\Omega_{d-1}^2
    \label{metricansatz}
\end{equation}
where $g_{ij}$ is a two-dimensional metric and $\phi(x^i)$ is the dilaton, which is a generic function of the two-dimensional coordinates only (indicated by lower case latin indices $i,j=0,1$). Using the ansatz (\ref{metricansatz}) in the braneworld cosmology action and integrating out the angular part we obtain the following dimensionally reduced action:
\begin{widetext}
\begin{equation}
    \begin{split}
        S=&-\frac{V_{d-1}}{16\pi G}\int_{\mathcal{M}}d^2x\sqrt{g}\left[\phi^{d-1}R+U(\phi)-4\pi G Z(\phi)F_{ij}F^{ij}\right]    -\frac{V_{d-1}}{8\pi G}\int_{\partial\mathcal{M}_\infty}du\sqrt{\gamma_{uu}}\left[\phi^{d-1}K^\infty+8\pi G Z(\phi)F^{ij}n_i^\infty A_j\right]\\
        &-\frac{V_{d-1}}{8\pi G}\int_{etw}dv\sqrt{h_{vv}}\left[\phi^{d-1}K^{etw}-(d-1)\phi^{\frac{d}{2}}r_e^{\frac{d}{2}-1}T_{etw}+8\pi G Z(\phi)F^{ij}n_i^{etw} A_j\right]
    \end{split}
    \label{totactionweyl}
\end{equation}
\end{widetext}
where we defined
\begin{equation}
    \begin{split}
        &U(\phi)=r_e^{d-2}\left[\frac{d(d-1)}{L^2_{AdS}}\phi+\frac{(d-1)(d-2)}{\phi}\right]\\
        &Z(\phi)=\frac{\phi^{2d-3}}{r_e^{d-2}}.
    \end{split}
\end{equation}
As we have already discussed, we are not interested in charge fluctuations, and therefore we can integrate out the gauge field in the path integral \cite{Brown:2018bms}. This sums up to solving Maxwell's equations and computing the bulk and boundary electromagnetic terms on-shell. After long but straightforward calculations, we obtain the effective action 
\begin{widetext}
\begin{equation}
    \begin{split}
        S_{eff}=&-\frac{V_{d-1}}{16\pi G}\int_{\mathcal{M}}d^2x\sqrt{g}\left[\phi^{d-1}R+U(\phi)-(d-1)(d-2)\frac{r_e^{d-2}}{\phi^{2d-3}}Q^2\right]
        -\frac{V_{d-1}}{8\pi G}\int_{\partial\mathcal{M}_\infty}du\sqrt{\gamma_{uu}}\phi^{d-1}K^\infty\\
        &-\frac{V_{d-1}}{8\pi G}\int_{etw}dv\sqrt{h_{vv}}\left[\phi^{d-1}K^{etw}-(d-1)\phi^{\frac{d}{2}}r_e^{\frac{d}{2}-1}T_{etw}\right]
    \end{split}
    \label{totactionweylgaugeonshell}
\end{equation}
\end{widetext}
where the charge parameter $Q$ is uniquely determined by the extremal horizon radius $r_e$ (or vice versa) \cite{Antonini:2019qkt}.

The dimensionally reduced effective action (\ref{totactionweylgaugeonshell}) is valid all the way to the asymptotic boundary and for any value of the charge parameter $Q$. In order to obtain the JT gravity action, we must take the near-extremal, near-horizon limit of equation (\ref{totactionweylgaugeonshell}) and regularize the asymptotic boundary to the region specified by equation (\ref{allcond}). Note that on-shell the dilaton $\phi$ plays the role of the radius of the $(d-1)$-dimensional sphere $S^{d-1}$. Therefore, in the extremal limit and at the black hole horizon $\phi=\phi_0=r_e$, and in the near-extremal, near-horizon limit it will vary slightly from this value. 

It is useful for our purposes to redefine the dilaton as $\Phi=\phi^{d-1}$. We can then split the redefined dilaton in a constant piece $\Phi_0=r_e^{d-1}$ and a dynamical field $\Phi_1$, i.e. $\Phi=\Phi_0+\Phi_1$. In the near-extremal, near-horizon limit $\Phi_1\ll \Phi_0$ holds, and the on-shell expression for $\Phi_1$ (which will be useful to derive the ETW particle trajectory in Appendix \ref{appendixd}) at leading order in $R_2^2/r_ez$ is
\begin{equation}
    \Phi_1\approx (d-1)r_e^{d-2}\frac{R_2^2}{z}.
    \label{onshelldilatonappendix}
\end{equation}
We can now expand the effective action (\ref{totactionweylgaugeonshell}) to first order in $\Phi_1/\Phi_0$, finally obtaining the result (\ref{JTaction}), where
\begin{widetext}
\begin{equation}
\begin{split}
    &S_{0}=-\frac{V_{d-1}\Phi_0}{16\pi G}\left[\int_\mathcal{M}d^2x\sqrt{g}R+2\int_{\partial\mathcal{M}_\infty}du\sqrt{\gamma_{uu}}K^\infty+2\int_{etw}dv\sqrt{h_{vv}}K^{etw}\right],\\
    &S_0^{etw}=\frac{(d-1)V_{d-1}\Phi_0^{\frac{d}{2(d-1)}}T_{etw}}{8\pi G}\int_{etw}dv\sqrt{h_{vv}}.
    \end{split}
    \label{s0terms}
\end{equation}
\end{widetext}
$S_0$ is a topological term and it is a constant by Gauss-Bonnet theorem\footnote{We remind that the boundary of the manifold $\mathcal{M}$ is $\partial\mathcal{M}=\partial\mathcal{M}_{\infty}\cup ETW$.}. $S_0^{etw}$ is proportional to the proper length of the ETW particle trajectory, and will give a contribution to the equations of motion of the ETW particle, as discussed in Appendix \ref{appendixd}.

\section{Schwarzian action and ETW particle trajectory}
\label{appendixd}

In this Appendix we will outline the steps needed to obtain the Schwarzian effective action (\ref{effactionsch}) governing the gravitational dynamics in the near-horizon region of the near-extremal black hole, and derive the NN boundary conditions that must be imposed at the location of the ETW particle, leading to the trajectory described in Section \ref{trajectorysection}.

As we have already pointed out, we must integrate out the dilaton $\Phi_1$ and compute the effective action (\ref{JTaction}) on-shell. The only non-vanishing and non-constant term is the GHY term for the regularized boundary. The metric is fixed to be the two-dimensional part of equation (\ref{nearmetric}), and the induced metric and the dilaton on the regualarized boundary must satisfy the DD boundary conditions (\ref{ddbc}). After parametrizing the trajectory of the regularized boundary in the bulk by $(\tau(u),z(u))$, the boundary condition for the metric leads to equation (\ref{boundaryparam}), while the one-form dual to the outward-pointing unit vector normal to the regularized boundary is given by $n^\infty_i=N(z'(u),-\tau'(u))$, where $N$ is an appropriate normalization factor that can be computed by imposing $g^{ij}n^\infty_in^\infty_j=1$. At second order in $\varepsilon$ the extrinsic curvature then reads
\begin{equation}
    K^\infty=\frac{1}{R_2}+\varepsilon^2 R_2\left\{\tan\left(\pi T\tau(u)\right),u\right\}+\mathcal{O}\left(\varepsilon^4\right).
\end{equation}
Using again the DD boundary conditions (\ref{ddbc}), the GHY term for the regularized boundary takes the form 
\small
\begin{equation}
    I_{GHY}\approx -\frac{V_{d-1}\phi_b}{8\pi G}\int_{\partial\mathcal{M}_\infty}du \left[\frac{1}{R_2\varepsilon^2}+R_2\left\{\tan\left(\pi T\tau(u)\right),u\right\}\right].
    \label{effactionnn}
\end{equation}
\normalsize
Note that the first term is divergent for $\varepsilon\to 0$. In order to cancel such constant divergent term, generally in JT gravity a counterterm
\begin{equation}
    I_{ct}=\frac{V_{d-1}}{8\pi G}\int_{\partial\mathcal{M}_\infty}du\sqrt{\gamma_{uu}}\Phi_1\frac{1}{R_2}
\end{equation}
is introduced in the action. We could do the same here. However, since we are working at small but finite cutoff $\varepsilon$, the first term in equation (\ref{effactionnn}) is a constant finite term that does not pose particular problems and does not contribute to the dynamics, so the introduction of the counterterm is not necessary. In either case, we can discard such term together with any other constant term in the on-shell action, obtaining the effective action (\ref{effactionsch}). Note that in the effective action $u\in[-\beta/2,\beta/2]$ if the ETW particle does not intersect the regularized boundary, while $u\in[-u_0,u_0]$ if the ETW particle intersects the regularized boundary at $u=\pm u_0$. In the tangent case of our interest, $u_0=\beta/2$ and therefore the integral in equation (\ref{effactionsch}) runs over the whole thermal circle, i.e. $u\in[-\beta/2,\beta/2]$.

Let us now focus on the ETW particle action. As we have mentioned, we want to impose NN boundary conditions at the location of the ETW particle. This implies that, when the effective action (\ref{JTaction}) is varied, the variations $\delta\sqrt{h_{vv}}$ and $\delta\Phi_1|_{etw}$ do not vanish. The total ETW particle term in the variation of the action (\ref{JTaction}) is given by the sum of the variation of the ETW particle action and boundary terms arising from the variation of the bulk action\footnote{Note that the last term of the first line arises from the variation of $S_0^{etw}$.} \cite{Goel:2020yxl}:
\begin{widetext}
\begin{equation}
\begin{split}
    \delta S_{tot}^{etw}=&-\frac{V_{d-1}}{8\pi G}\int_{etw}dv\Big(\Big\{n_{etw}^i\partial_i\Phi_1-\Phi_1 K^{etw}-(d-1)\Phi_0^{\frac{d}{2(d-1)}}T_{etw}\\
    &\left. \left.+\Phi_1\left[K^{etw}-\frac{d}{2}\Phi_0^{-\frac{d-2}{2(d-1)}}T_{etw}\right]\right\}\delta\left(\sqrt{h_{vv}}\right)+\sqrt{h_{vv}}\left[K^{etw}-\frac{d}{2}\Phi_0^{-\frac{d-2}{2(d-1)}}T_{etw}\right]\delta\Phi_1\right).
    \end{split}
    \label{braneactionvar}
\end{equation}
\end{widetext}
The NN boundary conditions we are looking for are obtained by imposing $\delta S_{tot}^{etw}=0$, which immediately leads to the boundary conditions (\ref{nnbcmet}) and (\ref{NNbcdil}).

The ETW particle trajectory can be obtained from equation (\ref{NNbcdil}). It is useful to parametrize the ETW particle trajectory using the bulk time $\tau$ (i.e. $v=\tau$, $z=z(\tau)$), which implies $n^{etw}_i=\tilde{N}(z'(\tau),-1)$, where again $\tilde{N}$ is a normalization factor. By substituting the expression for the outward-pointing normal unit vector $n^{i}_{etw}=g^{ij}n^{etw}_j$ and the on-shell expression for the dilaton (\ref{onshelldilatonappendix}) into equation (\ref{NNbcdil}), we obtain the ETW particle equation of motion (\ref{braneeom}).

Equation (\ref{braneeom}) cannot be solved analytically. However, the expression of the inversion point can be found explicitly by requiring the square root to vanish:
\begin{widetext}
\begin{equation}
    z_{max}=\frac{dr_eR_2^2T^2_{etw}}{2\left(r_e^2T^2_{etw}+4\pi^2 T^2 R_2^2\right)}\left[-1+\sqrt{1+\frac{\left(r_e^2T^2_{etw}+4\pi^2 T^2 R_2^2\right)\left(4-d^2R_2^2T^2_{etw}\right)}{d^2R_2^2r_e^2T^4_{etw}}}\right].
    \label{minrad}
\end{equation}
\end{widetext}
Note that $z_{max}\to 1/(2\pi T)$ as $T_{etw}\to 0$, i.e. the minimum radius of the ETW particle is the black hole horizon radius, which is in agreement with the higher dimensional result \cite{Antonini:2019qkt}.

In the near-horizon region of our interest, the condition $z\gg R_2^2/r_e$ holds. Since we are only interested in the case where the ETW particle crosses the regularized boundary and enters the near-horizon region, we can use the condition $z(\tau)\gg R_2^2/r_e$ to rewrite equation (\ref{braneeom}) as
\begin{equation}
    z'(\tau)\approx\pm R_2\frac{1-4\pi^2 T^2 z^2}{T_{etw}r_e z}\sqrt{1-4\pi^2T^2z^2-\frac{r_e^2 T_{etw}^2}{R_2^2}z^2}.
    \label{analyticbraneeom}
\end{equation}
Note that we could not drop the $(2\pi T z)^2$ terms, because even if they are small in the region where the regularized boundary sits (determined by equation (\ref{allcond})), they are of order one when the ETW particle is very close to the horizon, i.e. when $z(\tau)\gtrsim 1/(2\pi T)$. This approximate equation of motion does have an analytic solution, which is given by equation (\ref{analyticsol}).

\begin{figure*}
    \centering
    \subfloat[]{
        \includegraphics[width=0.45\linewidth]{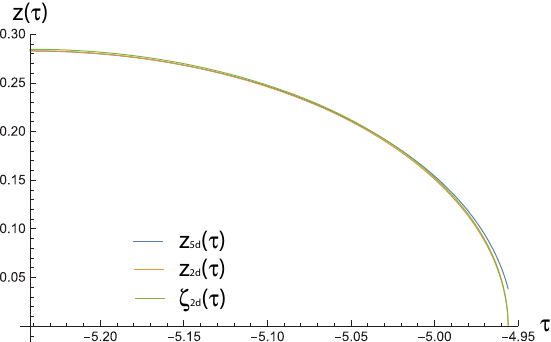}}
    \subfloat[]{
        \includegraphics[width=0.45\linewidth]{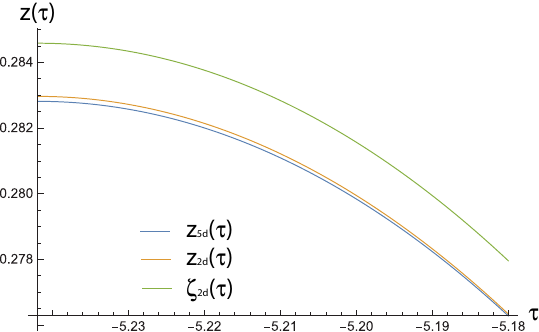}}
   \caption{\textbf{ETW particle and brane trajectories: expansion phase.} $d=4$, $L_{AdS}=1$, $r_+=100$, $r_-=99.9$, $T_{etw}=0.01$. The black hole is near-extremal and the tension is low. The brane enters the near-horizon region ($z(\tau)\gg  R_2^2/r_-\gtrsim R_2^2/r_e$). The exact numerical solution $\zeta_{2d}(\tau)$ of equation (\ref{braneeom}) and the approximate analytic solution $z_{2d}(\tau)$ (given by equation (\ref{analyticsol})) are in good agreement with each other, and with the numerical solution $z_{5d}(\tau)$ of the higher dimensional ETW brane equation of motion \cite{Antonini:2019qkt}. The agreement is particularly good in the near-horizon region (larger $z$), where our dimensionally reduced analysis is sensible, as expected. (a) Expansion phase between $z=z_{max}\approx 0.28$ and the asymptotic boundary at $z=0$. (b) Detail of the near-horizon region.}
    \label{trajectoryfigure}
\end{figure*}

We finally report in Figure \ref{trajectoryfigure} a plot of the expansion phase of the ETW particle trajectory for low tension in a near-extremal black hole for $d=4$, obtained from the approximate analytic solution (\ref{analyticsol}) (indicated by $z_{2d}(\tau)$) and by solving numerically equation (\ref{braneeom}) (indicated by $\zeta_{2d}(\tau)$). The trajectory of the ETW brane obtained by solving numerically the brane equation of motion in the full higher dimensional model \cite{Antonini:2019qkt} is also plotted (indicated by $z_{5d}(\tau)$). We remark how the trajectories are all in good agreement with each other, and particularly so in the near-horizon region where our dimensionally reduced two-dimensional effective action is sensible and approximately equivalent to the full five-dimensional braneworld cosmology action.

\end{appendices}

\bibliography{references}

\end{document}